\documentclass[twocolumn,showpacs,preprintnumbers,amsmath,amssymb]{revtex4-1}

\usepackage{CJK}

\usepackage{graphicx}

\usepackage{dcolumn}

\usepackage{bm}
\usepackage[normalem]{ulem}
\usepackage{color}

\usepackage{enumitem,kantlipsum}
\numberwithin{equation}{section}

\usepackage{epstopdf}

\newcommand{\ii}{{\rm i}}
\newcommand{\bx}{\mathbf{x}}

\newcommand{\bof}{\mathbf{f}}
\newcommand{\by}{\mathbf{y}}

\newcommand{\bq}{\mathbf{q}}
\newcommand{\bQ}{\mathbf{Q}}
\newcommand{\bv}{\mathbf{v}}

\newcommand{\br}{\mathbf{r}}

\newcommand{\bff}{\mathbf{f}}
\newcommand{\bu}{\mathbf{u}}
\newcommand{\bup}{\mathbf{u}_\perp}

\newcommand{\bkp}{\mathbf{k}_\perp}
\newcommand{\bfp}{\mathbf{f}_\perp}
\newcommand{\but}{\mathbf{u}_T}
\newcommand{\fut}{\mathbf{f}_T}
\newcommand{\ul}{u_L}

\newcommand{\bw}{\mathbf{w}}
\newcommand{\bbr}{\mathbf{r}}

\newcommand{\bk}{\mathbf{k}}

\newcommand{\bs}{\mathbf{s}}

\newcommand{\sep}{ \ \ \ , \ \ \ }

\newcommand{\beq}{\begin{equation}}
\newcommand{\eeq}{\end{equation}}
\newcommand{\beqn}{\begin{eqnarray}}
\newcommand{\eeqn}{\end{eqnarray}}
\newcommand{\pp}{\partial}
\newcommand{\dd}{{\rm d}}
\newcommand{\ee}{{\rm e}}

\newcommand{\fig}{Fig.\ }

\newcommand{\cO}{{\cal O}}

\newcommand{\la}{\langle}

\newcommand{\ra}{\rangle}

\newcommand{\vnab}{{\bf \nabla}}

\newcommand{\bew}{\begin{widetext}}
\newcommand{\ew}{\end{widetext}}
\newcommand{\nn}{\nonumber}

\begin{document}

\begin{CJK*}{UTF8}{gbsn}

\title{A novel nonequilibrium state of matter: a $d=4-\epsilon$ expansion study of Malthusian flocks}

\author{Leiming Chen (陈雷鸣)}
\email{leiming@cumt.edu.cn}
\affiliation{School of Physical science and Technology, China University of Mining and Technology, Xuzhou Jiangsu, 221116, P. R. China}
\author{Chiu Fan Lee}
\email{c.lee@imperial.ac.uk}
\affiliation{Department of Bioengineering, Imperial College London, South Kensington Campus, London SW7 2AZ, U.K.}
\author{John Toner}
\email{jjt@uoregon.edu}
\affiliation{Department of Physics and Institute for Fundamental
	Science, University of Oregon, Eugene, OR $97403$}

\begin{abstract}
We show that ``Malthusian flocks" -- i.e., coherently
moving collections of self-propelled entities
(such as living creatures) which are being ``born"
and ``dying" during their motion -- belong to a new
universality class in spatial dimensions $d>2$. We
calculate the universal exponents and scaling laws
of this new universality class to $O(\epsilon)$ in a
$d=4-\epsilon$ expansion, and find these are
different from the ``canonical" exponents previously
conjectured to hold for ``immortal" flocks (i.e., those
without birth and death) and shown to hold for
incompressible flocks with spatial dimensions in the
range of $2 < d \leq 4$. We also obtain a universal
amplitude ratio relating the damping of transverse
and longitudinal velocity and density fluctuations in
these systems. Furthermore, we find a universal
separatrix  in  real ($\br$)  space  between two regions in which
 the equal time density correlation
$\langle\delta\rho(\br, t)\delta\rho(0, t)\rangle$  has
opposite signs. Our expansion should be quite accurate in $d=3$, allowing precise quantitative comparisons between our theory, simulations, and experiments.
\end{abstract}
\maketitle
\end{CJK*}

\section{Introduction\label{intro}}

``Active matter", loosely defined as systems whose constituents have internal energy sources which drive motion, has been receiving intense attention in the physics community \cite{Active1,Active2,Active3,Active4}.
While one obvious motivation for this interest is its direct relevance to non-equilibrium physics and biophysics, active matter is also interesting because it exhibits a number of unusual phenomena. Among these is its ability to develop long-ranged orientational order in spatial
dimension $d=2$ \cite{Vicsek,TT1,Chate1,Chate2}, and the  ``anomalous hydrodynamics" exhibited by many of its ordered phases \cite{TT1,TT3,birdrev} even in spatial dimensions $d>2$. By ``anomalous hydrodynamics" we mean that the long-wavelength, long-time behavior of these systems can {\it not} be accurately described by a linear theory; instead, non-linear interactions between fluctuations must be taken into account, even to get the correct scaling laws. Indeed, it is the anomalous hydrodynamics in $d=2$ that makes the existence of long-ranged order possible \cite{TT1,TT3,birdrev}.

Of course, in addition to making active matter interesting, these intrinsically non-linear phenomena also makes it extremely difficult to treat analytically. How non-linear active matter is depends primarily on the symmetry of the state it is in, or, to borrow the language of equilibrium condensed matter physics, what ``phase" it is in. The most non-linear phase found so far is what is known as the ``polar ordered fluid" phase, which we will hereafter sometimes refer to as a ``flock". This is a phase of active (i.e., self-propelled) particles in which the only order is the alignment of the particles' directions of motion, which breaks rotation invariance. Rotation invariance is ``broken" because we consider systems whose underlying dynamics {\it is} rotation invariant. This aligning of the particles' motion is the ``polar order" of the phase's name; the absence of other types of order (in particular, {\it translational} order) is the reason we describe this phase as ``fluid".

As always, the hydrodynamic (i.e., long length and time scale) behavior of polar ordered active fluids   is determined by the symmetries and conservation laws of the system. Here, symmetries include not only the symmetries of the underlying microscopic dynamics, but the symmetries of the {\it state} as well. In particular, this means it depends on which symmetries are {\it broken} in the ordered state.  Again, in polar ordered active fluids, the broken symmetry is rotational invariance.

Much of the past work \cite{TT1,TT3,birdrev} on polar ordered active fluids  has focused on systems without momentum conservation, as is appropriate for active particles moving over a frictional substrate which can act as a momentum sink, but with number conservation. Hereafter we call these systems ``immortal flocks". For such systems, the density local number density $\rho$ of ``flockers" (i.e., self-propelled particles) is a hydrodynamic variable. This considerably complicates the hydrodynamic theory; in particular, it gives rise to six additional relevant non-linearities \cite{NL}, rendering the problem effectively intractable. All we know with any certainty about these systems is that they exhibit anomalous hydrodynamics in all spatial dimensions $d\le4$, and that this anomaly stabilizes long-ranged orientational order (or, equivalently, makes it possible for an arbitrarily large flock to have a non-zero average velocity) in $d=2$.  A plausible but unproven conjecture \cite{NL} makes it possible to obtain exact scaling exponents characterizing the long-distance, long time scaling behavior for this system in $d=2$. In other dimensions, in particular $d=3$, little beyond the existence of anomalous hydrodynamics can be said.

 Interestingly, one system about which more can be said is incompressible flocks  \cite{chen_njp_2018, chen_nc_2016}; i.e., polar ordered active fluids  in which the density is fixed, either by an infinitely stiff
equation of state, or by long-ranged forces. For these systems, it is possible to obtain exact exponents for all spatial dimensions; as for number conserving systems with density fluctuations, these prove to be anomalous for spatial dimensions $d$ in the range $2\le d\le4$. Specifically, there are three universal exponents characterizing the hydrodynamic behavior of these systems. One is  the ``dynamical exponent" $z$, which gives the scaling of hydrodynamic time scales $t(L_\perp)$ with length scale $L_\perp$ perpendicular to the mean direction of flock motion (i.e., the direction of the average velocity $\left<\bv\right>$); that is, $t(L_\perp)\propto L_\perp^z$. Likewise, the scaling of characteristic hydrodynamic length scales $L_\parallel$ {\it along} the direction of flock motion scale with those $L_\perp$ perpendicular to that direction is given by an ``anisotropy exponent" $\zeta$ defined via  $L_\parallel(L_\perp)\propto L_\perp^\zeta$. Finally,  fluctuations $\bu_\perp$ of the local velocity perpendicular to its mean direction define a ``roughness exponent" $\chi$ via   $\bu_\perp\propto L_\perp^\chi$. For incompressible flocks, these exponents are given by
\beq
z={2(d+1)\over5} \,\,, \,\,\, \zeta={d+1\over5}\,\,\,,\,\, \chi={3-2d\over5} \,\,,
\label{canon}
\eeq
for spatial dimensions satisfying $2  < d\le4$, by $z=2$, $\zeta=1$, and $\chi={2-d\over2}$ for $d>4$ (the latter range of $d$   is obviously only of interest for simulations).  For $d=2$, the static properties of the ordered phase can be mapped onto the (1+1)-dimensional Kardar-Parisi-Zhang model \cite{chen_nc_2016} and the exact scaling exponents are
$\zeta=2/3$ and $\chi=-1/3$, while the value of $z$ remains unknown.
We will hereafter refer to the exponents (\ref{canon}) as the ``canonical" exponents.

The exponents (\ref{canon}) were originally asserted \cite{TT1,TT3} to hold for compressible, number conserving flocks, but this was later shown to be incorrect \cite{NL}, due to the presence of  the aforementioned extra non-linearities associated with the conserved density. If one conjectures that those extra non-linearities, which are relevant near the unstable linear fixed point near $d=4$, are in fact {\it irrelevant} near the non-linear fixed point that controls the ordered phase in $d=2$, then one obtains the ``canonical" values
 \beq
z={6\over5} \,\,, \,\,\, \zeta={3\over5}\,\,\,,\,\, \chi=-{1\over5} \,\,,
\label{canond=2}
\eeq

In this paper, we will consider so-called  ``Malthusian flocks"\cite{Malthus}; that is,  polar ordered active fluids  with no conservation laws at all; in particular, {\it particle number} is {\it not} conserved. Such systems are readily experimentally realizable in experiments on a, e.g.,
growing bacteria colonies and cell tissues, and ``treadmilling''
molecular motor propelled biological macromolecules in
a variety of intracellular structures, including the cytoskleton,
and mitotic spindles, in which molecules are
being created and destroyed as they move.

In addition to describing biological and other active systems, our model for Malthusian flocks  may also be viewed as  a generic non-equilibrium $d$-dimensional $d$-component spin model  in which the spin vector space $\bs(\br)$ and the coordinate
space $\br$ are treated on an equal footing, and couplings between the two are allowed. In particular, terms like $\bs \cdot \nabla \bs$ and $(\nabla \cdot \bs)\bs$,
will be present in the EOM that describes such a generic non-equilibrium system. As a result, the fluctuations in the system can propagate spatially in a spin-direction-dependent manner, but the spins themselves are not moving. Therefore, there are no density fluctuations and the only hydrodynamic variable is the spin field,
the equation of motion (EOM) for which is exactly the same as the one we derive here for a Malthusian flock, with spin playing the role of the velocity field. (Of course, non-equilibrium spin systems in which the spins ``live" in real space in the sense described here will {\it not} map onto Malthusian flocks if those spins live on a lattice, due to the breaking of rotation invariance by the lattice itself. There {\it are}, however, ways of eliminating these ``crystal field" effects \cite{lattice}.)

 For Malthusian flocks, exact exponents can be obtained in $d=2$ \cite{Malthus}, and they again take on the  ``canonical" values (\ref{canond=2}) implied by (\ref{canon}) in $d=2$.

 Overall, the theoretical situation is therefore still quite unsatisfactory: we only have the scaling laws for
 flocks if they either are incompressible (which requires either infinitely strong, or infinitely ranged, interactions), or in $d=2$.
And in the cases in which we {\it do} know the exponents, their values are either
the canonical ones (\ref{canon}) \cite{Malthus,chen_njp_2018}, or those from the (1+1)-dimensional KPZ model \cite{chen_nc_2016}.

It would clearly be desirable to find the scaling laws and exponents of some  compressible three dimensional flocks, and to see if, as for incompressible flocks, they are also given by the canonical values (\ref{canon}).

In this paper, we do so for Malthusian flocks in $d>2$. Specifically, we study these systems in a $d=4-\epsilon$ expansion. We find that they belong to a new universality class which does {\it not} have the canonical exponents (\ref{canon}). Instead, we find, to leading order in $\epsilon$,
\beqn
&&z = 2 -  \frac{6 \epsilon}{11} +\cO(\epsilon^2) \,,\\
&&\zeta  = 1 -  \frac{3 \epsilon}{11} +\cO(\epsilon^2)   \,,\\
&&\chi = -1 + \frac{6 \epsilon}{11} +\cO(\epsilon^2)\, ,
\label{eps}
\eeqn
which the interested reader can easily check do {\it not} agree with the ``canonical" values (\ref{canon}) near $d=4$ (i.e., for small $\epsilon$). However, it should also be noted
that in $d=3$ these values are not very different from the canonical values (\ref{canon}); setting $\epsilon=1$ in  (\ref{eps}) gives $z=16/11=1.45454.....$, $\zeta=8/11= 0.72727272.....$ and $\chi=-5/11=- 0.45454....$, which are fairly close to the ``canonical" values (\ref{canon}), which give, in $d=3$,
 $z=8/5=1.6$, $\zeta=4/5= 0.8$ and $\chi=-3/5=- 0.6$.

 We have also estimated the exponents in $d=3$ by applying the one-loop (i.e., lowest order in perturbation theory) perturbative renormalization group recursion relations in arbitrary spatial dimensions. This approach, although strictly speaking an uncontrolled approximation, can easily be shown to give exponents for the $\cO(n)$ model critical point in $d=3$ that are at least as accurate as the first order in $d=4-\epsilon$  expansion with $\epsilon$ set to $1$.

And there is reason to believe that this approach may be even more accurate for our problem: this ``one-loop truncated" approach not only recovers the exact linear order in $\epsilon$ expansion results (\ref{eps}), but it also recovers the exact results (\ref{canond=2}) in $d=2$. Thus, while uncontrolled, this approach should provide a very effective interpolation formula for $d$ between $2$ and $4$, that should be quite accurate (indeed, probably more accurate than the $\epsilon$ expansion) in $d=3$.

Using this approach, we find
\beqn
\label{uncz}
&&z = 2 -  \frac{2(4-d)(4d-7)}{14d-23} \,,\\
\label{unczeta}
&&\zeta  = 1 -  \frac{(4-d)(4d-7)}{14d-23}  \,,\\
&&\chi = -1 + \frac{2(4-d)(4d-7)}{14d-23}\, ,
\label{uncchi}
\eeqn
which indeed recover our $\epsilon$ expansion results near $d=4$, and the exact results (\ref{canond=2}) in $d=2$, as the readers can verify for themselves.

In the physically interesting case $d=3$, these give
\beqn
\label{unczd=3}
&&z = {28\over19}\approx1.47 \,,\\
\label{unczetad=3}
&&\zeta  = {14\over19}\approx0.74  \,,\\
&&\chi = -{9\over19}\approx-0.47\, .
\label{uncchid=3}
\eeqn
These are our best numerical estimates of the values of these exponents in $d=3$. We suspect that they are accurate to $\pm1\%$, an error estimate which we will motivate in section \ref{sec:beyondepsilon} below. That is, the digits shown after the approximate equalities above are probably all correct.

These exponents
 govern the scaling behavior of the  experimentally measurable velocity correlation function:
 \bew
 \begin{eqnarray}
C_u(\br,t)&\equiv&\langle\bu_{\perp}(\br,t)\cdot\bu_{\perp}({\mathbf{0},0})\rangle=r_{\perp}^{2\chi}F_u\left({(|x-\gamma t|/\xi_x)\over (r_{\perp}/\xi_\perp)^{\zeta}},{(t/\tau)\over(r_{\perp}/\xi_\perp)^z}\right)\nonumber\\\nonumber\\
&\propto&\left\{
\begin{array}{ll}
r_{\perp}^{2\chi}\sep&(r_{\perp}/\xi_\perp)^{\zeta}\gg|x-\gamma t|/\xi_x\sep (r_{\perp}/\xi_\perp)^z\gg(|t|/\tau)\\\\
|x-\gamma t|^{2\chi\over\zeta}\sep&|x-\gamma t|/\xi_x\gg (r_{\perp}/\xi_\perp)^{\zeta}\sep|x-\gamma t|/\xi_x\gg (|t|/\tau)^{\zeta\over z}\\\\
|t|^{2\chi\over z}\sep&(|t|/\tau)\gg (r_{\perp}/\xi_\perp)^z\sep(|t|/\tau)\gg(|x-\gamma t|/\xi_x)^{z\over\zeta} \,,
\end{array}
\right.
\label{eq:Cu_scaling}
\end{eqnarray}
\ew
where $F_u$ is a universal scaling function (i.e., the same for all Malthusian flocks), $\gamma$ is a non-universal (i.e.,  system dependent) speed, $\xi_{\perp,x}$ are non-universal lengths, and $\tau$ is a non-universal time. We also note that fluctuations of the velocity field $\bu_\perp$ are always positively correlated, i.e., $C_u$ is always positive.

Density correlations also obey a scaling law involving the same universal exponents $z$, $\zeta$, and $\chi$, and non-universal lengths $\xi_{\perp,x}$ and time $\tau$:
 \bew
 \begin{eqnarray}
C_\rho(\br,t)&\equiv&\langle\delta\rho(\br,t)\delta\rho({\mathbf{0},0})\rangle=r_{\perp}^{2(\chi-1)}F_\rho\left({(|x-\gamma t|/\xi_x)\over (r_{\perp}/\xi_\perp)^{\zeta}},{(t/\tau)\over(r_{\perp}/\xi_\perp)^z}\right)\nonumber\\
&\propto&\left\{
\begin{array}{ll}
r_{\perp}^{2(\chi-1)}\sep&(r_{\perp}/\xi_\perp)^{\zeta}\gg|x-\gamma t|/\xi_x\sep (r_{\perp}/\xi_\perp)^z\gg(|t|/\tau)\\\\
|x-\gamma t|^{2(\chi-1)\over\zeta}\sep&|x-\gamma t|/\xi_x\gg (r_{\perp}/\xi_\perp)^{\zeta}\sep|x-\gamma t|/\xi_x\gg (|t|/\tau)^{\zeta\over z}\\\\
|t|^{2(\chi-1)\over z}\sep&(|t|/\tau)\gg (r_{\perp}/\xi_\perp)^z\sep(|t|/\tau)\gg(|x-\gamma t|/\xi_x)^{z\over\zeta} \,,
\end{array}
\right.
\label{rhoscale}
\end{eqnarray}
\ew
where we've defined $\delta\rho(\br,t)\equiv\rho(\br,t)-\rho_0$, with $\rho_0$ the mean density.

\begin{figure}
	\begin{center}
		\includegraphics[scale=.35]{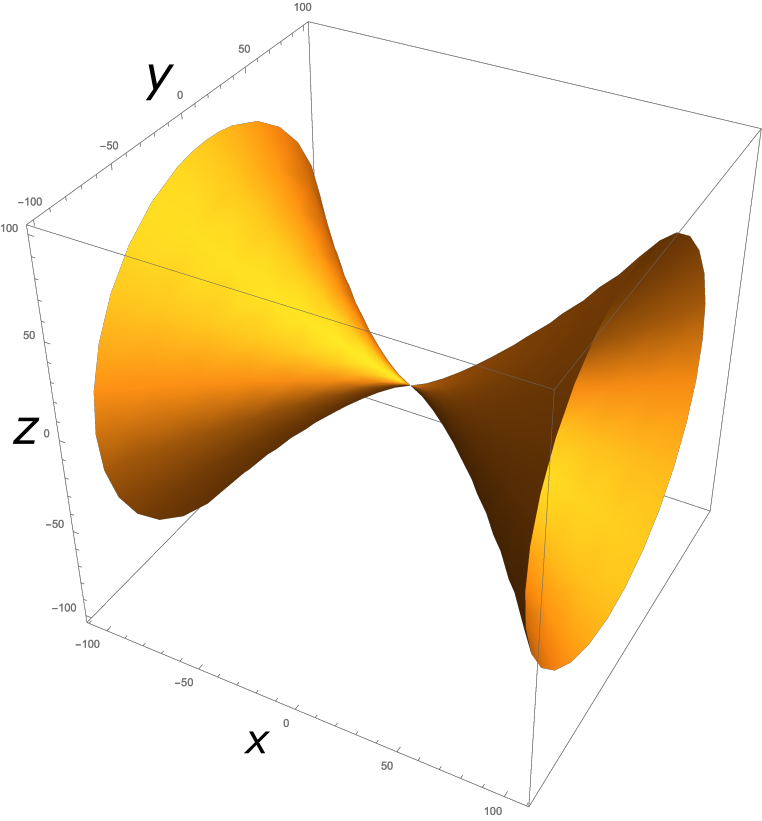}
	\end{center}
	\caption{In $d=3$, the equal time density correlations are
	positive inside the trumpet-shaped surface  defined by
	(\ref{Separ}).
	and negative outside it.}
	\label{fig:rho_correlation}
\end{figure}

{In contrast to the velocity-velocity correlation function,
$C_\rho$ can be positive or negative. Indeed, the
equal-time density correlation
is positive when $|x| / \xi_x > r_\perp/\xi_\perp$
for $|x|<\xi_x$ or $|x| / \xi_x > \left(r_\perp/\xi_\perp\right)^\zeta$ for $|x|<\xi_x$, and negative otherwise. Therefore, there is a separatrix
on which the equal time density correlations are exactly
zero (Fig.~\ref{fig:rho_correlation}). This separatrix is  given by
\beqn
{x\over\xi_x}
=\left\{
\begin{array}{ll}
\frac{r_\perp}{\xi_\perp}\sep&x<\xi_x\\\\
\left(\frac{r_\perp}{\xi_\perp}\right)^\zeta\sep&x>\xi_x
\end{array}
\right.
\label{Separ}
\eeqn
}

The remainder of this paper is organized as follows:
in section II, we review the derivation in \cite{Malthus} of the hydrodynamic EOM for Malthusian flocks. In section III, we develop the linearized theory of these equations. In section IV, we study non-linear effects using the dynamical renormalization group (DRG), and obtain the fixed points and  scaling laws governing the ordered phase. We also obtain a universal amplitude ratio. Section V summarizes our results and discusses their implications for experiments and simulations.
In  Appendix A, we present a simple DRG analysis that confirms the existence of the fixed point found in our more general treatment. Appendix B
presents the lengthy and arduous details of the full DRG calculation, which shows that the fixed point found by the simplified analysis is the {\it only} stable fixed point for this problem, at least to one-loop order.  We have also provided a list of useful formulae in Appendix C.

\section{Derivation of the equation of motion\label{EOM}}

We begin by deriving the equation of motion. This derivation is virtually identical to that done in reference \cite{Malthus}; we review it here simply to make this paper self-contained.
Our starting equation of motion for the velocity is identical to that of a  flock with number conservation \cite{TT1,TT3,birdrev, NL}:
\begin{eqnarray}
&&\partial_{t}
\bv+\lambda_1 (\bv\cdot\vnab)\bv+
\lambda_2 (\vnab\cdot\bv)\bv
+\lambda_3\vnab(|\bv|^2)
 =\nonumber \\&&
U(\rho, |\bv|)\bv -\vnab P_1 -\bv
\left( \bv \cdot \vnab  P_2 (\rho,|\bv|) \right)\nonumber \\&&+\mu_{B}\vnab
(\vnab
\cdot \bv) + \mu_{T}\nabla^{2}\bv +
\mu_{A}(\bv\cdot\vnab)^{2}\bv+\bof \,.
\label{vEOM}
\end{eqnarray}
In this equation,
$\lambda_i (i = 1 \to 3)$,
$U$, $\mu_{B,T,A}$ and the  ``pressures'' $P_{1,2}(\rho,
|\bv|)$ are, in general, functions of the flocker number density $\rho$ and the magnitude $|\bv|$ of the local velocity.
We will expand all of them to the order necessary to include all terms that are ``relevant" in the  sense of changing the long-distance behavior of the flock. 

This equation is derived purely from symmetry arguments \cite{TT1, TT2, NL}. However, each term in it has a simple physical interpretation, which we now give.

The $U(\rho, |\bv|)$ term is responsible for spontaneous flock motion. Our analysis will apply to an extremely large class of $U$'s; specifically, to all of those that satisfy $U(|\bv|<v_0)>0$, and  $U(|\bv|>v_0)< 0$ in the ordered phase. This last condition insures that in the absence of fluctuations, the flock will move at a speed $v_0$.

The diffusion constants $\mu_{B,T,A}$  reflect the tendency of  flockers to follow their neighbors.
The $\bof$ term is a random Gaussian white noise, reflecting  errors made by the flockers, with correlations:
\begin{eqnarray}
 \la f_{i}(\br,t)f_{j}(\br',t') \ra=2D
\delta_{ij}\delta^{d}(\br-\br')\delta(t-t')
\label{white noise}
\end{eqnarray}
where the noise strength $D$ is a constant hydrodynamic parameter (analogous to the temperature in an equilibrium system, as it sets the scale of fluctuations), and $i , j$ label
vector components.
The ``anisotropic pressure'' $P_2(\rho, |\bv|)$ in
(\ref{vEOM}) is only allowed due to the non-equilibrium nature of the
flock; in an equilibrium fluid such a term is forbidden by Pascal's
Law. This term reflects the fact that, once the system locally breaks rotation invariance by choosing a direction for the velocity $\bv$, there is no reason in an out of equilibrium system that the response of the system to a density gradient along the direction of flock motion need be identical to the response perpendicular to that direction.

Note that  (\ref{vEOM}) is {\it not}
Galilean invariant;  it holds only in the frame of the fixed medium through or on which  the creatures move, which we assume remains fixed. Situations in which the background medium is itself a fluid  which can flow (which are now referred to as "wet active matter") have been studied elsewhere \cite{Active1,Active2,Active3,Active4}.

We turn now to the EOM for $\rho$. In immortal flocks, this is just  the usual continuity equation of compressible fluid dynamics. For Malthusian flocks, the equation needs an additional term representing  the effects of birth and
death. As first noted by Malthus \cite{Malthus_1789}, {\it any}  collection of entities that is reproducing and dying can only reach a non-zero steady state population density if the death rate exceeds the birth rate for population densities greater than the steady state density, and the converse for population densities less than the steady state density \cite{Malthus_1789}. This ``Malthusian" condition implies that the net, local growth rate of number density in the absence of motion,
which we'll call $\kappa(\rho)$, which is just the local birth rate per unit volume minus the local death rate (also per unit volume), vanishes at some fixed point density $\rho_0$,
with larger densities decreasing (i.e., $\kappa(\rho > \rho_0) < 0)$, and smaller
densities increasing (i.e., $\kappa(\rho < \rho_0) > 0)$.

The EOM for the density is now simply:
\begin{eqnarray}
\partial_t\rho +\vnab\cdot(\bv\rho)=\kappa(\rho)~~.
\label{conservation}
\end{eqnarray}
Note that in the absence of birth and death, $\kappa(\rho) = 0$, and equation
(\ref{conservation}) reduces to the usual continuity equation, as it should, since
``flocker number" is then conserved.

Since birth and death quickly restore the
fixed point density $\rho_0$, departures of $\rho$ from $\rho_0$ are no longer hysdrodynamic variables (since a hydrodynamic variable is, by definition, slow). It can therefore, like all non-hydrodynamic variables, be expressed, at long time scales, as a purley local (in both space {\it and} time) function of the truly hydrodynamic varaiables (in our case, the velocity). To show this explicitly,  we will
write $\rho(\br, t)
= \rho_0 + \delta\rho(\br, t)$ and expand both sides of equation
(\ref{conservation}) to leading  order in $\delta\rho$. This gives
$\rho_0\vnab \cdot \bv \cong \kappa' (\rho_0)\delta\rho ,
$ where we've dropped the $\partial_t\rho$ term relative to the $\kappa' (\rho_0)\delta\rho$ term since we're interested in the hydrodynamic limit, in which the fields evolve extremely slowly.
This equation can be readily solved to give
\begin{eqnarray}
\delta\rho \cong {\rho_0 \vnab \cdot \bv \over  \kappa'
(\rho_0)} \equiv - {\Delta \mu_B\over \sigma}(\vnab
\cdot \bv) \label{rho-v}
\end{eqnarray}
where $\Delta \mu_B$ is a positive constant (since $\kappa'(\rho_0)<0$, because $\kappa(\rho > \rho_0) < 0$ and $ \kappa(\rho < \rho_0) > 0$) ,  and $\sigma$ (which must be positive for stability) is  the first expansion coefficient  for $P_1$ (i.e., the analog of the inverse compressibility in an equilibrium system).
We can now insert this solution (\ref{rho-v}) for $\delta\rho$ in terms of $\bv$ into the isotropic pressure $P_1$; the resulting EOM for $\bv$ is:
\begin{widetext}
\begin{eqnarray}
\partial_{t}
\bv+\lambda_1 (\bv\cdot\vnab)\bv+
\lambda_2 (\vnab\cdot\bv)\bv
+\lambda_3\vnab(|\bv|^2)
 &=&\nonumber
U(\rho, |\bv|)\bv -\bv
\left( \bv \cdot \vnab  P_2 (\rho,|\bv|) \right)\nonumber +\mu'_{B} \vnab
(\vnab
\cdot \bv) \\&&+ \mu_{T}\nabla^{2}\bv +
\mu_{A}(\bv\cdot\vnab)^{2}\bv+\bof~,
\label{vEOM2}
\end{eqnarray}
\end{widetext}
where we've defined $\mu'_B\equiv \mu_B+\Delta \mu_B$.

In the ordered state (i.e.,  in which $\left<\bv (\br, t) \right>= v_0 {\bf \hat x}$, where we've chosen the spontaneously picked direction of mean flock motion as our $x$-axis),
we can expand the $\bv$ EOM for small departures $ \bu (\br,t) \equiv u_x {\bf \hat x} + \bu_{\perp}(\br,t) $
of $\bv(\br,t) $ from uniform motion with speed $v_0$:
\begin{eqnarray}
\bv (\br, t) = (v_0+u_x)  {\bf \hat x} + \bu_{\perp}(\br,t) ~~,
\label{6}
\end{eqnarray}
where, henceforth  $x$ and  $\perp$ denote components along and perpendicular to the mean velocity, respectively.

In this hydrodynamic approach,
we're  interested only in fluctuations of $\bu(\br, t)$
that vary slowly in space and time. The component $u_x$ of the fluctuation of the velocity
{\it along} the direction of mean motion is {\it not} such a fluctuation. Rather, like the density fluctuation $\delta\rho$, it is a non-hydrodynamic or ``fast" variable. It therefore can be eliminated from the equations of motion in much the same manner as we just eliminated the density fluctuations.

The details of this elimination are a bit tricky, and are discussed in detail in \cite{NL}); here we will very briefly review the argument, as applied to our EOM (\ref{vEOM2}).

To focus on fluctuations in the magnitude of the velocity (which are, strictly speaking, the fast variable here, we
take the dot product of both sides of (\ref{vEOM2}) with $\bv$ itself. This gives
\begin{widetext}
\begin{eqnarray}
{1\over 2}\left(\partial_{t}|\bv|^2+(\lambda_1+2\lambda_3)(\bv\cdot\vnab)|\bv|^2\right) + \lambda_2(\vnab\cdot\bv)|\bv|^2&=& U(|\bv|)|\bv|^{2}-|\bv|^{2}\bv \cdot \vnab  P_2 +\mu'_B\bv\cdot\vnab
(\vnab\cdot \bv)
\nonumber \\&&+
\mu_{T}\bv\cdot\nabla^{2}\bv +\mu_A\bv\cdot\left((\bv\cdot\vnab)^{2}\bv\right)+\bv\cdot\bof
\label{v parallel elim}~.
\end{eqnarray}
\end{widetext}

In this hydrodynamic approach,
we're  interested only in fluctuations $\bu_\perp(\br, t)$ and $\delta \rho(\br, t)$
that vary slowly in space and time.
Hence, terms involving spatiotemporal derivatives of
$\bu_\perp(\br, t)$ and $\delta \rho(\br, t)$
are always negligible, in the hydrodynamic limit, compared to terms involving the same number of powers of fields with fewer spatiotemporal derivatives. Furthermore, the fluctuations
$\bu_\perp(\br, t)$ and $\delta \rho(\br, t)$ can themselves be shown to be small in the long-wavelength limit. Hence, we need only keep terms in (\ref{v parallel elim}) up to linear order in
$\bu_{\perp}(\br, t)$ and $\delta \rho(\br, t)$. The
$\bv\cdot\bof$ term can likewise be dropped.

These observations can be used to eliminate many  terms in equation (\ref{v parallel elim}), and solve for the quantity $U$; we obtain:
$U=\lambda_2 \vnab\cdot\bv
+\bv\cdot\vnab P_2$.
 Inserting this expression for $U$ back into equation (\ref{vEOM2}),
we find that $P_2$ and $\lambda_2$ cancel out of the $\bv$ EOM, leaving, ignoring irrelevant terms:
\begin{eqnarray}&&\partial_{t}
\bv+\lambda_1(\bv\cdot\vnab)\bv+\lambda_3 \vnab(|\bv|^2) =\mu_T\nabla^2 \bv
\nonumber \\&&
+\mu'_{B} \vnab(\vnab\cdot \bv) +
\mu_{A}(\bv\cdot\vnab)^{2}\bv
 + \bof ,
\label{vperpEOM}
\end{eqnarray}
This can be made into an EOM for $\bu_\perp$ involving only $\bu_\perp(\br, t)$
itself by projecting perpendicular to the direction of mean flock motion $\hat{\bx}$, 
and eliminating $u_x$ using $U=\lambda_2 \vnab\cdot\bv
+\bv\cdot\vnab P_2$ and
the expansion $
U\approx-\Gamma_1  \left(u_x +{u_\perp^2\over 2v_0}\right)- \Gamma_2 \delta \rho$,
where we've defined $\Gamma_1 \equiv -\left({\partial U
 \over \partial |\bv|}\right)_{\rho, 0}$ and
$\Gamma_2 \equiv - \left({\partial U
 \over \partial \rho}\right)_{|\bv|,  0}$, with
subscripts $0$ denoting functions of  $\rho$ and
$|\bv|$ evaluated at $\rho = \rho_0$ and $ |\bv|=v_0$. Doing this, and using (\ref{rho-v}) for $\rho$, we obtain:
\beqn
\nonumber
&&\pp_t\bu_\perp +\gamma\partial_x\bu_\perp+\lambda (\bu_\perp \cdot \nabla_\perp)\bu_\perp = \\
&&\mu_1 \nabla^2_\perp \bu_\perp + \mu_2 \nabla_\perp (\nabla_\perp \cdot \bu_\perp)
+\mu_x \pp_x^2 \bu_\perp +\bff_\perp
\ ,
\label{eq:unboosted}
\eeqn
where  we've defined  $\lambda\equiv\lambda_1^0$, $\gamma\equiv\lambda_1^0 v_0$,
$\mu_2\equiv \mu'^{0}_B+2v_0\lambda_3^0(\lambda_2^0-\Gamma_2\Delta \mu_B/\sigma)
/\Gamma_1$ , $\mu_x\equiv \mu_T^0+\mu_A^0 v_0^2$,
and $\mu_1 \equiv\mu^0_{T}$, where the superscripts $0$ denote coefficients evaluated at $\rho = \rho_0$ and $ |\bv|=v_0$. In writing (\ref{eq:unboosted}) we have ignored irrelevant terms which comes from the higher order expansion of the coefficients in $\delta\rho$ and $u_x+{u_\perp^2\over 2v_0}$ than the zeroth order.

Changing co-ordinates to a new Galilean frame $\br'$
 moving with respect to our original frame
(which, we remind the reader, is that of the fixed background medium through which the flock moves)
in the direction ${\bf \hat{x}}$ of mean flock motion at speed $\gamma$ -- i.e.,
\beq
\br' \equiv \br-\gamma t  {\bf \hat x} \, ,
\label{boost}
\eeq
we obtain
\beqn
\nonumber
\pp_t\bu_\perp +\lambda (\bu_\perp \cdot \nabla_\perp)\bu_\perp &=& \mu_1 \nabla^2_\perp \bu_\perp + \mu_2 \nabla_\perp (\nabla_\perp \cdot \bu_\perp)
\\
&&+\mu_x \pp_x^2 \bu_\perp +\bff_\perp
\ ,
\label{eq:main}
\eeqn
where we have dropped the prime in $\bbr$.

 This equation will be the basis of our remaining theoretical analysis. Note that to obtain correlations in the original (unboosted) coordinate system, we need to take into account the boost (\ref{boost}).

\section{Linear theory\label{lin}}

\subsection{Response functions}

In this section we treat the linear approximation to the model (\ref{eq:main}). Keeping only the linear terms in (\ref{eq:main}), and writing the resultant EOM in Fourier space, we obtain
\beqn
-\ii\omega \bu_\perp(\tilde{\bk})&=&-\mu_1 k^2_\perp \bu_\perp(\tilde{\bk}) - \mu_2\bk_\perp \left( \bk_\perp \cdot \bu_\perp(\tilde{\bk})\right)\nonumber\\ &&-\mu_xk_x^2\bu_\perp(\tilde{\bk}) +\bff_\perp(\tilde{\bk})
\ .
\label{EOMlinFT}
\eeqn
where $\tilde{\bk}\equiv (\bk, \omega)$, and
\begin{eqnarray}
\bu_\perp(\tilde{\bk})={1\over \left(\sqrt{2\pi}\right)^{d+1}}
\int \dd t\, \dd^d r\, \bu_\perp(\br,t)\ee^{\ii\left(\omega t-\bk\cdot\br\right)}\,.~~~~
\end{eqnarray}
	
The linear equation (\ref{EOMlinFT}) can be easily solved  by separating  $\bup$ into its component  along $\bkp$ (which we'll hereafter call ``longitudinal") and  its remaining $d-2$ components perpendicular to $\bkp$ (which we'll hereafter call ``transverse").  (We remind the reader that $\bup$ has only $d-1$ independent components, since it is by definition orthogonal to the mean direction of flock motion $\hat{\bx}$.)

That is, we write:
\beq
\bu_\perp(\tilde{\bk})= u_L(\tilde{\bk})\hat{\bk}_\perp  +\bu_T(\tilde{\bk})
\ ,
\label{uLTdecomp}
\eeq
with $\bkp\cdot\bu_T=0$ by definition.
These components $u_L$ and $\bu_T$ can be computed using
\beq
\ul(\tilde{\bk})=\hat{\bk}_\perp \cdot\bup(\tilde{\bk})
\label{uldef}
\eeq
and
\beq
u^T_i(\tilde{\bk})=P^\perp_{ij}(\bk)u^\perp_{j}(\tilde{\bk}) \ ,
\label{utdef}
\eeq
where we've defined the ``transverse projection operator"
\beq
P_{ij}^\perp(\bk)\equiv\delta^\perp_{ij} -\frac{k^\perp_{i}k^\perp_{j}}{k_\perp^2},
\label{Pdef}
\eeq
which projects any vector into the $(d-2)$-dimensional space orthogonal to both the direction of mean flock motion $\hat{\bx}$ 
and $\bkp$.
We can decompose {\it any} vector in the space orthogonal to $\hat{\bx}$, 
including, in particular,  the random force $\bfp$, in exactly the same way.

We can now easily rewrite the EOM (\ref{EOMlinFT}) for $\bup$ as decoupled equations for
$\ul$ and $\but$. To obtain the former, we take  the dot product of $\hat{\bk}_\perp$ 
 with (\ref{EOMlinFT}); this gives a closed EOM for $\ul$:
\beq
-\ii\omega \bu_L(\tilde{\bk}) =- \mu_L k_\perp^2 \bu_L(\tilde{\bk}) -\mu_x k_x^2 \bu_L(\tilde{\bk})+ \bff_L(\tilde{\bk})
\ ,
\label{uLEOMlin}
\eeq
where we have defined
\beq
\mu_L\equiv\mu_1+\mu_2 \ .
\label{muLdef}
\eeq
Likewise, acting on both sides of  (\ref{EOMlinFT}) with the transverse projection operator (\ref{Pdef}) gives a closed EOM for $\but$:

\beq
-\ii\omega \but (\tilde{\bk})= -\mu_1 k_\perp^2 \bu_T(\tilde{\bk}) -\mu_x k_x^2 \bu_T(\tilde{\bk})+ \bff_T(\tilde{\bk})
\ .
\label{uTEOMlin}
\eeq

Before proceeding to solve these two simple linear equations for $\ul$ and $\but$ in terms of  the forces $f_L$ and $\fut$, it is informative to first determine the eigenfrequencies $\omega(\bk)$ of the normal modes of this system.
These are clearly just
\beq
\omega_L(\bk)=-\ii\left(\mu_L k_\perp^2 +\mu_x k_x^2\right)
\label{omegaL}
\eeq
for the longitudinal mode, and
\beq
\omega_T(\bk)=-\ii\left(\mu_1 k_\perp^2 +\mu_x k_x^2\right)
\
\label{omegaT}
\eeq
for the transverse mode.
In order for the system to be stable, we must have the imaginary part $I_{L,T}(\omega(\bk))<0$ for both modes; this clearly requires that
\beq
\mu_{L,1,x}>0
\ .
\label{stability1}
\eeq
Note that this condition (\ref{stability1}) does {\it not} require $\mu_2>0$; using the definition (\ref{muLdef}) of $\mu_L$ in (\ref{stability1}) requires only that
\beq
\mu_2>-\mu_1
\ ,
\label{stability2}
\eeq
or, equivalently,
\beq
{\mu_2\over\mu_1}>-1
\ .
\label{stability3}
\eeq
This last condition for stability was noted in the associated short paper \cite{short}.

Now we turn to the solutions of the EOMs (\ref{uLEOMlin}) and (\ref{uTEOMlin}). These can be immediately read off:
\beqn
\ul(\tilde{\bk})=G_L(\tilde{\bk})f_L(\tilde{\bk}) \ ,\label{usol}
\\
\but(\tilde{\bk})=G_T(\tilde{\bk})\fut(\tilde{\bk}) \ ,
\label{usot}
\eeqn
where
we've defined the longitudinal and transverse ``propagators"
\beqn
G_L(\tilde{\bk}) &=& \frac{1}{-\ii \omega +\mu_L k_\perp^2 +\mu_x k_x^2} \ ,
\\
G_T(\tilde{\bk}) &=& \frac{1}{-\ii \omega +\mu_1 k_\perp^2 +\mu_x k_x^2} \ .
\eeqn
These propagators will also have an important role to play in our DRG analysis later.

The solutions (\ref{usol}, \ref{usot}) for $\ul$ and $\but$ can be summarized in a single equation using the relations (\ref{uldef}) and (\ref{utdef}) between $\bup$ and its components $\ul$ and $\but$, along with the analogous relations between $\bff_\perp$ and $f_L$ and $\fut$; we obtain
\beq
u^\perp_i(\tilde{\bk}) =G_{ij}(\tilde{\bk})f^\perp_j(\tilde{\bk})\ ,
\label{usoltensor}
\eeq
where
\beq
G_{ij}(\tilde{\bk})\equiv L^\perp_{ij}(\bkp)G_L(\tilde{\bk}) +P^\perp_{ij}(\bkp)G_T(\tilde{\bk})\,,
\eeq
and we have defined the ``longitudinal projection operator"
\beq
L^\perp_{ij}(\bkp)\equiv \frac{k^\perp_ik^\perp_j}{k_\perp^2} \ ,
\label{Ldef}
\eeq
which projects any vector along $\bkp$.

\subsection{Velocity correlation functions}

Using (\ref{usoltensor}), we obtain the autocorrelations:
\beqn
\left<u^\perp_i(\tilde{\bk}) u^\perp_j(\tilde{\bk}')\right>
&=&G_{im}(\tilde{\bk})G_{jn}(\tilde{\bk}')
\left<f^\perp_m(\tilde{\bk}) f^\perp_n(\tilde{\bk}')\right>
\nonumber\\
&=&2DC_{ij}(\tilde{\bk})\delta(\bk+\bk')\delta(\omega+\omega')
\ ,\nonumber\\
\label{ucorrtensor}
\eeqn
where in the second equality we have used the correlations of the noise
in Fourier space:
\beqn
\left<f^\perp_m(\tilde{\bk}) f^\perp_n(\tilde{\bk}')\right>
=2D\delta_{mn}\delta(\bk+\bk')\delta(\omega+\omega')\,,~~~~
\eeqn
and we've defined
\beqn
C_{ij}(\tilde{\bk})\equiv
L_{ij}^\perp(\bk)|G_L(\tilde{\bk})|^2 +P_{ij}^\perp(\bk)|G_T(\tilde{\bk})|^2\,.~~~
\eeqn
In writing (\ref{ucorrtensor}), we have also made liberal use of the property shared by  both projection operators $L^\perp_{ij}$ and $P^\perp_{ij}$ that their squares are themselves.

Transforming the above correlation function back to spatio-temporal domain, we obtain the velocity correlation in real space and time.
First let's calculate the equal-time correlation function:
\beqn
&&\left<\bu_\perp(0, \br)\cdot \bu_\perp(0,{\bf 0})\right>
\nonumber\\
&=& \frac{1}{\left(2\pi\right)^{d+1}}\int\dd\omega\dd\omega'\dd^dk\dd^dk'
\left<\bu_\perp(\tilde{\bk})\cdot \bu_\perp(\tilde{\bk}')\right>\ee^{\ii\bk\cdot\br}
\nonumber\\
&=&{2D\over \left(2\pi\right)^{d+1}}\int \dd\omega\dd^dk\,\ee^{\ii\bk\cdot\br}
\left[\frac{1}{\omega^2+\left(\mu_L k_\perp^2+\mu_x k_x^2\right)^2 } +\right.\nonumber\\
&&\left.~~~~~~~~~~~~~~~\frac{d-2}{\omega^2+\left(\mu_1 k_\perp^2+\mu_x k_x^2\right)^2 } \right]
\nonumber\\
&=&D[U_L(\br)+(d-2)U_T(\br)]
\, .\label{uuc1}
\eeqn
where
\beqn
U_L(\br)={1\over \left(2\pi\right)^d}\int \dd^dk\,
\frac{\ee^{\ii\bk\cdot\br}}{\mu_L k_\perp^2+\mu_x k_x^2}\, ,\label{Ul1}\\
U_T(\br)={1\over \left(2\pi\right)^d}\int \dd^dk\,
\frac{\ee^{\ii\bk\cdot\br}}{\mu_1 k_\perp^2+\mu_x k_x^2}\, .
\eeqn
Clearly, $U_{L,T}(\br)$ satisfy the anisotropic Poisson equations:
\beqn
\left(\mu_L \nabla_\perp^2 + \mu_x \pp_x^2 \right) U_L(\br) =-\delta^d (\br)\,,\\
\left(\mu_1 \nabla_\perp^2 + \mu_x \pp_x^2 \right) U_T(\br) =-\delta^d (\br)\,.
\label{anisPoiss}
\eeqn
The solutions to the above equations are, for $d>2$,
\beqn
U_L(\br) =\frac{\left(\frac{\mu_L}{\mu_x} x^2+r_\perp^2\right)^{(2-d)/2}}{S_d(d-2) \sqrt{\mu_x\mu_L}}\ ,\label{Ul2}\\
U_T(\br) =\frac{\left(\frac{\mu_1}{\mu_x} x^2+r_\perp^2\right)^{(2-d)/2}}{S_d(d-2) \sqrt{\mu_x\mu_1}}\ ,
\eeqn
where $S_d$ is the surface area of a $d$-dimensional unit sphere. Inserting the above results into Eq. (\ref{uuc1}) we get
\beqn
\left<\bu_\perp(t, \br)\cdot \bu_\perp(t,{\bf 0})\right>\propto r^{-(d-2)}
\, .\label{u_perp_ET}
\eeqn

Now we calculate the temporal correlation. Setting the spatial distance to zero in the correlation function to get
\beqn
&&\left<\bu_\perp(t, \mathbf{0})\cdot \bu_\perp(0,{\bf 0})\right>
\nonumber\\
&=& \frac{1}{\left(2\pi\right)^{d+1}}\int\dd\omega\dd\omega'\dd^dk\dd^dk'
\left<\bu_\perp(\tilde{\bk})\cdot \bu_\perp(\tilde{\bk}')\right>\ee^{-\ii\omega t}
\nonumber\\
&=&{2D\over \left(2\pi\right)^{d+1}}\int \dd\omega\dd^dk\, \ee^{-\ii\omega t}
\left[\frac{1}{\omega^2+\left(\mu_L k_\perp^2+\mu_x k_x^2\right)^2 } +\right.\nonumber\\
&&\left.~~~~~~~~~~~~~~~\frac{d-2}{\omega^2+\left(\mu_1 k_\perp^2+\mu_x k_x^2\right)^2 } \right]
\nonumber\\
&=&{D\over \left(2\pi\right)^d}\int \dd^dk\,
\left[\frac{\ee^{-(\mu_L k_\perp^2+\mu_x k_x^2)|t|}}{\mu_L k_\perp^2+\mu_x k_x^2} +\right.\nonumber\\
&&\left.~~~~~~~~~~~~~~~\frac{(d-2)\ee^{-(\mu_1 k_\perp^2+\mu_x k_x^2)|t|}}{\mu_1 k_\perp^2+\mu_x k_x^2 } \right]
\nonumber\\
&=&|t|^{-{d-2\over 2}}D\int {\dd^dq\over \left(2\pi\right)^d}\,
\left[\frac{\ee^{-(\mu_L q_\perp^2+\mu_x q_x^2)}}{\mu_L q_\perp^2+\mu_x q_x^2} +\right.\nonumber\\
&&\left.~~~~~~~~~~~~~~~\frac{(d-2)\ee^{-(\mu_1 q_\perp^2+\mu_x q_x^2)}}{\mu_1 q_\perp^2+\mu_x q_x^2 } \right]
\nonumber\\
&\propto& |t|^{-{d-2\over 2}}
\, ,\label{}
\eeqn
where in the  penultimate equality we have made the change of vectorial variable, $\bq=|t|^{1\over 2}\bk$,  while in the ultimate proportionality we have  used the fact that the integral  over $\bq$ is a finite constant (i.e., independent of time $t$).

We can easily generalize these results to arbitrary spatio-temporal separations. We start with
\beqn
&&\left<\bu_\perp(t, \br)\cdot \bu_\perp(0,{\bf 0})\right>
\nonumber\\
&=& \frac{1}{\left(2\pi\right)^{d+1}}\int\dd\omega\dd\omega'\dd^dk\dd^dk'
\left<\bu_\perp(\tilde{\bk})\cdot \bu_\perp(\tilde{\bk}')\right>
\ee^{\ii\left(\bk\cdot\br-\omega t\right)}
\nonumber\\
&=&{2D\over \left(2\pi\right)^{d+1}}\int \dd\omega\dd^dk\,
\ee^{\ii\left(\bk\cdot\br-\omega t\right)}
\left[\frac{1}{\omega^2+\left(\mu_L k_\perp^2+\mu_x k_x^2\right)^2 } +\right.\nonumber\\
&&\left.~~~~~~~~~~~~~~~\frac{d-2}{\omega^2+\left(\mu_1 k_\perp^2+\mu_x k_x^2\right)^2 } \right]
\, .\label{uuc2}
\eeqn
Changing the variables of integration from $\bk$ and $\omega$ to $\bQ$ and $\Upsilon$:
\beq
\bk_\perp=\bQ_\perp/r_\perp \,\,\,\,, \,\,\,k_x=Q_x/r_\perp \,\,\,\,, \,\,\,\omega=\Upsilon/ r_\perp^2 \,,
\label{scale1}
\eeq
we obtain
\beqn
\left<\bu_\perp(t, \br) \cdot \bu_\perp(0,{\bf 0})\right>=r_\perp^{-(d-2)}H_u\left({x\over r_\perp}, {t\over r_\perp^2}\right)\nonumber
\\
\propto\left\{
\begin{array}{ll}
	r^{-(d-2)},&r\gg |t|^{1\over 2}\\
	|t|^{-{(d-2)\over 2}},&|t|\gg r_\perp^2
\end{array}
\right.,~~~~~
\label{Linearuu}
\eeqn
where we've defined the scaling function
\beqn
H_u(a,b)\equiv{2D\over \left(2\pi\right)^{d+1}}\int \dd\Upsilon\dd^dQ\,
\ee^{\ii\left[\bQ_\perp\cdot{\bf \hat{r}}_\perp+Q_x a-\Upsilon b\right]}\times
\nonumber\\
\left[\frac{1}{\Upsilon^2+\left(\mu_L Q_\perp^2+\mu_x Q_x^2\right)^2 }+
\frac{d-2}{\Upsilon^2+\left(\mu_1 Q_\perp^2+\mu_x Q_x^2\right)^2 }\right]
\,.\nonumber\\
\eeqn

\subsection{Density correlations\label{Sec:dencorr_linear}}

Although it is not a ``soft mode" of Malthusian flocks, since it is not conserved in these systems, the density $\rho$ nonetheless exhibits long-ranged spatio-temporal correlations by virtue of being enslaved to the slow $u_L$ field via (\ref{rho-v}). Using this relation in Fourier space, we obtain
\beq
\left<\delta\rho(\tilde{\bk})\delta\rho(\tilde{\bk}') \right>= \frac{2D' k_\perp^2\delta(\bk+\bk')\delta(\omega+\omega')}{\omega^2 +(\mu_L k_\perp^2 +\mu_x k_x^2)^2}\,,
\label{rhocorrFT}
\eeq
where we've defined
\beq
D'\equiv D \left({\rho_0\over\kappa'(\rho_0)}\right)^2 \,.
\label{D'def}
\eeq

The spatio-temporal correlations can be calculated by Fourier transforming (\ref{rhocorrFT}) back to real space and time.
In particular, the equal time correlation function is
\beqn
&&\la \delta \rho (0, \br) \delta \rho (0, \mathbf{0})\ra\nonumber\\
&=&{1\over \left(2\pi\right)^{d+1}}\int\dd\omega\dd\omega'\dd^dk\dd^dk'
\left<\delta\rho(\tilde{\bk})\delta\rho(\tilde{\bk}') \right>
\ee^{\ii\bk\cdot\br}
\nonumber\\
&=&{1\over \left(2\pi\right)^d}\int \dd^dk\
\frac{ D' k_\perp^2\ee^{\ii\bk\cdot\br}}{\mu_L k_\perp^2+\mu_x k_x^2 } 
\ ,
\eeqn
where in the last equality we have  used (\ref{rhocorrFT}). To calculate this correlation function we write
\beq
\la \delta \rho (t, \br) \delta \rho (t, 0)  \ra = -D'\nabla_\perp^2 U_L(\br)\,,
\label{rhocorr1}
\eeq
where $U_L(\br)$ is given in (\ref{Ul2}). Inserting (\ref{Ul2}) into the above expression gives
\beqn
\nonumber
&&\la \delta \rho (t, \br) \delta \rho (t, 0)  \ra
\\
\nonumber
&=&
\left({\rho_0\over\kappa'(\rho_0)}\right)^2{ D\over S_d}\sqrt{\mu_x^{d-1}\over\mu_L}\left[{
{\mu_L(d-1)x^2-\mu_xr_\perp^2}\over\left(\mu_L x^2+\mu_xr_\perp^2\right)^{(2+d)/2}}\right]
\\
&\propto &r^{-d}
\ .\label{rhocorr2}
\eeqn

In particular, for $d=3$, we have
\beq
\la \delta \rho (t, \br) \delta \rho (t, 0) \ra  \sim r^{-3}
\ .
\eeq

It is clear from (\ref{rhocorr2}) that the equal time correlation function of the density fluctuation $\delta\rho$ vanishes on  the surface
\beq
x=\pm\left(\sqrt{\mu_x\over\mu_L(d-1)}\right)r_\perp\,\label{Sepa_linear}
\eeq
which, in $d=3$, is a cone.
For $|x| > \sqrt{\mu_x/\mu_L(d-1)} r_\perp$,
$\la \delta \rho (t, \br) \delta \rho (t, 0) \ra$ is  positive;
otherwise, the correlation is negative.

The qualitative shape of the regions of positive
and negative density correlations can be understood
heuristically as follows.
We first recall that in the hydrodynamic limit, we can
ignore velocity fluctuations in the $x$ direction.
Hence, equation
(\ref{rho-v}) implies that
$\delta \rho \propto \nabla_\perp \cdot \bu_\perp$.
That is,
a positive $\delta \rho(\br_0)$, results from a positive
divergence of $\bu_\perp$ at $\br_0$.
Therefore, a positive $\delta \rho(\br_0)$
will occur if, e.g., $u_y(\br_0-\epsilon \hat{\by})>u_y(\br_0+\epsilon \hat{\by})$,
where  $\epsilon$ is a small distance.
Since we know that the equal-time correlation of $\bu_\perp$ is always positive,
we expect in this situation that $u_y(\br_0-\epsilon \hat{\by})>u_y(\br_0+\epsilon \hat{\by})$  will remain positive even if $\br_0$ is shifted along the $x$ direction. Therefore,  we expect that, more often than not,  $\delta  \rho(t, A \hat{\bx})>0$ if  $\delta \rho(t,{\bf 0})>0$. Thus, this case will make a positive contribution to $\la \delta \rho(t,{\bf 0})\delta  \rho(t, A \hat{\bx})\ra$ where $A$ is any positive or negative number.

 One can make a similar argument for the case
in which $\delta\rho(\br_0)<0$, and conclude that
usually $\delta  \rho(t, A \hat{\bx})<0$ if
$\delta \rho(t,{\bf 0}) <0$. Thus, this case will also make a positive contribution to $\la \delta \rho(t,{\bf 0})\delta  \rho(t, A \hat{\bx})\ra$.

This explains the positive region of the density correlation.
Now, as the equal-time density correlation function is in the form of the Laplacian of a function (\ref{rhocorr1}), the overall spatial integral of the correlation function must be zero. Therefore, there must be a separatrix that separates the positive region  and the negative region, which is the region  roughly perpendicular to the $x$ direction.

In section \ref{exp}, we will show that the shape of this separatrix will be modified if we go beyond the linear theory.

Now we turn to the temporal correlations:
\beqn
&&\la \delta \rho (t, {\mathbf 0}) \delta \rho (0, {\mathbf 0})  \ra \nonumber\\
&=&{1\over \left(2\pi\right)^{d+1}}\int\dd\omega\dd\omega'\dd^dk\dd^dk'
\left<\delta\rho(\tilde{\bk})\delta\rho(\tilde{\bk'}) \right>\ee^{-\ii\omega t}
\nonumber\\
&=&{1\over \left(2\pi\right)^{d+1}}\int \dd\omega\dd^dk\
\frac{ 2D' k_\perp^2\ee^{-\ii\omega t}}
{\omega^2+\left(\mu_L k_\perp^2+\mu_x k_x^2\right)^2 }\nonumber\\
&=& \int \frac{\dd^d \bk}{(2\pi)^d}  \frac{ D' k_\perp^2 \ee^{-(\mu_L k_\perp^2+\mu_x k_x^2 )|t|}}{\mu_L k_\perp^2+\mu_x k_x^2 }\nonumber\\
&=& |t|^{-d/2}\int \frac{\dd^d \bq}{(2\pi)^d}  \frac{ D' q_\perp^2 \ee^{-(\mu_L q_\perp^2+\mu_x q_x^2 )}}{\mu_L q_\perp^2+\mu_x q_x^2 }\nonumber
\\
&\propto& |t|^{-d/2}
\ ,
\eeqn
where, again, in the  penultimate equality we have made the change of variable, $\bq=|t|^{1\over 2}\bk$,  and in the ultimate proportionality we have  used the fact that the integral  over $\bq$ is a finite constant (i.e., independent of time $t$).

For arbitrary spatio-temporal separations, the correlation function is given by
\beqn
&&\la \delta \rho (t, \br) \delta \rho (0, \mathbf{0})\ra\nonumber\\
&=&{1\over \left(2\pi\right)^{d+1}}\int\dd\omega\dd\omega'\dd^dk\dd^dk'
\left<\delta\rho(\tilde{\bk})\delta\rho(\tilde{\bk}') \right>
\ee^{i\left(\bk\cdot\br-\omega t\right)}
\nonumber\\
&=&{1\over \left(2\pi\right)^{d+1}}\int \dd\omega\dd^dk
\frac{2D' k_\perp^2\ee^{i\left(\bk\cdot\br-\omega t\right)}}
{\omega^2+\left(\mu_L k_\perp^2+\mu_x k_x^2\right)^2 }
\,.
\eeqn
Making the changes of variables of integration prescribed by (\ref{scale1})
we obtain
\beqn
\la \delta \rho (t, \br) \delta \rho (0, \mathbf{0})\ra&=&r_\perp^{-d}H_\rho\left({x\over r_\perp}, {t\over r_\perp^2}\right)\nonumber
\\
&\propto&\left\{
\begin{array}{ll}
r^{-d},&r\gg |t|^{1\over 2}\\
|t|^{-{d\over 2}},&|t|\gg r^2
\end{array}
\right.,
\label{Linearrho}
\eeqn
where we've defined the scaling function
\beqn
H_\rho(u,v)\equiv{2D'\over \left(2\pi\right)^{d+1}}\int \dd\Upsilon\dd^dQ
\frac{Q_\perp^2\ee^{\ii\left[\bQ_\perp\cdot{\bf \hat{r}}_\perp+Q_x u-\Upsilon v\right]}}
{\Upsilon^2+\left(\mu_L Q_\perp^2+\mu_x Q_x^2\right)^2 }\,.\nonumber\\
\eeqn

In any spatial dimension $d$, these correlations decay too rapidly to give rise to giant number fluctuations (GNF) \cite{Chate2,GNF}; that is, they are {\it not} sufficiently long-ranged to make the rms number fluctuations $\delta N\equiv\sqrt{\la (N-\la N\ra)^2\ra}$ in a large region grow more rapidly than the square root of the mean number $\sqrt{\la N\ra}$. However, they are sufficiently long-ranged to make $\delta N$ depend on the shape of the region in which the particle number $N$ is being counted \cite{toner_jcp19}.

Unfortunately, as we will see in the next section, these scaling laws, in particular the power law with which correlations decay with distance $r$, are changed by non-linear effects, leading to a more rapid decay which eliminates this shape dependence. Nonetheless, the strange power law dependence of density correlations persists (albeit with different exponents than found here in the linear theory), and still displays universal exponents which can be readily measured in experiments and simulations.

\section{Nonlinear effects and dynamic RG analysis\label{nonlin}}

\subsection{Full nonlinear equation of motion in Fourier space}
We write the full model (\ref{eq:main}) in Fourier space in tensor form:
\beqn
-\ii\omega u_i^{\perp}(\tilde{\bk})&=&-\left(\mu_1 k^2_{\perp}+\mu_xk_x^2\right)u_i^{\perp}(\tilde{\bk})+f_i^{\perp}(\tilde{\bk})
\nonumber\\
&&- \mu_2k_i^{\perp} \left( \bk_{\perp} \cdot \bu_{\perp}(\tilde{\bk})\right)
-{\ii \lambda\over \left(\sqrt{2\pi}\right)^{d+1}}\times\nonumber\\
&&\int_{\tilde{\bq}} \left[ \bu_{\perp} (\tilde{\bk}-\tilde{\bq})\cdot {\bq}_{\perp}\right] u_i^{\perp}(\tilde{\bq})\ ,
\label{eq:main1}
\eeqn
where $\tilde{\bq}\equiv (\bq,\Omega)$ and $\int_{\tilde{\bq}} \equiv
	\int \dd\Omega\ \dd^dq$.
Going through essentially the same calculation as the one which leads to (\ref{usoltensor}) we get
\beqn
u^\perp_i(\tilde{\bk})=G_{ij}(\tilde{\bk})\left\{ f^\perp_j(\tilde{\bk})
-{\ii \lambda\over\left(\sqrt{2\pi}\right)^{d+1}}\times\right.\nonumber\\
\left.\int_{\tilde{\bq}} \left[ \bu_{\perp} ({\bk}_{\perp}-{\bq}_{\perp})\cdot {\bq}_{\perp}\right] u_j^{\perp}(\tilde{\bq})\right\}
\ .
\label{usoltensor1}
\eeqn

\subsection{Dynamical Renormalization Group I: recursion relations\label{sec:rr}}

To probe what happens for $d>2$, we use a DRG analysis together with the $\epsilon$-expansion method to one-loop level 	 \cite{FNS}. Readers interested in a more complete and pedagogical  discussion of the DRG are referred to \cite{FNS} for the details of this general approach, including the use of Feynmann graphs in it.

First we decompose the Fourier modes $\bup(\tilde{\bk})$ into a rapidly varying part $\bup^>(\tilde{\bk})$  and  a slowly varying part $\bup^<(\tilde{\bk})$ in (\ref{eq:main1}). The rapidly varying part is supported in the momentum shell $-\infty <k_x<\infty$,
$\Lambda\ee^{-\dd\ell}<k_{\perp}<\Lambda$, where $\dd\ell$ is an infinitesimal
and $\Lambda$ is the ultraviolet cutoff. The slowly varying part is supported in $-\infty <k_x<\infty$, $0<k_{\perp}<\Lambda\ee^{-\dd\ell}$.

The  DRG procedure then consists of two steps. In step 1, we eliminate $\bup^>(\tilde{\bk})$ from (\ref{eq:main1}).
We do this by solving (\ref{usoltensor1}) iteratively for $\bup^>(\tilde{\bk})$. This solution is a series of $\lambda$
which depends on $\bup^<(\tilde{\bk})$. We substitute this solution into (\ref{eq:main1}) and average over  the short wavelength components $\bff^>(\tilde{\bk})$ of the noise $\bff$, which gives a closed EOM  for $\bup^<(\tilde{\bk})$. Step 2, rescale the length and time as the following
\beq
\br_\perp \mapsto  \ee^{\dd\ell} \br_\perp
\ , \
x \mapsto  \ee^{\zeta\dd\ell} x
\ , \
t \mapsto  \ee^{z \dd\ell} t
\ , \
\bu_\perp \mapsto  \ee^{\chi \dd\ell} \bu_\perp
\,,
\eeq
which restores the ultraviolet cutoff back to $\Lambda$. We reorganize the resultant EOM so that it has the same form as (\ref{eq:main1}) but with various coefficients renormalized.

\begin{figure}
	\begin{center}
 		\includegraphics[scale=.6]{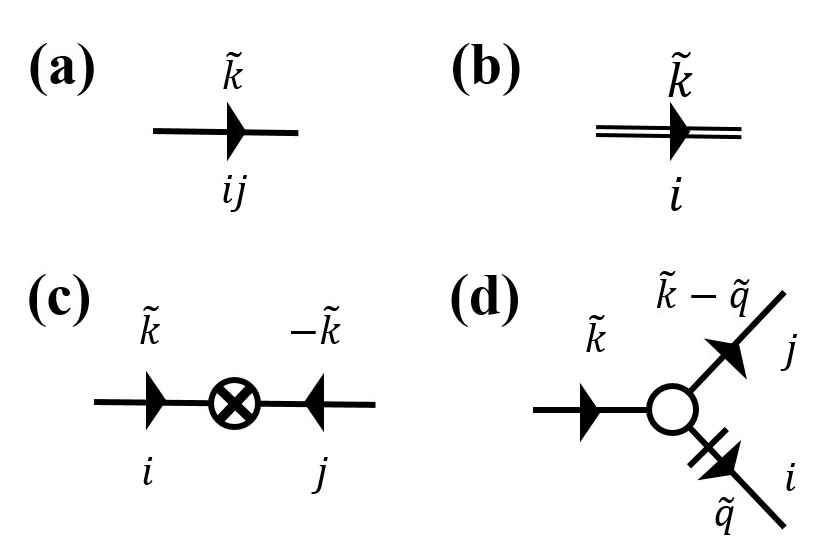}
	\end{center}
	\caption{Rules of graphical representation: (a) $=G_{ij}(\tilde{\bk})$;	(b) $=u^{\perp}_i(\tilde{\bk})$; (c) $=2DC_{ij}(\tilde{\bk})$; (d) the nonlinear term proportional to $=-\ii\lambda_1$; the slash represents a factor $q^{\perp}_j$.}
		\label{FR}
\end{figure}

The calculation of the renormalization of the coefficients arising from the process of eliminating $\bup^>(\tilde{\bk})$ can be represented by graphs. The basic rules for the graphical representation are illustrated in Fig.~{\ref{FR}}.

\begin{figure}
	\begin{center}
		 \includegraphics[scale=.5]{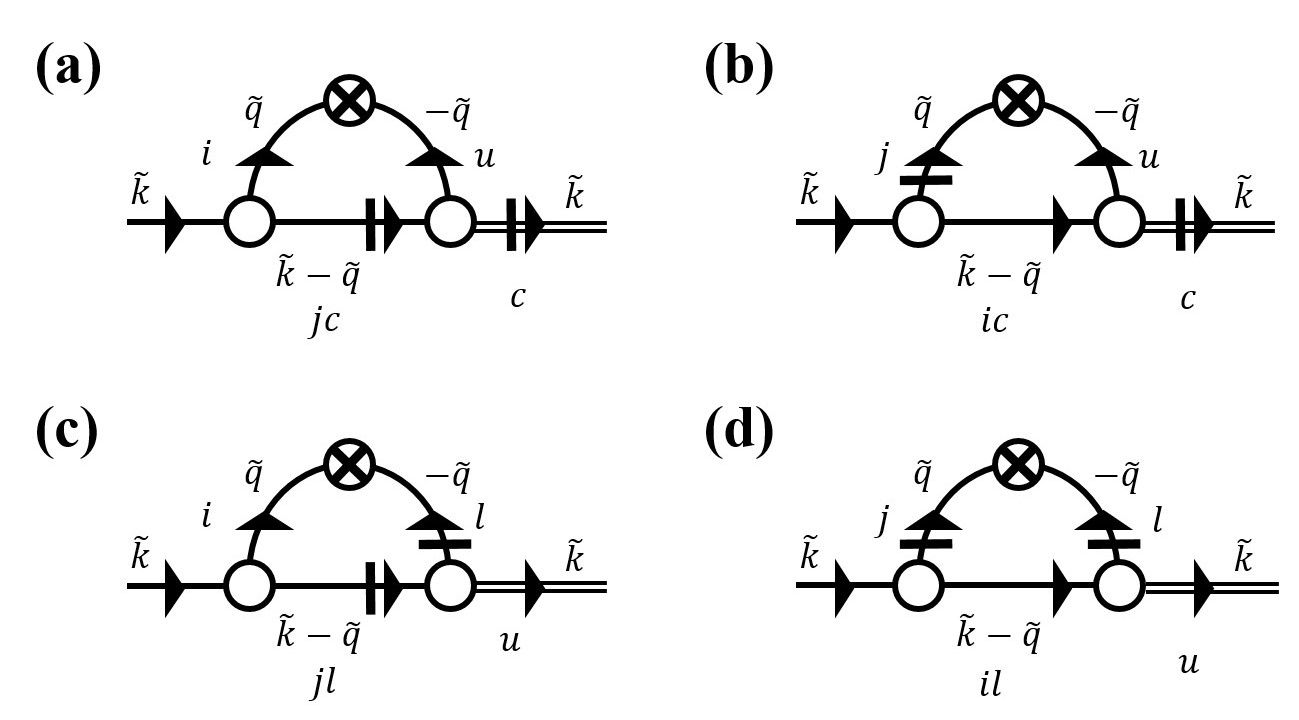}
	\end{center}
	\caption{Graphical representation of the correction to the linear terms in  the equation of motion (\ref{eq:main1}).}
	\label{fig:prop}
\end{figure}

\begin{figure}
	\begin{center}
		 \includegraphics[scale=.5]{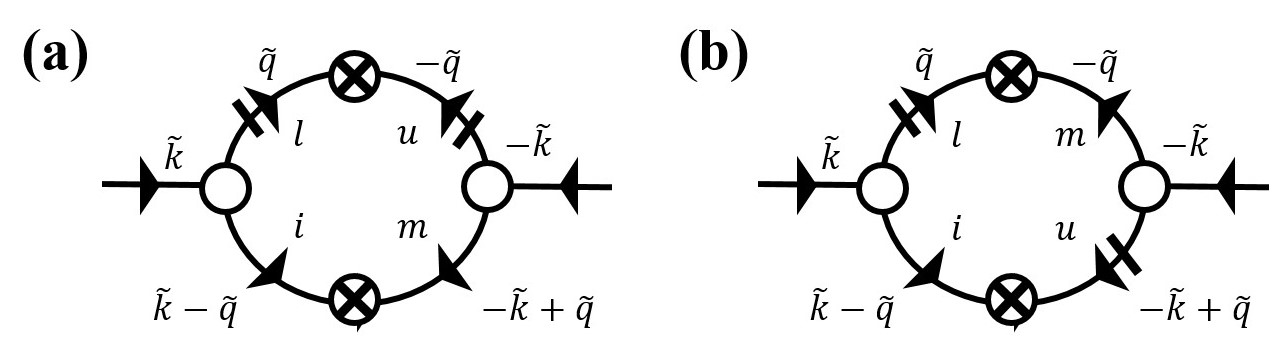}
	\end{center}
	\caption{Graphical representation of the correction to the noise correlator $\left<f_\ell(\tilde\bk)f_u(-\tilde\bk)\right>$.}
	\label{fig:noise}
\end{figure}

Following these rules and the prescription of \cite{FNS},  the renormalization of the linear terms and the noise to one-loop order are represented by the graphs in Fig.~\ref{fig:prop} and Fig.~\ref{fig:noise}, respectively. For example, Fig.~\ref{fig:prop}a represents a linear term in the EOM for $u_j^\perp$  given by
\beqn
-{2D\lambda^2 k_u^{\perp}u^{\perp}_c(\tilde{\bk})\over \left(2\pi\right)^{d+1}}
\int_{\tilde{q}}^>(k_i^\perp-q_i^\perp) C_{iu}(\tilde{\bq})G_{jc}(\tilde{\bk}-\tilde{\bq})\,,\nonumber\\\label{proa}
\eeqn
where
\beqn
\int_{\tilde{\bq}}^>&\equiv&\int_{-\infty}^{+\infty}\dd \Omega\int_{-\infty}^{+\infty}\dd q_x
\int_{\ee^{-\dd\ell}\Lambda<q_{\perp}<\Lambda}\dd^{d-1}q_{\perp}
\,.~\nonumber\\
\eeqn
By expanding the integrand to $O(k)$ we show in appendix \ref{sec:profiga} that (\ref{proa}) gives contributions to the two linear terms $k_{\perp}^2u^{\perp}_j(\tilde{k})$ and $k_j^{\perp}k_c^{\perp}u^{\perp}_c(\tilde{k})$, which lead  respectively to the renormalization of $\mu_{1}$  and $\mu_{2}$.

We iterate the DRG procedure repeatedly, which leads to the following flow equations of the coefficients to one-loop order:
\beqn
\label{eq:D}
\frac{1}{D}\frac{\dd D}{\dd \ell}&=&
z-2\chi-d+1-\zeta +g_1G_D(g_2)\,,
\\
\label{eq:L}
\frac{1}{\lambda}\frac{\dd \lambda}{\dd \ell}&=&
z+\chi-1\,,
\\
\label{eq:mux}
\frac{1}{\mu_x}\frac{\dd \mu_x}{\dd \ell}&=&
z-2\zeta\,,
\\
\label{eq:mu1}
\frac{1}{\mu_1}\frac{\dd \mu_1}{\dd \ell}&=&
z-2+ g_1G_{\mu_1}(g_2)\,,
\\
\label{eq:mu2}
\frac{1}{\mu_2}\frac{\dd \mu_2}{\dd \ell}&=&
z-2+g_1G_{\mu_2}(g_2)
\ ,
\eeqn
where we've defined
\beq
	g_1 \equiv \frac{D\lambda^2}{\sqrt{\mu_x \mu_1^5}} \frac{S_{d-1}}{(2\pi)^{d-1}} \Lambda^{d-4}
	\sep
	g_2 \equiv \frac{\mu_2 }{\mu_1}
	\ ,
	\label{eq:def_g}
	\eeq
where $S_{d-1}$ is the surface area of a $d-1$-dimensional unit sphere,
and $G_{D,\mu_1,\mu_2}$ are all functions of $g_2$. They are
\bew
\beqn
\label{eq:GD0}
G_D(g_2)&\equiv&{(d-2)\over2(d-1)}\frac{1}{g_2^2}
\left[1+\frac{1}{ \sqrt{g_2+1}}-\frac{2 \sqrt{2}}{ \sqrt{g_2+2}} \right]\\&=&
\frac{1}{g_2^2}
\left[\frac{1}{3}+\frac{1}{3 \sqrt{g_2+1}}-\frac{2 \sqrt{2}}{3 \sqrt{g_2+2}}
\right]\,\,\,, \,\,\,\,\,\,\,\, \,\,\,\, \,\,\,\,  (d=4)
\label{GD}
\\G_{\mu_1}(g_2)&\equiv& \frac{2}{d^2-1}  \Bigg(\frac{2d^2-6d+3}{32}+\frac{(d+3)\sqrt{2}}{g_2^2 (g_2+2)^{3/2}}-\frac{1}{g_2^2}-\frac{d+1}{2 g_2^2 \sqrt{g_2+1}}+\frac{d-3}{2 g_2}+\frac{d+15}{2 \sqrt{2} g_2 (g_2+2)^{3/2}}
\nn
\\
&&+\frac{3}{\sqrt{2} (g_2+2)^{3/2}}
-\frac{d+1}{4 g_2 \sqrt{g_2+1}}+\frac{3-d}{2 \sqrt{2} g_2 \sqrt{g_2+2}}
\Bigg)\label{Gen1}\\
&=&
\frac{2}{15}  \Bigg(\frac{11}{32}+\frac{7 \sqrt{2}}{g_2^2 (g_2+2)^{3/2}}-\frac{1}{g_2^2}-\frac{5}{2 g_2^2 \sqrt{g_2+1}}+\frac{1}{2 g_2}+\frac{19}{2 \sqrt{2} g_2 (g_2+2)^{3/2}}+\frac{3}{\sqrt{2} (g_2+2)^{3/2}}\nn
\\
&&-\frac{5}{4 g_2 \sqrt{g_2+1}}-\frac{1}{2 \sqrt{2} g_2 \sqrt{g_2+2}}
\Bigg)\,\,\,, \,\,\,\,\,\,\,\, \,\,\,\, \,\,\,\,  (d=4)
\eeqn
\ew
\bew
\beqn
G_{\mu_2}(g_2)&\equiv&\frac{2 }{(d^2-1)g_2}\Bigg(-\frac{(3d-1) \sqrt{2}}{g_2^2 (g_2+2)^{3/2}}+\frac{(d-1)}{g_2^2}+\frac{(d+1)}{2 g_2^2 \sqrt{g_2+1}}+\frac{(d^2-4d+3)\sqrt{2}}{4g_2 \sqrt{g_2+2}}-\frac{(d^2-7d+8)}{4g_2}
\nn
\\
&&+\frac{d+1}{64 (g_2+1)^{3/2}}
+\frac{(13-15d)}{2 \sqrt{2} g_2 (g_2+2)^{3/2}}-\frac{3(d-1)}{\sqrt{2} (g_2+2)^{3/2}}+\frac{(d+1)}{4 g_2 \sqrt{g_2+1}}+\frac{2d^2-9d+11}{32}
\Bigg)\label{Gen2}\\
&=&
\frac{2 }{15g_2}\Bigg(-\frac{11 \sqrt{2}}{g_2^2 (g_2+2)^{3/2}}+\frac{3}{g_2^2}+\frac{5}{2 g_2^2 \sqrt{g_2+1}}+\frac{3 \sqrt{2}}{4g_2 \sqrt{g_2+2}}+\frac{1}{g_2}+\frac{5}{64 (g_2+1)^{3/2}}\nn
\\
&&-\frac{47}{2 \sqrt{2} g_2 (g_2+2)^{3/2}}-\frac{9}{\sqrt{2} (g_2+2)^{3/2}}+\frac{5}{4 g_2 \sqrt{g_2+1}}+\frac{7}{32}
\Bigg)\,\,\,. \,\,\,\,\,\,\,\, \,\,\,\, \,\,\,\,  (d=4)
\label{Gen3}
\eeqn
\ew

The fact that there are no graphical corrections to $\lambda$ is not an accident, nor an artifact of our one loop approximation.
Rather, it is a consequence of the fact that $\lambda$ is ``protected" by a pseudo-Galilean symmetry.
That is, the EOM is invariant under the substitutions: $\bx_\perp \mapsto \bx_\perp +t\lambda \bw$ and $\bu_\perp \mapsto \bu_\perp+\bw$ for some arbitrary constant  vector  $\bw$ perpendicular to the mean velocity $\la \bu \ra$. Since this {\it exact} symmetry involves $\lambda$, and the renormalization group preserves the underlying symmetries of the problem, it follows that $\lambda$ can {\it not} be renormalized (except trivially by rescaling): its graphical corrections {\it must} vanish in {\it any} dimension $d$.

The absence of graphical corrections to $\mu_x$, on the other hand, {\it is} likely an artifact of the one-loop approximation, which we will discuss in later sections.

Note that the appearance of negative powers of $g_2$ in the  expressions (\ref{eq:GD0})-  (\ref{Gen3}) is somewhat misleading: despite those negative powers, none of  these functions diverges at $g_2=0$; in fact, these singularities all cancel, and $G_{D}$, $G_{\mu_1}$, and $G_{\mu_2}$ are all smooth, analytic, and finite for all finite $g_2$ ({\it including} $g_2=0$), that satisfy the stability constraint $g_2>-1$.

From Eqs (\ref{eq:D}-\ref{eq:mu2}) we obtain the closed  flow equations for $g$'s:
\beqn
\frac{1}{g_1} \frac{\dd g_1}{\dd \ell} &=&
\epsilon +g_1(G_D -\frac{5}{2} G_{\mu_1}) \equiv \epsilon+g_1 G_{g_1}(g_2)\nonumber\\
\label{dg1/dl}\label{eq:G1}
\\
\label{eq:G2}
\frac{1}{g_2} \frac{\dd g_2}{\dd \ell} &=&
g_1(G_{\mu_2}-G_{\mu_1}) \equiv g_1 G_{g_2}(g_2)
\,,\label{dg2/dl}
\eeqn
where
\bew
\beqn
G_{g_1}(g_2)&=&\frac{(-10d^2+30d-15)}{32(d^2-1)}+\frac{(d^2-d+8)}{2(d^2-1) g_2^2}-\frac{(2d^2+3d+11) \sqrt{2}}{(d^2-1) g_2^2 (g_2+2)^{3/2}}+\frac{(d+3)}{2(d-1) g_2^2 \sqrt{g_2+1}}+\frac{(15-5d)}{2(d^2-1) g_2}\nonumber
\\
&&-\frac{ \sqrt{2}(4d^2-9d+97)}{4(d^2-1) g_2 (g_2+2)^{3/2}}+\frac{(5d-45)}{2 \sqrt{2}(d^2-1) (g_2+2)^{3/2}}+\frac{5}{4(d-1) g_2 \sqrt{g_2+1}}
\label{gendGg1}
\\
&=& -\frac{11}{96}+\frac{2}{3 g_2^2}-\frac{11 \sqrt{2}}{3 g_2^2 (g_2+2)^{3/2}}+\frac{7}{6 g_2^2 \sqrt{g_2+1}}-\frac{1}{6 g_2}-\frac{25}{6 \sqrt{2} g_2 (g_2+2)^{3/2}}-\frac{5}{6 \sqrt{2} (g_2+2)^{3/2}}\nonumber
\\
&&+\frac{5}{12 g_2 \sqrt{g_2+1}}\,\,\,, \,\,\,\,\,\,\,\, \,\,\,\, \,\,\,\,  (d=4)
\label{d=4Gg1}
\eeqn
\ew
\bew
\beqn
G_{g_2}(g_2)&=&\left({2\over d^2-1}\right)\Bigg( \frac{(1-3d)\sqrt{2}}{g_2^3 (g_2+2)^{3/2}}+\frac{(d-1)}{g_2^3}+\frac{(d+1)}{2 g_2^3 \sqrt{g_2+1}}
+\frac{(1-19d)}{2\sqrt{2} g_2^2 (g_2+2)^{3/2}}+\frac{(d^2-4d+3)}{2\sqrt{2} g_2^2 \sqrt{g_2+2}}-\frac{(d^2-7d+4)}{4 g_2^2}\nonumber
\\
&&+\frac{3(d+1)}{4 g_2^2 \sqrt{g_2+1}}-\frac{3\sqrt{2}}{2 (g_2+2)^{3/2}}-\frac{(9+7d)}{2\sqrt{2} g_2 (g_2+2)^{3/2}}+\frac{(2d^2-25d+59)}{32 g_2}
+\frac{d+1}{64 g_2 (g_2+1)^{3/2}}+\frac{(d+1)}{4 g_2 \sqrt{g_2+1}}\nonumber
\\
&&+\frac{(d-3)}{2 \sqrt{2} g_2 \sqrt{g_2+2}}-\frac{(2d^2-6d+3)}{32}\Bigg)
\label{gendGg2}
\\
&=&
 -\frac{22 \sqrt{2}}{15 g_2^3 (g_2+2)^{3/2}}+\frac{2}{5 g_2^3}+\frac{1}{3 g_2^3 \sqrt{g_2+1}}
-\frac{5}{\sqrt{2} g_2^2 (g_2+2)^{3/2}}+\frac{1}{5\sqrt{2} g_2^2 \sqrt{g_2+2}}+\frac{4}{15 g_2^2}\nonumber
\\
&&+\frac{1}{2 g_2^2 \sqrt{g_2+1}}-\frac{\sqrt{2}}{5 (g_2+2)^{3/2}}-\frac{37}{15\sqrt{2} g_2 (g_2+2)^{3/2}}-\frac{3}{80 g_2}
+\frac{1}{96 g_2 (g_2+1)^{3/2}}+\frac{1}{6 g_2 \sqrt{g_2+1}}
\label{d=4Gg2}
\nonumber
\\
&&+\frac{1}{15 \sqrt{2} g_2 \sqrt{g_2+2}}-\frac{11}{240}\,\,\,. \,\,\,\,\,\,\,\, \,\,\,\, \,\,\,\,  (d=4)
\eeqn
\ew

Finding the fixed points of these two flow equations is equivalent to finding the fixed points of the original flow equations of the coefficients.  We turn to this calculation in the next subsection.

\begin{figure}
	\begin{center}
 \includegraphics[scale=.6]{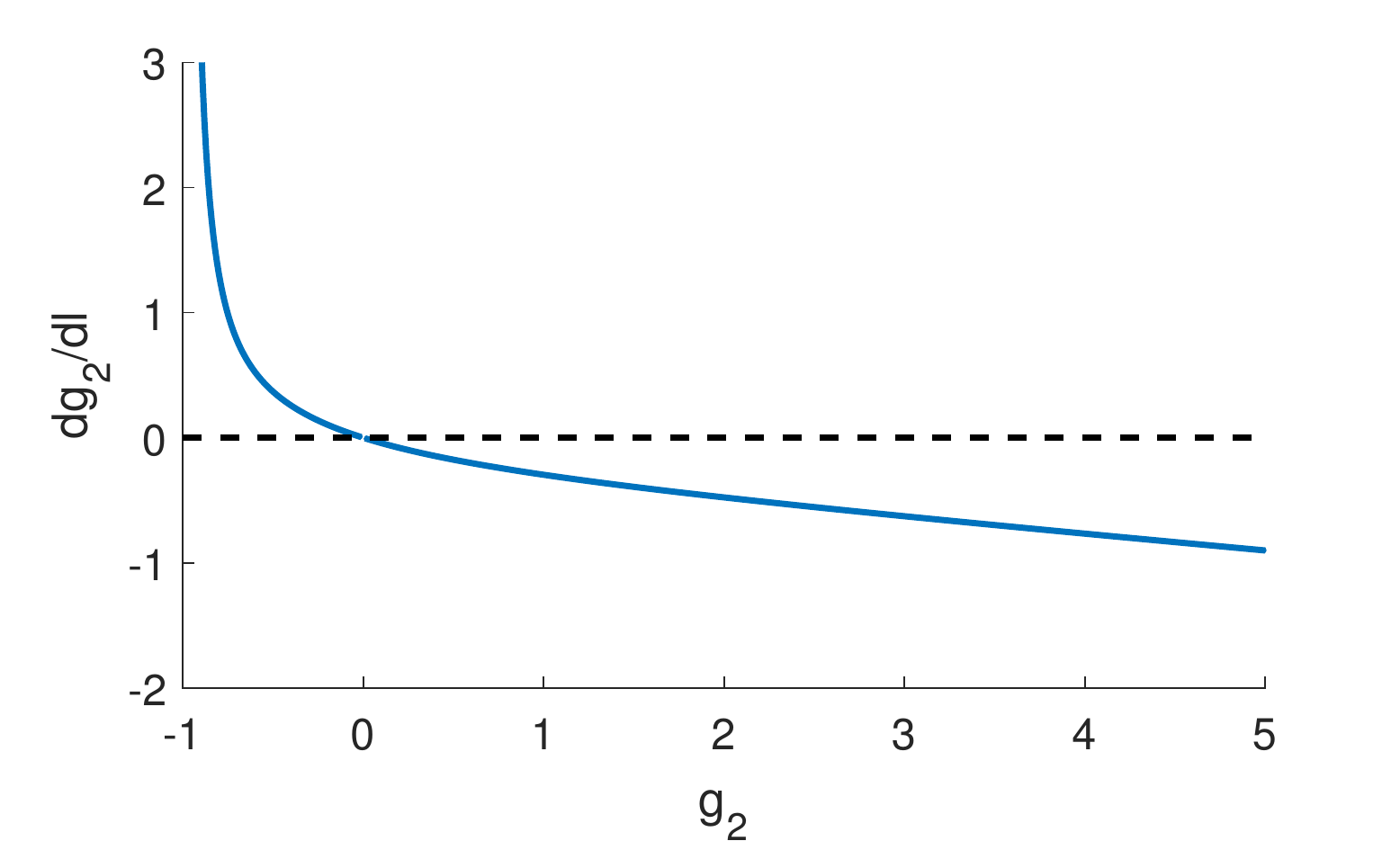}
	\end{center}
	\caption{Plot of ${\dd g_2\over \dd\ell}$ versus $g_2$ for $\epsilon =1$ and  $g_1$ fixed at 64/11, showing that  ${\dd g_2\over \dd\ell}$ vanishes only at $g_2=0$.  The plot only changes by a constant multiplicative factor if we change the value of $g_1$, so for all values of $g_1$, ${\dd g_2\over \dd\ell}$ vanishes only at $g_2=0$.}
	\label{fig:G_{g_2}}
\end{figure}

\begin{figure}
	\begin{center}
		 \includegraphics[scale=.48]{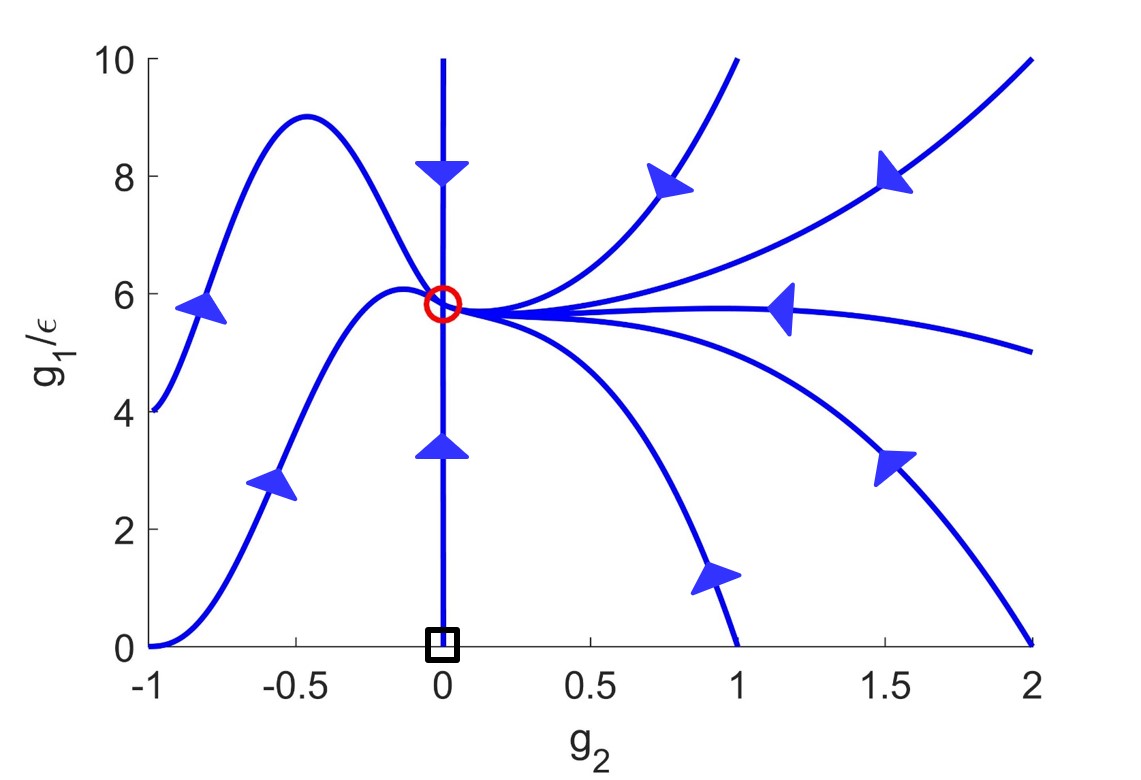}
	\end{center}
	\caption{RG flow of the coefficients $g$'s at $\epsilon =1$. The stable fixed point (red circle) is at $g_1^* = 64 \epsilon/11$ and $g_2^* = 0$, while the unstable Gaussian fixed point is denoted by the black square.	}
	\label{fig:RGflow}
\end{figure}

\subsection{Renormalization Group fixed points in the $\epsilon$-expansion}

We now seek fixed points of these recursion relations to linear order in $\epsilon$. To this order, it is sufficient to evaluate the graphical corrections $G_{g_{1,2}}$ in precisely $d=4$. We start with the recursion relations (\ref{eq:G2}) for $g_{2}$. In (\fig \ref{fig:G_{g_2}}) we plot ${\dd g_2\over \dd \ell}$ versus $g_2$ for fixed $g_1$ and spatial dimension $d=4$. As shown in the figure, the only point at which ${\dd g_2\over \dd \ell}$ vanishes is $g_2=0$. Hence, the fixed points in the $(g_1,g_2)$ plane must lie at $g_2=0$. Furthermore, since ${\dd g_2\over \dd \ell}>0$ for $g_2<0$, and ${\dd g_2\over \dd \ell}<0$ for $g_2>0$, these fixed points at  $g_2=0$ are stable with respect to $g_2$.
(We will later demonstrate  more thoroughly  the stability of these fixed points.)

The fact that the fixed points are at $g_2=0$ may seem like rather miraculous result, given the complexity of the recursion relations (\ref{dg1/dl},\ref{dg2/dl}) for $g_{1,2}$ (note the hideous expressions for $G_{g_1,g_2}$). In fact, it is quite simple to show that, at one-loop order, there must be a fixed point at $g_2=0$. This is because if we take $g_2=0$ initially, which is equivalent to taking $\mu_2=0$ initially, then the propagators and correlation functions simplify so much that it becomes quite easy to show that $\mu_2$ cannot be generated {\it at one-loop order}  in this limit upon renormalization. This argument is presented in appendix \ref{Sec:mu_2=0}. We note here that we do {\it not} expect this result to persist to higher loop orders. We will discuss the implications of this in section \ref{Sec:highloop}.

To find the value of $g_1$ at these fixed points we take  the slightly tricky limit $g_2\to0$ in our expression (\ref{d=4Gg1}) for $G_{g_1}$ in $d=4$. This gives
\beq
G_{g_1}(g_2=0)=-{11\over64} \,.
\label{Gg1(g2=0)}
\eeq

Inserting this value of $G_{g_1}$ into the recursion relation (\ref{dg1/dl}) for $g_1$ and finding the values of $g_1$ at which ${\dd g_1\over \dd \ell}=0$ (a value that we'll refer to as $g_1^*$) gives two solutions: $g_1^*=0, g_2^*=0$, which is just the Gaussian fixed point, and obviously unstable, and a stable  non-Gaussian fixed point at:
\beq
g_1^* = \frac{64 }{11}\epsilon +\cO(\epsilon^2) \sep g_2^* =0
\,.
\label{FP}
\eeq

Note that the value of $g_1$ at this non-Gaussian fixed point is $\cO(\epsilon)$, so  our perturbation theory, which is valid for small $g_1$, should be accurate for small $\epsilon$; i.e., for spatial dimensions near the upper critical dimension $d=4$. This validity for small $\epsilon$ is, of course, a standard feature of all $\epsilon$ expansions.

To demonstrate the stability of this fixed point, we show that small departures from it decay to zero upon renormalization. Specifically, we linearize the recursion relations (\ref{dg1/dl}) and (\ref{eq:G2}) around the fixed point, writing
\beq
g_1(\ell)=g_1^* +\delta g_1(\ell)
\label{delg1def}
\eeq
and expanding the recursion relations (\ref{dg1/dl}) and (\ref{eq:G2}) to linear order in $\delta g_1(\ell)$ and $g_2(\ell)$. This leads to the recursion relations:
\beqn
{d\delta g_1\over d\ell}&=&-\epsilon\delta g_1(\ell)-{160\over 121}\epsilon^2 g_2 \\
{dg_2\over d\ell}&=&-{16\over 33}\epsilon g_2(\ell) \,,
\label{rrlineps}
\eeqn
from which it is obvious that the fixed point (\ref{FP}) is stable.

The full renormalization group flows in the $g_1$-$g_2$ plane for small $\epsilon$ are illustrated in (\fig \ref{fig:RGflow}).

\subsection{Scaling exponents}

With the location of the fixed point (\ref{FP}) in hand, we can now easily find the universal  scaling exponents governing the behavior of all properties (in particular, correlation functions) of Malthusian flocks.

The most direct way to do this is to choose the heretofore arbitrary RG rescaling exponents -- that is, the dynamical exponent $z$, the anisotropy exponent $\zeta$, and the roughness exponent $\chi$ -- to keep all of the other important parameters  (i.e., the noise strength $D$, the diffusion constants $\mu_{x,1}$ \cite{foot1}, and the convective nonlinearity $\lambda$) fixed.

Keeping the noise strength $D$ fixed leads, via (\ref{eq:D}), to the condition
\beq
\label{eq:Dfix}
z-2\chi-d+1-\zeta +g_1^*G_D^*=0
\,,
\eeq
where we've defined $G_D^*\equiv G_D(g_2=0)$; i.e., the value of $G_D(g_2)$ at the fixed point $g_2=0$. From our  expression (\ref{GD}) for $G_D(g_2)$, it is relatively simple to take the limit $g_2\to0$  and obtain, in $d=4$,
\beq
G_D^*= G_D(g_2=0)={1\over16} \,.
\label{GD*}
\eeq
Inserting this, the fixed point value (\ref{FP}) of $g_1^*$, and $d=4-\epsilon$ into (\ref{eq:Dfix}) gives
\beq
\label{eq:cond1}
z-2\chi-\zeta =3-{15\over11}\epsilon
\,.
\eeq

From the above, we can obtain two more linear conditions on our three exponents $z$, $\zeta$ and $\chi$, by requiring that $\mu_x$ and $\lambda$ remain fixed. The former condition leads to
\beq
\label{muxfix}
z=2\zeta \,,
\eeq
while the latter implies
\beq
z+\chi=1 \,.
\label{lambdafix}
\eeq

The three linear equations (\ref{eq:cond1}), (\ref{muxfix}), and (\ref{lambdafix}) are easily solved to give:
\beqn
z&=&2-\frac{6 \epsilon}{11} +\cO(\epsilon^2)
\label{eq:z_ep}
\\
\chi &=& -1 + \frac{6 \epsilon}{11} +\cO(\epsilon^2)
\label{eq:chi_ep}
\\
\zeta  &=& 1 -  \frac{3 \epsilon}{11} +\cO(\epsilon^2)
\label{eq:zeta_ep}
\ .
\eeqn
To the best of our knowledge, the above fixed point and the associated scaling exponents  characterize a previously undiscovered universality class.

\begin{figure}
	\begin{center}
		\includegraphics[scale=.62]{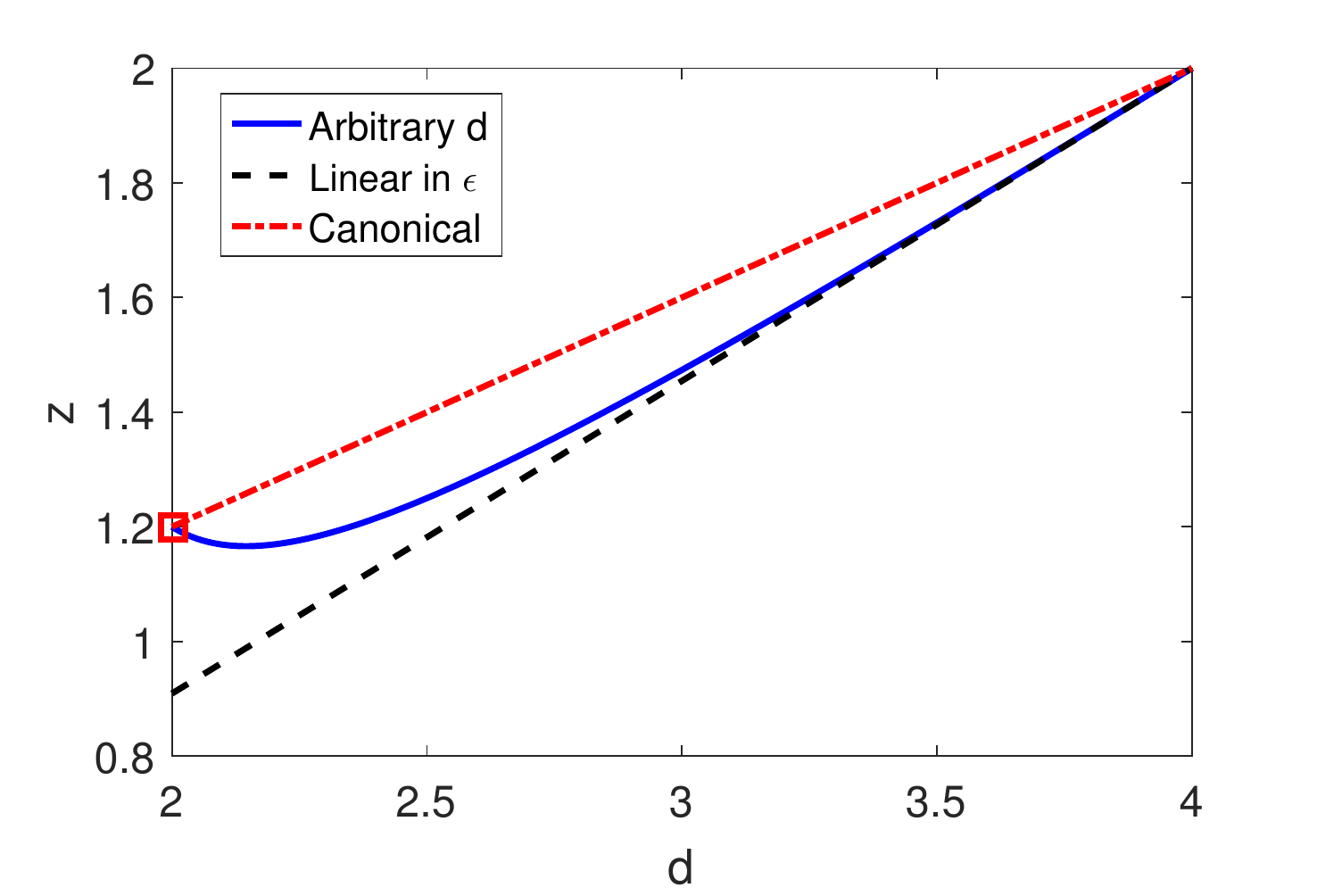}
	\end{center}
	\caption{Graphical summary of our results for the  dynamic exponent $z$  as a function of the spatial dimension $d$. The result (\ref{eq:z_ep}) based on the $\epsilon$-expansion method to $\cO(\epsilon)$ is shown by the broken black line, while the extrapolation to arbitrary $d$ based on our one-loop result is shown in the blue line (\ref{uncz2}), which converges to the known exact value (red square) in 2D.  The dashed red line is the ``canonical" value $z={2(d+1)\over5}$.
	}
	\label{fig:exponent}
\end{figure}

\subsection{Beyond linear order in $\epsilon$}
\label{sec:beyondepsilon}

Our results so far are based on a one-loop calculation, which
 is exact to linear order in $\epsilon$. However, since all of our  expressions  for $G_{D,\mu_1,\mu_2}$ are evaluated for general $d$, one can potentially extrapolate our results to arbitrary $d$ based on our one-loop calculation,  ignoring higher loop graphs. We must emphasize that this is a uncontrolled approximation, since the higher loop graphs are of higher order in $g_1$, but $g_1$ is {\it not} small at the fixed point once $d$ is far from $4$. Nonetheless, there are two limits in which this approach will recover exact results:

\noindent 1) near $d=4$, where it will automatically recover the exact $4-\epsilon$ expansion results we've just obtained, and

\noindent 2) in precisely $d=2$, where, as we'll show below, this approach reproduces the known exact ``canonical" exponents (\ref{canon}) \cite{Malthus}.

 Given these constraints, it's quite likely that the exponents obtained by this uncontrolled approximation  are extremely close
to the actual values.

 This truncated one-loop calculation for general $d$ now proceeds in much the same way as our small $\epsilon$ approach.
We start by noting that once again, as for $d$ near $4$, ${dg_2\over d\ell}$ looks like figure \ref{fig:G_{g_2}}; in particular, ${\dd g_2\over \dd \ell}>0$ for $g_2<0$, and ${\dd g_2\over \dd \ell}<0$ for $g_2>0$. Hence, as for small $\epsilon$, in our current uncontrolled one-loop approximation, we again have two fixed points, which are both at $g_2=0$, and  of  which  again only the non-Gaussian one is stable.
(We will do a more thorough analysis of the stability of this fixed point for general $d$ later.)

Since the fixed point value $g_2^*$ of $g_2$ is zero, we again only need the values of $G_{g_{1,2}}$ at $g_2=0$, but now for general $d$. With a bit more assistance from le Marquis de l'H\^{o}pital, we find, for the non-Gaussian fixed point,
\beqn
G_{g_1}^*&=&G_{g_1}(g_2=0)={23-14d\over64(d-1)} \,,
\label{G_g1 gen d}
\\
G_{g_2}^*&=&G_{g_2}(g_2=0)={5(4-d-d^2)\over64(d^2-1)} \,.
\label{G_g2 gen d}
\eeqn

Using the first of these in the recursion relations (\ref{dg1/dl}) for $g_1$,
and expressing the fixed point values of $g_1^*$ and $G_{\mu_1}^*$ in terms of $d$ (instead of $\epsilon$), we have
\beqn
g_1^* &=& \frac{64 (4-d) (d-1)}{14 d-23} \,.\label{FP1}
\eeqn

To demonstrate the stability of this fixed point, we show that small departures from it decay to zero upon renormalization. Specifically, we linearize the recursion relations (\ref{dg1/dl}) and (\ref{eq:G2}) around the fixed point, writing
\beq
g_1(\ell)=g_1^* +\delta g_1(\ell)
\label{delg1def}
\eeq
and expanding the recursion relations (\ref{dg1/dl}) and (\ref{eq:G2}) to linear order in $\delta g_1(\ell)$ and $g_2(\ell)$. This leads to the recursion relations:
\beqn
{d\delta g_1\over d\ell}&=&(d-4)\delta g_1(\ell) +\left(g_1^*\right)^2G'_{g_1}\left(g_2=0\right)g_2\nonumber
\\
&=&(d-4)\delta g_1(\ell)\nonumber
\\
&&
-{160(d-1)\left(3d^2-8d-1\right)\over (14d-23)^2(d+1)}(4-d)^2g_2\nonumber\\
\label{rrlingend1}
\\
{dg_2\over d\ell}&=&G_{g_2}(g_2=0)g_1^*g_2\nonumber\\
&=&\left({5(4-d-d^2)\over(d+1)(14d-23)}\right)\epsilon g_2\,.
\label{rrlingend2}
\eeqn
Because $d^2+d-4>0$ for all spatial dimensions $d$ in the range of interest $2\le d\le 4$, it is obvious from  (\ref{rrlingend1}, \ref{rrlingend2}) that the fixed point (\ref{FP1}) is stable.

For this uncontrolled one-loop approximation the full renormalization group flows in the $g_1$-$g_2$ plane still looks qualitatively like  \fig \ref{fig:RGflow}.

We can now  determine the scaling exponents $z$, $\zeta$, and $\chi$, as we did in the $\epsilon$ expansion, by choosing them to keep $D$, $\mu_x$, and $\lambda$ fixed. This leads to the same conditions (\ref{eq:Dfix}), (\ref{muxfix}) and (\ref{lambdafix}) as in the $\epsilon$ expansion, but now in (\ref{eq:Dfix}) we use the value
\beq
G_D^*=G_D(g_2=0)={3(d-2)\over32(d-1)}
\label{GD gen d}
\eeq
which arises from our one-loop truncation in arbitrary dimension $d$. Solving these three linear equations  (\ref{eq:Dfix}), (\ref{muxfix}) and (\ref{lambdafix}) for the three exponents now gives
\beqn
\label{uncz2}
&&z = 2 -  \frac{2(4-d)(4d-7)}{14d-23} \,,\\
\label{unczeta2}
&&\zeta  = 1 -  \frac{(4-d)(4d-7)}{14d-23}  \,,\\
&&\chi = -1 + \frac{2(4-d)(4d-7)}{14d-23}\, ,
\label{uncchi2}
\eeqn
which are the results in general dimension $d$ quoted in the introduction.

As noted earlier, these exponents (\ref{uncz2}), (\ref{unczeta2}), and (\ref{uncchi2}), in addition to automatically recovering the exact linear order in $\epsilon=4-d$ behavior that we found earlier, also become exact in $d=2$. The reason for this is simple: as can be seen by inspecting our one-loop recursion relations, they correctly recover the {\it exact}  fact that, in $d=2$, $\lambda$, $\mu_x$, and $D$ are unrenormalized graphically. Keeping them fixed therefore leads to three very simple linear equations for $z$. $\zeta$, and $\chi$, whose solutions are the ``canonical" exponents (\ref{canond=2}).

Why are these parameters exactly unrenormalized in $d=2$? For the non-linearity  $\lambda$, this is because it is unrenormalized in {\it any} dimension due to the pseudo-Galilean invariance of (\ref{eq:main}), for which we have given a detailed argument in section \ref{sec:rr}.

The absence of graphical corrections to $\mu_x$ is, as we'll discuss below, highly likely an artifact of the one-loop approximation, {\it except} in $d=2$, where it becomes exact because the sole non-linearity in the problem -- namely, the $\lambda (\bu_\perp \cdot \nabla_\perp)\bu_\perp$ term in (\ref{eq:main}) -- becomes a total $y$-derivative: $\lambda (\bup\cdot\nabla)\bup=\lambda u_y\pp_y u_y{\bf \hat{y}}=\lambda (\pp_y u_y^2/2){\bf \hat{y}}$. This implies that this non-linearity can only generate terms in the EOM that involve $y$ derivatives. Since the $\mu_x$ term only involves $x$ derivatives, it cannot be renormalized in $d=2$.

In our one-loop calculation, the graphical correction $G_D(g_2)$ to
$D$ vanishes  due to the explicit factor of $d-2$ in our expression  (\ref{eq:GD0}) for $G_D$. The presence of this factor is not an accident; rather, it reflects the same fact that implied $D$ is unrenormalized:  in $d=2$, the non-linearity can only generate terms involving $y$ derivatives. Since the noise correlation $D$ has weight at $\bq={\bf 0}$, it cannot be renormalized,  to {\it all} orders in a loop expansion. The factor of $d-2$ in equation (\ref{eq:GD0}) for $G_D$ is simply explicit confirmation of this fact at one-loop order.

To summarize, all of the properties required to obtain the canonical exponents (\ref{canond=2}) in $d=2$ are correctly reproduced by the uncontrolled, truncated one-loop approach.  This is why it reproduces the exact exponents in $d=2$.

Note that the predicted values of the scaling exponents in 3D obtained from these two approaches ($\epsilon$ expansion and one-loop in arbitrary $d$) are in fact very  quantitatively similar
(\fig \ref{fig:exponent}). For example, the value of $z$ obtained from the $\epsilon$ expansion in $d=3$, obtained from equation (\ref{eq:z_ep}) by setting $\epsilon=1$, is $z_\epsilon={16\over11}$, while that obtained from our uncontrolled one-loop approximation is $z_u={28\over19}$. The difference between these is $z_u-z_\epsilon={4\over209}$, which is only ${1\over77}$ of $z_u$. The other exponents are comparably close. Furthermore, since we know that the uncontrolled exponents approach the exact answer in $d=2$, they are probably closer to the exact answer in $d=3$ than the difference between themselves and the $\epsilon$ expansion result. We thereby conclude that the values given by the uncontrolled approximation in $d=3$, namely
\beqn
\label{unczd=3}
&&z = {28\over19}\approx1.47 \,,\\
\label{unczetad=3}
&&\zeta  = {14\over19}\approx0.74  \,,\\
&&\chi = -{9\over19}\approx-0.47\, .
\label{uncchid=3}
\eeqn
 are likely accurate to $\pm1\%$. As noted in the introduction, this implies that the digits shown after the approximate equalities above are probably all correct.

\subsection{Beyond one-loop order\label{Sec:highloop}}

In this section, we discuss what  features of the above results are artifacts of the one-loop truncation. Aside from small quantitative corrections to the precise values of the exponents, which we have just argued are small, there are two more significant changes that we expect will occur in a higher order calculation (which, we should emphasize, we have {\it not} done!).

The first of these is that the diffusion constant $\mu_x$ will no longer be unrenormalized at higher order. We expect this to be the case because there is no symmetry that ``protects" $\mu_x$ from renormalizing. Its failure to renormalize at one-loop order is therefore to some extent simply a coincidence, and almost certainly an artifact of the one-loop approximation. In this respect, its failure to renormalize is very similar to the result in $\epsilon$-expansions for $\phi^4$ theories of phase transitions \cite{Ma} that the
critical
exponent $\eta$ is zero to $\cO(\epsilon)$. As is well known, $\eta$ becomes non-zero at $\cO(\epsilon^2)$; or, equivalently, at two-loop order. We are quite confident that the same thing is true of renormalization of $\mu_x$.

The most important qualitative consequence of this is that the scaling relation
\beq
z=2\zeta\sep (\rm{one-loop})
\label{zzetascaling}
\eeq
which emerges at one-loop order from the requirement that $\mu_x$ remain fixed upon renormalization will no longer hold, since $\mu_x$ will now get graphical corrections.

However, the analogy just noted with critical phenomena strongly suggests
that the corrections to (\ref{zzetascaling})
will be very small in $d=3$. The exponent $\eta$ in $\phi^4$ theories is typically of order $\eta \sim0.01 - 0.02$, so it seems reasonable to expect the corrections to  (\ref{zzetascaling}), which also arise only at two-loop order in a problem with a critical dimension of $4$, to be comparable in magnitude. So, although it is an artifact of the one-loop approximation,  (\ref{zzetascaling}) probably holds to within a few percent. But as a matter of principle,  (\ref{zzetascaling}) is {\it not} an exact scaling relation.

The second change that will occur at higher loop order is that the fixed point will no longer be at $\mu_2\ne0$. This is because, as for the renormalization of $\mu_x$, there is no symmetry that prevents a non-zero $\mu_2$ from being generated, even  when the initial (bare) $\mu_2=0$.

As a result, the recursion relation for $g_2$ near $g_1=0$ will, at two loop order, become
\beq
 \frac{\dd g_2}{\dd \ell} =
 g_1 G_{g_2}(g_2)g_2+g_1^2 H(g_2)
\,,
\label{twoloop}
\eeq
where $H(g_2)$ is a function of $g_2$
that will presumably be even more formidable than $G_{g_2}(g_2)$. More importantly, it will be non-zero at $g_2=0$. Expanding the right hand side of (\ref{twoloop}) for small $g_2$ and $\epsilon$ gives
\beq
\frac{\dd g_2}{\dd \ell} =
g_1g_2G_{g_2}(g_2=0)+g_1^2 H(g_2=0)
\,,
\label{twoloop2}
\eeq
where the alert reader will recognize the first term on the right hand side from our linearized recursion relation (\ref{dg2/dl}) to one-loop order. Solving for the fixed point value $g_2^*$ of $g_2$ by setting $\frac{\dd g_2}{\dd \ell} =0$ and $g_1=g_1^*$ gives
\beq
g_2^*=-{H(g_2=0)\over G_{g_2}(g_2=0)}g_1^*=\cO(\epsilon)\,,
\label{g2*}
\eeq
where in the last equality we have used the fact that $g_1^*=\cO(\epsilon)$. Thus $g_2^*$ is non-zero, and $\cO(\epsilon)$, once higher loop corrections are taken into account. Unfortunately, it is impossible to say much more about the {\it value} of $g_2^*$, other than that it is non-zero, without actually doing the two-loop calculation necessary to determine the function $H(g_2)$ in equation (\ref{twoloop}). We have not attempted this formidable calculation, and so can say no more except note that $g_2^*$ will be non-zero. (Frustratingly, we can not even determine its sign!)

This has  experimental consequences, because, as we'll show in section \ref{exp} below, the value of $g_2^*$ determines  a universal amplitude ratio that appears in the velocity correlation function.
\newline

\section{Experimental consequences}
\label{exp}

\subsection{Scaling laws for velocity and density  correlations}

The scaling exponents $z$, $\zeta$, and $\chi$ just determined control the scaling properties of velocity and density correlations, as embodied in  equations (\ref{eq:Cu_scaling}) and (\ref{rhoscale}) for the velocity and density autocorrelations, respectively. This can be seen by using the ``trajectory integral matching formalism" \cite{TIMF}, which is simply a fancy way of describing the process of undoing all of the variable and coordinate rescaling done in the renormalization group process. This implies, for example, that the velocity autocorrelation function
\bew
\beq
C_u\Big(r_\perp, x-\gamma t, t; \{D_0, \mu_{x0}, \mu_{10}, \mu_{20}, \lambda_0\}\Big)\equiv\langle\bu_{\perp}(\br,t)\cdot\bu_{\perp}({\mathbf{0},0})\rangle
\label{Cvdef}
\eeq
\ew
of the original system (whose parameters -- the ``bare" parameters -- are denoted by the subscript $0$) can be related to that of the system after a renormalization group time $\ell$ has elapsed via
\bew
\beqn
&&C_u\Big(r_\perp, x-\gamma t, t; \big\{D_0, \mu_{x0}, \mu_{10}, \mu_{20}, \lambda_0\big\} \Big)=\exp\Bigg[2\int_0^{\ell}\chi(\ell') d\ell'\Bigg]\nonumber
\\&&
 \ \  \times C_u\left(r_\perp e^{-\ell}, (x-\gamma t)\exp\bigg(-\int_0^{\ell}\zeta(\ell') d\ell'\bigg), t\exp\bigg(-\int_0^{\ell}z(\ell') d\ell'\bigg); \big\{D(\ell), \mu_x(\ell), \mu_1(\ell), \mu_2(\ell), \lambda(\ell)\big\}\right)\,.\label{TIMF}
\eeqn
\ew

In this relation the combination $x-\gamma t$ appears rather than $x$ due to the boost (\ref{boost}) we performed to obtain the model equation (\ref{eq:main}) which we actually used for the renormalization group.

The relation (\ref{TIMF}) holds for an {\it arbitrary} choice of the rescaling exponents $\chi(\ell)$, $\zeta(\ell)$, and $z(\ell)$; they need not be the special choice (\ref{eq:z_ep}), (\ref{eq:chi_ep})  and (\ref{eq:zeta_ep})
that we made earlier to produce fixed points. Indeed, as our notation suggests, we can even choose different values for these exponents at different renormalization
group times $\ell$.
We will take advantage of this freedom to use (\ref{TIMF}) to derive the scaling relation  (\ref{eq:Cu_scaling}). We will do so by choosing the rescaling exponents $\chi(\ell)$, $\zeta(\ell)$, and $z(\ell)$ according to the following scheme: for $\ell<\ell^*$, where $\ell^*$ is the renormalization group time at which $g_{1,2}$ get close to their fixed point values,
we will choose these exponents so that {\it at} $\ell=\ell^*$, the parameters $D(\ell^*)$, $\mu_x(\ell^*)$, and $\mu_1(\ell^*)$ take on the values $D(\ell^*)=D_{\rm{ref}}$, $\mu_x(\ell^*)=\mu_1(\ell^*)=\mu_{\rm{ref}}$, where $D_{\rm{ref}}$ and $\mu_{\rm{ref}}$ are some reference values that we are free to choose. Note that we have deliberately chosen to to make $\mu_x(\ell^*)=\mu_1(\ell^*)$.

Since there are three free scaling exponents at our disposal, and an equal number (three) of parameters that we wish to force to take on values of our choosing, we can always find a choice of the rescaling exponents $\chi(\ell)$, $\zeta(\ell)$, and $z(\ell)$  (even with the assumption that these exponents are constant for $\ell > \ell^*$) that will achieve the target values $D_{\rm{ref}}$ and $\mu_{\rm{ref}}$.

The precise values of $D_{\rm{ref}}$ and $\mu_{\rm{ref}}$ that we choose are unimportant; what is important is that we choose the same values {\it no matter what the initial (bare) values} $D_0$, $\mu_{x0}$, $\mu_{10}$, $\mu_{20}$,  and $\lambda_0$ of the parameters were in our original model. The only ``memory" of these original parameters on the right hand side of (\ref{TIMF}) will therefore be contained in the value of $\ell^*$ \cite{irrel}.

Once we have fixed $D(\ell^*)$, $\mu_x(\ell^*)$, and $\mu_1(\ell^*)$, the parameters $\mu_2(\ell^*)$ and $\lambda(\ell^*)$ are also determined, the former by the relation $g_2={\mu_2\over\mu_1}$, the latter by the definition (\ref{eq:def_g}) of $g_1$. Combining this fact with  $g_1(\ell^*)=g_1^*$ and $g_2(\ell^*)=g_2^*$, (which follows our definition of $\ell^*$ as the renormalization group time at which we get close to the fixed point), we have that
\beq
\mu_2(\ell^*)=g_2^*\mu_{\rm{ref}} \sep \lambda(\ell^*)=\sqrt{g_1^*\mu_{\rm{ref}}^3(2\pi)^{d-1}\Lambda^{4-d}\over D_{\rm{ref}}S_{d-1}} \, ,
\label{mu2lambdaref}
\eeq
Hereafter we also refer to $\lambda(\ell^*)$ as $\lambda_{\rm ref}$.

Note finally that the value of $\ell^*$ at which we get close to the fixed point is unaffected by our arbitrary choice of
the rescaling exponents $\chi(\ell)$, $\zeta(\ell)$, and $z(\ell)$, since these do not enter the recursion relations for $g_{1,2}$.

For $\ell>\ell^*$, we will choose the rescaling exponents $\chi(\ell)$, $\zeta(\ell)$, and $z(\ell)$ to take on the values (\ref{eq:z_ep}), (\ref{eq:chi_ep})  and (\ref{eq:zeta_ep}) that we showed earlier keep all of the parameters fixed once $g_{1,2}$ have flowed to their fixed point values. For the remainder of this subsection, we will refer to these values of
$\chi$, $\zeta$, and $z$ as the ``fixed point" values $\chi_{\rm{FP}}$, $\zeta_{\rm{FP}}$, and $z_{\rm{FP}}$.

 This choice will, for all $\ell>\ell^*$, keep all of the parameters fixed at the ``reference" values we have just described.

With this choice of $\chi(\ell)$, $\zeta(\ell)$, and $z(\ell)$, we can rewrite equation (\ref{TIMF}) as
\bew
\beqn
&&C_u \Big(r_\perp, x-\gamma t, t; \{D_0, \mu_{x0}, \mu_{10}, \mu_{20}, \lambda_0\} \Big)=\exp\Bigg[2\int_0^{\ell_*}\chi(\ell') d\ell'+2\chi_{FP}(\ell-\ell_*)\Bigg]\nonumber\\&&
\ \ \times C_u\Bigg(r_\perp e^{-\ell}, (x-\gamma t)\exp\bigg(-\int_0^{\ell_*}\zeta(\ell') d\ell'-\zeta_{FP}(\ell-\ell_*)\bigg), t\exp\bigg(-\int_0^{\ell_*}z(\ell') d\ell'-z_{FP}(\ell-\ell_*)\bigg);\nonumber
\\
&&\ \ \ \big\{D_{\rm{ref}}, \mu_{\rm{ref}}, \mu_{\rm{ref}},g_2^*\mu_{\rm{ref}}, \lambda_{\rm{ref}}\big\}\Bigg)\,.
\label{TIMF2}
\eeqn
\ew
To derive our scaling law (\ref{eq:Cu_scaling}) for $C_u$, we simply apply this relation (\ref{TIMF2}) at particular value of $\ell$, which we'll call $\ell(r_\perp)$, determined by
\beq
\label{lrpdef}
e^{-\ell(r_\perp)}r_\perp=a\equiv{1\over\Lambda} \,.
\eeq

Setting $\ell=\ell(r_\perp)$ on the right hand side of (\ref{TIMF2}) gives
\bew
\beqn
C_u\Big(r_\perp, x-\gamma t, t; \big\{D_0, \mu_{x0}, \mu_{10}, \lambda_0\big\}\Big)&=&Ar_\perp^{2\chi_{FP}} C_u\Bigg(a, a{({|x-\gamma t|}/\xi_x)\over(r_\perp/\xi_\perp)^{\zeta_{FP}}}, {(t/\tau)\tau_0\over(r_\perp/\xi_\perp)^{z_{FP}}}; \big\{D_{\rm{ref}}, \mu_{\rm{ref}}, \mu_{\rm{ref}},g_2^*\mu_{\rm{ref}}, \lambda_{\rm{ref}}\big\}\Bigg)\nonumber\\
&\equiv&r_{\perp}^{2\chi_{FP}}F_u\left({(|x-\gamma t|/\xi_x)\over (r_{\perp}/\xi_\perp)^{\zeta_FP}},{(t/\tau)\over(r_{\perp}/\xi_\perp)^{z_FP}}\right)\,,
\label{TIMF3}
\eeqn
\ew
where we've defined
scaling function
\bew
\beq
F_u\equiv AC_u\Bigg(a, a{(|x-\gamma t|/\xi_x)\over(r_\perp/\xi_\perp)^{\zeta_{FP}}}, {(t/\tau)\tau_0\over(r_\perp/\xi_\perp)^{z_{FP}}}; \big\{D_{\rm{ref}}, \mu_{\rm{ref}}, \mu_{\rm{ref}},g_2^*\mu_{\rm{ref}}, \lambda_{\rm{ref}}\big\}\Bigg)\,,
\label{Fudef}
\eeq
\ew
the constant
\beq
A\equiv a^{-2\chi_{FP}}\exp\Bigg[2\int_0^{\ell_*}(\chi(\ell')-\chi_{FP}) d\ell'\Bigg]\,,
\label{Adef}
\eeq
and
the non-universal ``non-linear lengths" $\xi_{\perp,x}$ to satisfy
\beq
\xi_\perp=e^{\ell^*}a \,,
\label{xipdef}
\eeq
and
\beqn
{\xi_\perp^{\zeta_{FP}}\over\xi_x}=\exp\bigg[\int_0^{\ell^*}(\zeta_{FP}-\zeta(\ell)) d\ell\bigg]a^{\zeta_{FP}-1} \,,
\label{xixcond}
\eeqn
and the non-universal ``non-linear time" $\tau$ to satisfy
\beqn
{\xi_\perp^{z_{FP}}\over\tau}\tau_0=\exp\bigg[\int_0^{\ell^*}(z_{FP}-z(\ell)) d\ell\bigg]a^{z_{FP}} \,.
\label{taucond}
\eeqn
Here the  value of the characteristic time $\tau_0$ is not arbitrary, but set by the cutoff length $a$ and the $\mu_1$ of the rescaled system, namely $\mu_{\rm ref}$. Specifically it is given by
\beq
\tau_0={a^2\over\mu_{\rm ref}}\,.\label{tau_0}
\eeq
Note that, like the  reference values of  other parameters, this characteristic time is the same for all systems, regardless of the bare values of the parameters.

Because the parameters appearing in $C_u$ on the right hand side of (\ref{Fudef}), namely,   $a$, $\tau_0$, $D_{\rm{ref}}$, $ \mu_{\rm{ref}}$, $g_2^*\mu_{\rm{ref}}$, and $ \lambda_{\rm{ref}}$ are all independent of the initial system under consideration, the scaling function  $F_u$ is, as claimed in the introduction, a universal function of its arguments
${(|x-\gamma t|/\xi_x)\over(r_\perp/\xi_\perp)^{\zeta_{FP}}}$ and ${(t/\tau)\over(r_\perp/\xi_\perp)^{z_{FP}}}$, up to the non-universal multiplicative factor $A$, which is given by (\ref{Adef}).

 This concludes our derivation of the scaling law  for velocity correlations.
 The derivation of the density correlations  is almost identical. The only
 difference lies in the field rescaling. Since $\delta\rho$ is enslaved to
 $\bu_\perp$ by the relation (\ref{rho-v}), the recaling exponent for
 $\delta\rho$ is $\chi-1$ instead of $\chi$, the recaling exponent of
 $\bu_\perp$. Therefore, in analogy to (\ref{TIMF}), we get the following
 relation between density correlations in the original system and the
 rescaled system:
\bew
\beqn
&&C_\rho\Big(r_\perp, x-\gamma t, t; \big\{D_0, \mu_{x0}, \mu_{10}, \mu_{20}, \lambda_0\big\} \Big)=\exp\Bigg[2\int_0^{\ell}\chi(\ell')-1 d\ell'\Bigg]\nonumber
\\&&
 \ \  \times C_\rho\left(r_\perp e^{-\ell}, (x-\gamma t)\exp\bigg(-\int_0^{\ell}\zeta(\ell') d\ell'\bigg), t\exp\bigg(-\int_0^{\ell}z(\ell') d\ell'\bigg); \big\{D(\ell), \mu_x(\ell), \mu_1(\ell), \mu_2(\ell), \lambda(\ell)\big\}\right)\,.\label{}
\eeqn
\ew
From here on the derivation is virtually identical to that of the velocity
correlations, which we will not repeat. The final result is given by   (\ref{rhoscale}) in the introduction.

\subsection{Calculation of the non-linear lengths and times}

There are two independent ways of calculating the non-linear lengths and times appearing in the scaling functions (\ref{TIMF3}) and (\ref{rhoscale}) just derived. One way is to continue with the RG approach just presented. We take this approach in the next subsection. An alternative approach, which we present as a check on the RG approach, is to calculate perturbative corrections to the linear theory and calculate the length and time scales on which they become appreciable. These length and time scales prove to be precisely  the lengths $\xi_\perp$, and $\xi_x$, and the time $\tau$.

We'll begin here with the RG calculation; then, in the next subsection, we'll present  the perturbation theory approach.

\subsubsection{RG calculation}

The conditions (\ref{xixcond}) and (\ref{taucond}) can be solved for the non-linear length $\xi_\perp$ and  non-linear time $\tau$, giving
\beq
\xi_x=a\left( \frac{\xi_\perp}{a}\right)^{\zeta_{FP}}\exp\bigg[-\zeta_{FP}\ell^*+
\int_0^{\ell^*}\zeta(\ell) d\ell\bigg] \,,
\label{xix}
\eeq
and
\beq
\tau=\tau_0\left({\xi_\perp\over a}\right)^{z_{FP}}\exp\bigg[-z_{FP}\ell^*+\int_0^{\ell^*}z(\ell) d\ell\bigg]\,.
\label{tau}
\eeq
Using our expression (\ref{xipdef}) for $\xi_\perp$ in these expressions simplifies them to
\beq
\xi_x=a\exp\bigg[\int_0^{\ell^*}\zeta(\ell) d\ell\bigg] \,,
\label{xix2}
\eeq
and
\beq
\tau=\tau_0\exp\bigg[\int_0^{\ell^*}z(\ell) d\ell\bigg]\,.
\label{tau2}
\eeq
\newline

The alert reader may be alarmed by the apparent dependence of $\xi_\perp$ and $\tau$ in the arbitrary choices of $\zeta(\ell)$ and $z(\ell)$.  But those choices are not {\it completely} arbitrary, since they must lead to the parameters $D$ and $\mu_{1,x}$ flowing to their reference values.
This requirement proves to constrain the very integrals
that appear in (\ref{xix2}) and (\ref{tau2}) to (non-universal) values that are determined entirely by the bare parameters of the model
Likewise, the non-universal overall scale factor $A$ in the correlation function (\ref{Fudef}), while apparently dependent on our arbitrary choice of the velocity rescaling exponent $\chi(\ell)$, in fact does not, and is, instead, also determined solely by the non-universal values of the bare parameters of the model, as we'll show now.

The requirement that $\mu_x(\ell)$ and $\mu_1(\ell)$ reach equality at $\ell=\ell^*$ constrains the integral of $\zeta(\ell)$ in (\ref{xix2}). To see this, consider the recursion relations for $\mu_x$ and $\mu_1$. In complete generality, to arbitrary order in perturbation theory, these can be written:
\beqn
\label{eq:muxgen}
\frac{1}{\mu_x}\frac{\dd \mu_x}{\dd \ell}&=&
z-2\zeta(\ell)+ Y_x(g_1(\ell), g_2(\ell))\,,
\\
\label{eq:mu1gen}
\frac{1}{\mu_1}\frac{\dd \mu_1}{\dd \ell}&=&
z-2+Y_1(g_1(\ell), g_2(\ell))
\ .
\eeqn
To one-loop order, $Y_x=0$  and $Y_1=g_1G_{\mu_1}(g_2)$; here we'll use this more general form to demonstrate that our conclusion is {\it not} an artifact of the one-loop approximation, or, indeed, any approximation at all.

The recursion relations (\ref{eq:muxgen})and (\ref{eq:mu1gen}) taken together imply that the logarithm of the ratio ${\mu_x\over\mu_1}$ obeys the recursion relation
\bew
\beq
\label{eq:muxovermu1}
\frac{\dd }{\dd \ell}\ln\left({\mu_x\over\mu_1}\right)=\frac{1}{\mu_x}\frac{\dd \mu_x}{\dd \ell}-\frac{1}{\mu_1}\frac{\dd \mu_1}{\dd \ell}=
2(1-\zeta(\ell))+ Y_x(g_1(\ell), g_2(\ell))-Y_1(g_1(\ell), g_2(\ell))\,.
\eeq
\ew
The solution of this is
\bew
\beq
\ln\bigg[\left({\mu_x(\ell)\over\mu_1(\ell)}\right)\left({\mu_{10}\over\mu_{x0}}\right)\bigg]=2\ell-2\int_0^{\ell}\zeta(\ell') d\ell'+\int_0^{\ell}[Y_x(g_1(\ell'), g_2(\ell'))-Y_1(g_1(\ell'), g_2(\ell'))] d\ell' \,.
 \label{ratsol}
 \eeq
 \ew
 Evaluating both sides of this expression at $\ell=\ell^*$, and recalling that, by construction, $\mu_x(\ell^*)=\mu_1(\ell^*)$, gives
 \beq
 \ln\left({\mu_{10}\over\mu_{x0}}\right)=2\ell^*-2\int_0^{\ell^*}\zeta(\ell) d\ell+2\Phi(g_{10}, g_{20})\,,
 \label{rat*}
 \eeq
 where we've defined
 \beqn
 \Phi(g_{10}, g_{20})\equiv{1\over2}\int_0^{\ell^*}[Y_x(g_1(\ell'), g_2(\ell'))\nonumber\\
 -Y_1(g_1(\ell'), g_2(\ell'))] d\ell' \,.
 \label{Phidef}
 \eeqn
 Note that, as our notation suggests, $\Phi$ is completely determined by the bare values $g_{10,20}$ of $g_{1,2}$; in particular, it is {\it independent} of the arbitrary choice of the rescaling exponents $\chi(\ell)$, $\zeta(\ell)$, and $z(\ell)$. This is because the recursion relations for
 $g_{1,2}$ are independent of those exponents; so their solutions $g_{1,2}(\ell)$
are determined entirely by the initial conditions $g_{1,2}(\ell=0)=g_{10,20}$. Once those solutions are determined, the integrand in (\ref{Phidef}) is also fully determined (since it depends only on
$g_{1,2}(\ell)$). Furthermore, the limits on the integral are completely determined by $g_{10,20}$
 as well, since $\ell^*$ is. Hence, $\Phi$ is completely determined by  $g_{10,20}$ , as claimed.

 The condition (\ref{rat*}) can be rewritten as
 \beq
 \int_0^{\ell^*}\zeta(\ell) d\ell= {1\over2}\ln\left({\mu_{x0}\over\mu_{10}}\right)+\ell^*+\Phi(g_{10}, g_{20})\,.
\label{zetaintsol}
 \eeq
 Using this in (\ref{xix2}) gives
 \beq
 \xi_x=ae^{\ell^*}e^\Phi\sqrt{\mu_{x0}\over\mu_{10}}=\xi_\perp e^\Phi\sqrt{\mu_{x0}\over\mu_{10}}\,,
 \label{xix3}
 \eeq
 where in the last equality we have used our expression (\ref{xipdef}) for $\xi_\perp$.

Note that the ratio of $\xi_x$ to $\xi_\perp$ implied by (\ref{xix3})
depends only on $g_{10,20}$ and $\mu_{x0,10}$, and not at all on the exact choice of the functional dependence  rescaling exponent $\zeta(\ell)$ on $\ell$; {\it any} choice that leads to $\mu_x(\ell^*)=\mu_1(\ell^*)$ gives the same answer.

A similar argument can be applied to the time scale $\tau$. We start by solving the recursion relation (\ref{eq:mu1gen}) for $\mu_1(\ell^*)$:
\beq
\ln\left({\mu_{\rm{ref}}\over\mu_{10}}\right)= \int_0^{\ell^*}z(\ell) d\ell-2\ell^*+\Phi_{\mu_1}(g_{10},g_{20})\,,
\label{mu1sol}
\eeq
 where we've defined
 \beq
 \Phi_{\mu_1}(g_{10}, g_{20})\equiv\int_0^{\ell^*}Y_1(g_1(\ell), g_2(\ell))d\ell \,.
 \label{Phimu1def}
 \eeq
 Note that, like $\Phi$, $\Phi_{\mu_1}$ is completely determined by the bare values $g_{10,20}$, and is {\it independent} of the arbitrary choice of the rescaling exponents $\chi(\ell)$, $\zeta(\ell)$, and $z(\ell)$ for the same reasons as before: both integrand and the limits of integration in (\ref{Phimu1def}) depend only on $g_{10,20}$. Solving (\ref{mu1sol}) for $\int_0^{\ell^*}z(\ell) d\ell$ gives
 \beq
 \int_0^{\ell^*}z(\ell) d\ell=\ln\left({\mu_{\rm{ref}}\over\mu_{10}}\right)+2\ell^*-\Phi_{\mu_1}(g_{10},g_{20})\,.
\label{intzsol}
\eeq

 Inserting this result into (\ref{tau2}), and using (\ref{xipdef}) and (\ref{tau_0}) gives
 \beq
 \tau={\xi_\perp^2\over \mu_{10}}e^{-\Phi_{\mu_1}}\,.
 \label{tau3}
 \eeq

 If the bare parameter $g_{10}$ is small, then, up to factors of $\cO(1)$, we can take $\Phi$ and $\Phi_{\mu_1}$ to be zero, which reduces (\ref{xix3}) to
 \beq
\xi_x=\xi_\perp\sqrt{\mu_{x0}\over\mu_{10}}\,,
\label{xixsol}
\eeq
and (\ref{tau3}) to
\beq
 \tau={\xi_\perp^2\over \mu_{10}}\,.
 \label{tau4}
 \eeq
 We can also determine $\xi_\perp$ in this limit by noting that, for small $g_1$, the recursion relation (\ref{eq:G1}) for $g_1$ becomes simply
\beq
\frac{\dd g_1}{\dd \ell} =
\epsilon g_1 \,,
 \eeq
 which is easily solved to give
 \beq
 g_1(\ell)=g_{10}e^{\epsilon\ell} \,.
 \eeq
 Setting $ g_1(\ell^*)=1$ and solving for $ e^{\ell^*}$ gives
 \beq
 e^{\ell^*}=(g_{10})^{-{1\over\epsilon}} \,.
 \eeq
 Using our expression (\ref{eq:def_g}) for $g_1$, evaluated with the bare parameters, this gives
 \beq
 e^{\ell^*}=\Lambda\left({\mu_{x0}\mu_{10}^5\over D_0^2\lambda_0^4}\right)^{1\over2\epsilon} \,.
 \label{eell*}
 \eeq
Using this in turn in (\ref{xipdef}) gives
\beq
\xi_\perp=\left({\mu_x^0\mu_{10}^5\over D_0^2\lambda_0^4}\right)^{1\over2\epsilon}\,.
\label{xip}
\eeq
Note that $\xi_\perp$ is independent of the ultraviolet cutoff $\Lambda$ in this case, which is to be expected, since the divergent renormalization of the parameters is an infrared phenomenon.

In $d=3$, where $\epsilon=1$, this becomes
\beq
\xi_\perp={\sqrt{\mu_{x0}\mu_{10}^5}\over D_0\lambda_0^2}\sep d=3 \,.
\eeq
Using this in (\ref{xixsol}) and (\ref{tau4}) gives respectively
\beq
\xi_x={\mu_{x0}\mu_{10}^2\over D_0\lambda_0^2}\,,~~~~
\tau={\mu_{x0}\mu_{10}^4\over D_0^2\lambda_0^4}\sep d=3 \,.
\eeq
In $d=2$ ($\epsilon=2$), we obtain
\beq
\xi_\perp={(\mu_{x0}\mu_{10}^5)^{1\over4}\over \sqrt{D_0}\lambda_0}\sep d=2 \,,
\eeq
and
\beq
\xi_x={(\mu_{x0}\mu_{10})^{3\over4}\over \sqrt{D_0}\lambda_0} \,.~~~~
\tau={(\mu_{x0}\mu_{10}^3)^{1\over2}\over {D_0}\lambda_0^2}\sep d=2\,.
\eeq

We now turn to the last remaining concern about the scaling form of the correlation function $C_u$. We will now show that this is also independent of the arbitrary rescaling choices, and we'll also calculate it for small bare coupling $g_{10}$.

To do this, we see from (\ref{TIMF3}) that we need to calculate the value of  the integral $\int_0^{\ell^*}2\chi(\ell)d\ell$. We can obtain this by integrating the recursion relation (\ref{eq:D}) for the noise strength $D$ from $\ell=0$ to $\ell^*$, and using the fact that  $D(\ell^*)=D_{\rm ref}$. This gives
\bew
\beq
\ln\left({D_{\rm ref}\over D_0}\right)=\int_0^{\ell^*}[z(\ell) -2\chi(\ell)-\zeta(\ell)]d\ell+(1-d)\ell^*+\Phi_{D}(g_{10},g_{20})\,,
\label{Dsol}
\eeq
\ew
where we've defined
 \beq
 \Phi_{D}(g_{10}, g_{20})\equiv\int_0^{\ell^*}g_1(\ell)G_D(g_2(\ell))d\ell \,,
 \label{PhiDdef}
 \eeq
which again, like all of our $\Phi$'s, depends {\it only} on the bare couplings $g_{10,20}$.

Solving (\ref{Dsol}) for the integral $\int_0^{\ell^*}2\chi(\ell)d\ell$ gives
\bew
\beq
\int_0^{\ell^*}2\chi(\ell)d\ell=\int_0^{\ell^*}[z(\ell)-\zeta(\ell)]d\ell+\ln\left({D_0\over D_{\rm ref}}\right)+(1-d)\ell^*+\Phi_{D}(g_{10},g_{20})\,.
\label{chiintsol1}
\eeq
\ew
Using our results (\ref{intzsol}) and (\ref{zetaintsol}) for $\int_0^{\ell^*}z(\ell)d\ell$ and $\int_0^{\ell^*}\zeta(\ell)d\ell$ respectively, this becomes
\bew
\beq
\int_0^{\ell^*}2\chi(\ell)d\ell=\ln\left({D_0\mu_{\rm ref}\over \sqrt{\mu_{10}\mu_{x0}}D_{\rm ref}}\right)+(2-d)\ell^*+\Phi_{A}(g_{10},g_{20})\,,
\label{chiintsol2}
\eeq
\ew
where we've defined
\beq
\Phi_A\equiv\Phi_D-(\Phi_{\mu_1}+\Phi) \,.
\label{PhiAdef}
\eeq
Using this and our expression (\ref{xipdef}) for $\xi_\perp$ in equation (\ref{Adef}) for $A$ gives
\beq
A=a^{-2\xi_{FP}}\left({\xi_\perp\over a}\right)^{(2-d-2\chi_{\rm FP})}{\mu_{\rm ref} D_0\over\sqrt{\mu_{10}\mu_{x0}}D_{\rm ref}}e^{\Phi_A} \,.
\label{A}
\eeq
It is clear from this expression that, as claimed earlier, the value of $A$ depends only on the parameters of the bare model, not on the arbitrary choice of rescaling exponents.

For a model with the bare coupling $g_{10}\ll1$, we can set $\Phi_A=0$ in (\ref{A}), and obtain an explicit expression for $A$ in terms of the bare parameters:
\beq
A=a^{-2\xi_{FP}}\left({\xi_\perp\over a}\right)^{(2-d-2\chi_{\rm FP})}{\mu_{\rm ref} D_0\over\sqrt{\mu_{10}\mu_{x0}}D_{\rm ref}}\,.
\label{Aexp}
\eeq

Arguments virtually identical to those just presented can be used to show the scaling form of the density correlation function given by (\ref{rhoscale}).

The lengths $\xi_\perp$, and $\xi_x$, and the time $\tau$, have significance beyond their appearance in the scaling laws (\ref{eq:Cu_scaling}) and (\ref{rhoscale}): they are also the `non-linear lengths and time'. By this, we mean that, if all of the distances $r_\perp$, $|x-\gamma t|$, and the time $t$, are much smaller than the corresponding length or time -- that is, if $r_\perp \ll\xi_\perp$, $|x-\gamma t|\ll\xi_x$, and  $t\ll\tau$, the linear theory results of section III will apply. This can be seen either by a renormalization group argument, or perturbation theory.

 The renormalization group argument starts with the general trajectory integral matching expression (\ref{TIMF}). We then note that, if all the  conditions $r_\perp \ll\xi_\perp$, $|x-\gamma t|\ll\xi_x$, and  $t\ll\tau$ are satisfied, we can always choose to evaluate the right hand side at a value of $\ell<\ell^*$ at which  all the arguments $r_\perp e^{-\ell}$, $(x-\gamma t)\exp\bigg(-\int_0^{\ell}\zeta(\ell') d\ell'\bigg)$, and $t\exp\bigg(-\int_0^{\ell}z(\ell')\bigg)$ are microscopic. The $C_u$ on the right hand side of  (\ref{TIMF}) can then simply be  treated as a finite constant
since it is evaluated at short distances and times, and so will be unaffected by any   infrared divergences.

We'll now illustrate this
for the special case $r_{\perp}\ll\xi_{\perp}$, $x=0$, $t=0$. Again we choose $\ell=\ln\left(r_{\perp}/a\right)$. We will also choose $\chi$, $z$, and $\zeta$ to keep $\mu_{1,2,x}$ and $D$ fixed at their initial values. Since $\ell\ll\ell^*$  and  $r_{\perp}\ll\xi_{\perp}$, $g_1(\ell)$ is small if $g_{10}$ is small. Therefore, the graphical corrections in (\ref{eq:D},\ref{eq:mu1},\ref{eq:mu2})
are negligible. Then our special choices of the scaling exponents become
 \beq
 \chi={2-d\over 2}\,~~~~z=2\,,~~~~\zeta=1\,.
 \eeq
Inserting these exponents and $\ell=\ln\left(r_{\perp}/a\right)$ into (\ref{TIMF}) we obtain
\beqn
 &&C_u\Big(r_\perp, 0, 0; \big\{D_0, \mu_{x0}, \mu_{10}, \mu_{20}\big\} \Big)\nonumber\\
 &=&\left(r_{\perp}\over a\right)^{2-d}C_u\left(a, 0, 0; \big\{D_0, \mu_{x0}, \mu_{10}, \mu_{20}\big\}\right)\nonumber\\
 &\propto& r_{\perp}^{2-d}\,,
\eeqn
which agrees with the linear theory (\ref{Linearuu}). This argument can be easily extended  to correlation functions with more general spatial and temporal  separations, and to the density-density correlation function. Therefore, the conclusion that the linear theory results of section (\ref{lin}), in particular equations (\ref{Linearuu}) and (\ref{Linearrho}) for the velocity-velocity and density-density correlation functions,   hold if all distances and times are short compared to the corresponding non-linear lengths or times; that is, if the  conditions $r_\perp\ll\xi_\perp$, $|x-\gamma t|\ll\xi_x$, and  $t\ll\tau$ are satisfied.

\subsubsection{Perturbation theory approach}
In this subsection, we present the perturbation theory approach to the calculation of the non-linear lengths and time. We remind the reader that we do this by calculating perturbative corrections to the linear theory and finding the length and time scales on which they become appreciable. These prove to be precisely the lengths $\xi_\perp$, and $\xi_x$, and the time $\tau$ that we have just derived from the RG approach, thereby confirming the validity of that approach.

The perturbation theory calculation is very similar to the step 1 of the  DRG procedure which we described in section \ref{sec:rr}, and can also be represented by graphs. Here we focus on the correction to $\mu_1$ obtained from one particular graph, \fig \ref{fig:prop}a, which we have evaluated in detail in appendices \ref{sec:profiga} and \ref{sec:profiga0}.
Using different one-loop graphs, or considering renormalization of different parameters,  will lead to the same estimates of the non-linear lengths and times, up to factors of $O(1)$. We also simplify our calculation by  considering the case $\mu_2=0$; taking a non-zero $\mu_2$ only modifies the lengths $\xi_\perp$, and $\xi_x$, and the time $\tau$ by an $O(1)$ multiplicative factor.

Our strategy is to crudely estimate the nonlinear  lengths $\xi_\perp$ and $\xi_x$, and the nonlinear time $\tau$ by using their inverses as infra-red cutoffs of the infra-red divergent integrals that appear in a perturbation theory calculation of the renormalized $\mu_1$. We'll then determine the  values of $\xi_\perp$, $\xi_x$, and  $\tau$ as the values of these cutoffs for which the correction to $\mu_1$  becomes comparable to its bare value $\mu_{10}$. As mentioned earlier, we would get the same values for $\xi_\perp$, $\xi_x$, and  $\tau$ had we chosen to apply this logic to one of the parameters (i.e.,  $D$ or $\mu_x$) instead.

The graph \fig \ref{fig:prop}a represents a correction to $\pp_t u_j^\perp$ of the form
\beqn
&&\Delta (\pp_tu_j^\perp)_{\mu, a}\nonumber\\
&=&-{2D_0\lambda_0^2k_u^{\perp}u^{\perp}_c(\tilde{\bk})\over \left(2\pi\right)^{d+1}}
\int_{\tilde{q}}(k_i^\perp-q_i^\perp) C_{iu}(\tilde{\bq})G_{jc}(\tilde{\bk}-\tilde{\bq})\nonumber\\
&\equiv& -2D_0\lambda_0^2k_u^{\perp}u^{\perp}_c(\tilde{\bk})\left[ (I^{\mu, a}_1)_{cju}(\tilde{\bk}) - (I^{\mu, a}_2)_{cju}(\tilde{\bk})\right]\,,
\nonumber\\
\label{perturb}
\eeqn
where $\tilde{\bk}=(\omega,\bk)$, $\tilde{\bq}=(\Omega,\bq)$,
\beqn
 (I^{\mu, a}_1)_{cju} (\tilde{\bk}) &\equiv&
{k_i^\perp \over \left(2\pi\right)^{d+1}}\int_{\tilde{q}}
C_{iu}(\tilde{\bq})G_{jc}(\tilde{\bk}-\tilde{\bq})\,,\label{perturb_ia1}
\\
 (I^{\mu, a}_2)_{cju} (\tilde{\bk}) &\equiv&{1\over \left(2\pi\right)^{d+1}}\int_{\tilde{q}}
\label{perturb_ia2}
q_i^\perp C_{iu}(\tilde{\bq})G_{jc}(\tilde{\bk}-\tilde{\bq})\,,~~\\
G_{jc}(\tilde{\bq})&\equiv&G_T(\tilde{\bq})\delta^{\perp}_{jc}\nonumber\\
&=&\frac{\delta^{\perp}_{jc}}{-\ii\Omega+\mu_{10} q_\perp^2 +\mu_{x0} q_x^2}\,,
\label{perturb_gjc}\\
C_{iu}(\tilde{\bq})&\equiv& \mid G_T(\tilde{\bq})\mid^2\delta^{\perp}_{iu}
\nonumber\\
&=& \frac{\delta^{\perp}_{iu}}{\Omega^2 +\left(\mu_{10} q_\perp^2 +\mu_{x0} q_x^2\right)^2}\label{perturb_cau}\,,
\label{}
\eeqn
and the superscripts ``$\mu, a$"  indicate that this correction comes from the renormalization of the $\mu$ terms due to  \fig\ref{fig:prop}(a). Note that we have replaced all of the parameters $\lambda$, $D$, $\mu_1$, and $\mu_x$ by their bare values $\lambda_0$, $D_0$, $\mu_{10}$, and $\mu_{x0}$, since we are now doing perturbation theory, rather than the renormalization group.

Inserting (\ref{perturb_gjc},\ref{perturb_cau}) into (\ref{perturb_ia1},\ref{perturb_ia2}) we get
\beqn
 (I^{\mu, a}_1)_{cju} (\tilde{\bk}) &=& {k_u^\perp\delta^{\perp}_{jc}\over (2\pi)^{d+1}} \int_{\tilde{\bq} }  \mid G_T(\tilde{\bq})\mid^2G_T(\tilde{\bk}-\tilde{\bq})\nonumber\\\,,\label{}
\\
 (I^{\mu, a}_2)_{cju} (\tilde{\bk}) &=&{\delta^{\perp}_{jc}\over (2\pi)^{d+1}}\int_{\tilde{\bq} } q_u^\perp \mid G_T(\tilde{\bq})\mid^2G_T(\tilde{\bk}-\tilde{\bq})\nonumber\\\,.~~\label{}
\eeqn

Since (\ref{perturb}) has already a factor $k_u^\perp$ in front of it, and we are only interested in terms of $O(k^2)$ (since only these will be relevant at small $\bk$, that being the order of the  $\mu_1$ terms in the EOM),  to get relevant contributions to the linear terms of the EOM we can set the external frequency $\omega=0$ in $(I^a_{1,2})_{cju}(\tilde{\bk})$, and expand both of them to $O(k)$. This gives
\bew
\beqn
(I^{\mu, a}_1)_{cju} (\tilde{\bk}) &=& {k_u^\perp\delta^{\perp}_{jc}\over (2\pi)^{d+1}} \int_{\tilde{\bq} }  \mid G_T(\tilde{\bq})\mid^2G_T(-\tilde{\bq})
={k_u^\perp\delta^{\perp}_{jc}\over (2\pi)^{d+1}}\int_{\tilde{\bq} } {1\over (\Omega^2 +\left(\mu_{10} q_\perp^2 +\mu_{x0} q_x^2\right)^2)(\ii\Omega+\mu_{10} q_\perp^2 +\mu_{x0} q_x^2)}\ , \nonumber\\
\label{perturb_ia11}
\\
(I^{\mu, a}_2)_{cju} (\tilde{\bk})&=&{2\mu_{10}k_\ell^\perp\delta^{\perp}_{jc}\over (2\pi)^{d+1}}\int_{\tilde{\bq}}
q_u^\perp q_\ell^\perp\mid G_T(\tilde{\bq})\mid^2\left[G_T(-\tilde{\bq})\right]^2 ={2\mu_{10}k_\ell^\perp\delta^{\perp}_{jc}\over (2\pi)^{d+1}} \int_{\tilde{\bq} } {q_u^\perp q_\ell^\perp\over \bigg(\Omega^2 +\left(\mu_{10} q_\perp^2 +\mu_{x0} q_x^2\right)^2\bigg)^2}\, .
\nonumber\\ \label{perturb_ia21}
\eeqn
\ew
To calculate the non-linear length along $\perp$ directions, $\xi_{\perp}$, we  impose an infra-red cutoff $|\bq|_{\rm min}=\xi_\perp^{-1}$ on the $\bq_\perp$ integrals in this expression. That is, we define
\beq
\int_{\tilde{\bq}}\ \
\equiv\int_{-\infty}^{\infty}\dd\Omega\int_{-\infty}^{\infty}\dd q_x
\int_{\Lambda>{|\bq_\perp|>\xi_\perp^{-1}}}\dd^{d-1}q_\perp\,.
\eeq

The integrals over $\Omega$ and $q_x$ in (\ref{perturb_ia11}) and (\ref{perturb_ia21})
are straightforward, particularly if done in that order (i.e., integrating first over $\Omega$, then over $q_x$). The results are:
\beqn
&&(I^{\mu, a}_1)_{cju} (\tilde{\bk})
\nonumber\\
&=&\frac{k_u^\perp \delta^{\perp}_{jc}}{16\sqrt{\mu_{x0} \mu_{10}^3}}{1\over (2\pi)^{d-1}} \int_{\Lambda>{|\bq_\perp|>\xi_\perp^{-1}}}\frac{\dd^{d-1}q_\perp}{q_\perp^3}
\nonumber\\
&=&\frac{k_u^\perp\delta^{\perp}_{jc}}{16\sqrt{\mu_{x0} \mu_{10}^3}}{S_{d-1}\over (2\pi)^{d-1}}{1\over 4-d}
\left(\xi_\perp^{4-d}-\Lambda^{d-4}\right)\nonumber\\
&\approx&\frac{k_u^\perp\delta^{\perp}_{jc}}{16\sqrt{\mu_{x0} \mu_{10}^3}}{S_{d-1}\over (2\pi)^{d-1}} {\xi_\perp^{4-d}\over 4-d}\,,\label{perturb_ia12}
\eeqn
and
\beqn
&&(I^{\mu, a}_2)_{cju}(\tilde{\bk})\nonumber\\
&=&\frac{3k_\ell^\perp\delta^{\perp}_{jc}}{ 32\sqrt{\mu_{x0} \mu_{10}^3}}{1\over (2\pi)^{d-1}} \int_{\Lambda>{|\bq_\perp|
 >\xi_\perp^{-1}}}\frac{q^\perp_uq^\perp_\ell\dd^{d-1}q_\perp}{q_\perp^5}\nonumber\\
&=&\frac{3k_u^\perp\delta^{\perp}_{jc}}{32(d-1)\sqrt{\mu_{x0} \mu_{10}^3}}{1\over (2\pi)^{d-1}}\int_{\Lambda>{|\bq_\perp|
 >\xi_\perp^{-1}}}\frac{\dd^{d-1}q_\perp}{q_\perp^3}\nonumber\\
&\approx&\frac{3k_u^\perp\delta^{\perp}_{jc}}{32(d-1)\sqrt{\mu_{x0} \mu_{10}^3}}{S_{d-1}\over (2\pi)^{d-1}}
{\xi_\perp^{4-d}\over 4-d}\label{perturb_ia22}\,,
\eeqn
where, in the penultimate equality, we have used
\beq
\int\dd\Xi_{\bq_\perp}  \,\,q^\perp_uq^\perp_\ell=S_{d-1}{q_\perp^2\over d-1}\,,
\label{angleave}
\eeq
where $\int\dd\Xi_{\bq_\perp}$ denotes an integral over the $(d-1)$-dimensional solid angle associated with $\bq_\perp$. In the ultimate equality, we have
 used $\xi_{\perp}^{-1}\ll\Lambda$.

Inserting (\ref{perturb_ia12}, \ref{perturb_ia22}) into (\ref{perturb}) we obtain a correction to $\pp_t u_j$ given by
\bew
\beq
\Delta (\pp_tu_j)_{\mu,a}=-\frac{D_0\lambda_0^2}{\sqrt{\mu_{x0} \mu_{10}^3}}{S_{d-1}\over (2\pi)^{d-1}}
{\xi_\perp^{4-d}\over 4-d}\left[{1\over 8}-{3\over 16(d-1)}\right] k_\perp^2 u_j\,.
\eeq
\ew

From the form of this correction (i.e., the fact that it is proportional to $k_\perp^2 u_c$), we recognize this as a  perturbative correction to $\mu_1$:
\beq
(\Delta\mu_1)_{\mu,a}=\frac{D_0\lambda_0^2}{\sqrt{\mu_{x0} \mu_{10}^3}}{S_{d-1}\over (2\pi)^{d-1}}
{\xi_\perp^{4-d}\over 4-d}\left[{1\over 8}-{3\over 16(d-1)}\right]\,.
\eeq
Equating this correction to the bare $\mu_{10}$ gives, ignoring factors of $O(1)$,
\beq
\xi_{\perp}\propto\left(\mu_{x0}^0\mu_{10}^5\over D_0^2\lambda_0^4\right)^{1\over 2(4-d)}\,.
\eeq
This agrees with our earlier RG result (\ref{xip}).

To calculate the non-linear length $\xi_x$ along $x$, we now introduce $\xi_x^{-1}$ as an infrared cutoff on the integrals over $q_x$, and allow $\bq_\perp$ and $\Omega$ to run free. That is, we set the limits on our integrals as follows:
\beqn
\int_{\tilde{\bq}}\ \
\equiv 2\int_{-\infty}^{\infty}\dd\Omega\int_{|\bq_\perp|<\infty}\dd^{d-1}q_\perp
\int_{\xi_x^{-1}}^{\infty}\dd q_x
\,.
\eeqn
where the factor of $2$ takes into account the fact that the Brillouin zone in $q_x$, with this infrared cutoff, consists of two disjoint sections, one running from $\xi^{-1}$ to $\infty$, the other running from $-\infty$ to $-\xi^{-1}$ . These two regions make exactly equal contributions; hence the factor of $2$ above.

Then we obtain for the integrals in (\ref{perturb_ia11},\ref{perturb_ia21}):
\beqn
&&(I^{\mu, a}_1)_{cju} (\tilde{\bk})\nonumber\\
&=&{k_u^\perp\delta^{\perp}_{jc}\over 2(2\pi)^d}\int_{\xi_x^{-1}}^\infty\dd q_x
\int\frac{\dd^{d-1}q_\perp}{\left(\mu_{x0}q_x^2+\mu_{10}q_\perp^2\right)^2}\nonumber\\
&=&{k_u^\perp\delta^{\perp}_{jc} S_{d-1}\over (2\pi)^d}{H_1(d)\over 2}
\mu_{x0}^{d-5\over 2}\mu_{10}^{1-d\over 2}
\int_{\xi_x^{-1}}^\infty q_x^{d-5}\dd q_x\nonumber\\
&=&k_u^\perp\delta^{\perp}_{jc}{S_{d-1}\over (2\pi)^d}{H_1(d)\over 2}\mu_{x0}^{d-5\over 2}\mu_{10}^{1-d\over 2}
{\xi_x^{4-d}\over 4-d}\,,\label{perturb_ia13}
\eeqn
and
\beqn
&& (I^{\mu, a}_2)_{cju}(\tilde{\bk})\nonumber\\
&=&{\mu_{10}k^\perp_\ell\delta^{\perp}_{jc}\over 2(2\pi)^d}\int_{\xi_x^{-1}}^\infty\dd q_x
\int_{|\bq_\perp|<\infty}\frac{q^\perp_uq^\perp_\ell\dd^{d-1}q_\perp}
{\left(\mu_{x0}q_x^2+\mu_{10}q_\perp^2\right)^3}
\nonumber\\&=&{\mu_{10}k^\perp_u\delta^{\perp}_{jc}\over 2(d-1)(2\pi)^d}\int_{\xi_x^{-1}}^\infty\dd q_x
\int_{|\bq_\perp|<\infty}\frac{q_\perp^2\dd^{d-1}q_\perp}
{\left(\mu_{x0}q_x^2+\mu_{10}q_\perp^2\right)^3}
\nonumber\\
&=&{ S_{d-1}\mu_{10}k^\perp_u\delta^{\perp}_{jc}\over (2\pi)^d}{H_2(d)\over 2(d-1)}\mu_{x0}^{d-5\over 2}\mu_{10}^{1-d\over 2}
\int_{\xi_x^{-1}}^\infty q_x^{d-5}\dd q_x\nonumber\\
&=&{S_{d-1}\mu_{10}k^\perp_u\delta^{\perp}_{jc}\over (2\pi)^d}{H_2(d)\over 2(d-1)}\mu_{x0}^{d-5\over 2}\mu_{10}^{1-d\over 2}
{\xi_x^{4-d}\over 4-d}\,,\label{perturb_ia23}
\eeqn
where $H_{1,2}(d)$ are finite, $O(1)$ constants given by
\beq
H_1(d)\equiv\int_0^\infty{y^{d-2}\dd y\over \left(1+y^2\right)^2} ={\pi\over4}(d-3)\sec\left({\pi d\over2}\right)\,,~
\label{h1}
\eeq
and
\beq
H_2(d)\equiv\int_0^\infty{y^d\dd y\over \left(1+y^2\right)^3}={\pi\over16}(d-3)(1-d)\sec\left({\pi d\over2}\right)\,.
\label{h2}
\eeq
Note that, appearances to the contrary, neither $H_1(d=3)$ nor $H_2(d=3)$ vanishes; instead $H_1(3)=1/2$ and $H_2(3)=1/4$, as can be verified either by taking the singular limit $d\to3$ in (\ref{h1}) and (\ref{h2}), or by evaluating the corresponding integrals in exactly $d=3$.

Inserting (\ref{perturb_ia13},\ref{perturb_ia23}) into (\ref{perturb}) we obtain the perturbative correction to $\mu_1$:
\beqn
&&(\Delta\mu_1)_{\mu,a}\nonumber\\
&=&D_0\lambda_0^2\mu_{x0}^{d-5\over 2}\mu_{10}^{1-d\over 2}{S_{d-1}\over 2(2\pi)^d}
{\xi_x^{4-d}\over 4-d}\left[H_1(d)-{H_2(d)\over (d-1)}\right]\,.\nonumber\\
\label{perturb_mu1x}
\eeqn
Equating this correction to $\mu_1$ (\ref{perturb_mu1x}) to its bare value $\mu_{10}$ gives
\beq
\xi_x=\left(\mu_{x0}\mu_{10}^5\over D_0^2\lambda_0^4\right)^{1\over 2(4-d)} \sqrt{\mu_{x0}\over\mu_{10}}\times O(1)=\xi_\perp\sqrt{\mu_{x0}\over\mu_{10}}\times O(1)\,,
\eeq
which agrees with our earlier RG result (\ref{xixsol}).

Now we turn to the non-linear time scale $\tau$. We impose a lower limit $1/\tau$ on the frequency integral in (\ref{perturb}) and let the wave vectors be completely free:
\beq
\int_{\tilde{\bq}}\ \
\equiv 2 \int_{\tau^{-1}}^\infty\dd\Omega\int_{|\bq_\perp|<\infty}\dd^{d-1}q_\perp
\int_{-\infty}^{\infty}\dd q_x
\,,
\eeq
 where, much as in our treatment of the integral over $q_x$ earlier, the factor of $2$ takes into account the fact that the region of integration over $\Omega$, with this
infrared cutoff, consists of two disjoint sections, one running from $\tau^{-1}$ to $\infty$, the other running from
 $-\infty$ to $-\tau^{-1}$ . These two regions also make exactly equal contributions; hence the factor of $2$ above.

In this case we do the integral over wave vectors first. We obtain
\beqn
(I^{\mu, a}_1)_{cju} (\tilde{\bk})
&=&{2 H_3(d)\over (2\pi)^{d+1}} k_u^\perp\delta_{jc}^\perp
\mu_{x0}^{-{1\over 2}}\mu_{10}^{1-d\over 2}
\int_{1\over\tau}^\infty \omega^{{d\over 2}-3}\dd \omega\nonumber\\
&=&{4 H_3(d)\over (2\pi)^{d+1}} k_u^\perp\delta_{jc}^\perp
\mu_{x0}^{-{1\over 2}}\mu_{10}^{1-d\over 2}
{\tau^{4-d\over 2}\over 4-d}\,,\label{perturb_ia14}
\eeqn
and
\beqn
 (I^{\mu, a}_2)_{cju}(\tilde{\bk})
&=&{4 H_3(d)\over (2\pi)^{d+1}d} k_u^\perp\delta_{jc}^\perp
\mu_{x0}^{-{1\over 2}}\mu_{10}^{1-d\over 2}
\int_{1\over\tau}^\infty \omega^{{d\over 2}-3}\dd \omega\nonumber\\
&=&{8H_3(d)\over (2\pi)^{d+1}d} k_u^\perp\delta_{jc}^\perp
\mu_{x0}^{-{1\over 2}}\mu_{10}^{1-d\over 2}
{\tau^{4-d\over 2}\over 4-d}\,,\label{perturb_ia24}\,
\eeqn
where $H_{3}(d)$ is a finite, $O(1)$ constant given by
\beqn
H_3(d)&\equiv&\int_{|\bQ|<\infty}{Q^2\dd^d Q\over \left(1+Q^4\right)^2} ={(2-d)\over16}S_d\pi\sec\left({\pi d\over4}\right)\,.\nonumber\\
\eeqn

Inserting (\ref{perturb_ia14},\ref{perturb_ia24}) into (\ref{perturb}) we obtain the perturbative correction to $\mu_1$:
\beq
(\Delta\mu_1)_{\mu,a}={8D_0\lambda_0^2\mu_{x0}^{-{1\over 2}}\mu_{10}^{1-d\over 2}\over (2\pi)^{d+1}}
{\tau^{4-d\over 2}\over 4-d} \left({d-2\over d}\right)H_3(d)\,.
\label{}
\eeq
Equating this to $\mu_{10}$ gives
\beq
\tau=\left(\mu_{x0}\mu_{10}^5\over D_0^2\lambda_0^4\right)^{1\over (4-d)} {1\over{\mu_{10}}}\times O(1)={\xi_\perp^2\over{\mu_{10}}}\times O(1)\,,
\eeq
which agrees with (\ref{tau4}). This completes our calculation of the crossover length and time scales between linear and non-linear theories using perturbation theory.\\

	\subsection{Universal amplitude ratio\label{amp}}
The fact that there is a  fixed point  value of $g_2$, even if it's not zero at higher loop orders, implies an experimentally observable universal amplitude ratio. Specifically, it is the ratio of the damping of the transverse and the longitudinal modes, obtained as follows.
For the longitudinal mode, we look at the equal-time correlation:
\beq
C_L(t=0,\bk)=\frac{D(\bk)}{\mu_L(\bk) k_\perp^2 +\mu_x(\bk) k_x^2}
\ ,
 \eeq
 where we have explicitly shown the dependencies of the coefficients on the wavevector $\bk$.
A similar expression can be obtained for $C_T$.

If one considers a generic direction of ${\bf k}$ (i.e., any direction for which $k_\perp^\zeta \geq k_x$, we can ignore the second term in the above denominator and the ratio of $C_T(t=0,\bk)/C_L(t=0,\bk)$ is thus
\beq
\lim_{\bk_\perp \rightarrow 0}\frac{\mu_L(\bk)}{\mu_1(\bk)}  =
1+g_2^*  = 1+\cO(\epsilon^2)
 \ ,
\eeq
which is a universal number.\\

\subsection{Separatrix between positive and negative density correlations\label{}}

In section \ref{Sec:dencorr_linear} we have shown using linear theory that the sign of the equal time density correlation function depends on the spatial difference between the two correlating points $\br$. Specifically, in $\br$-space the positive and the negative regions of the density correlations are separated by a cone-shaped locus given by (\ref{Sepa_linear}), which, up to a $O(1)$ factor, can be rewritten in term of the non-linear lengths as
\beq
{|x|\over\xi_x}={r_\perp\over\xi_\perp}\,.\label{Sepa_linear1}
\eeq
For $x/\xi_x\gg r_\perp/\xi_\perp$ the density correlations are positive, while for $x/\xi_x\ll r_\perp/\xi_\perp$ they become negative. This result only holds for small distances (i.e., $x\ll\xi_x$ and $r_\perp\ll\xi_\perp$) since linear theory is
only valid at short length scales.

At large distances we expect a similar separatrix in $\br$-space which separates the regions with different signs of the density correlations. The scaling form of these correlations (\ref{rhoscale}) shows that sign of the equal time correlations is   determined by the ratio $|x|/\xi_x\over \left(r_\perp/\xi_\perp\right)^\zeta$ instead of $|x|/\xi_x\over \left(r_\perp/\xi_\perp\right)$. This implies at large distances
(i.e., $x\gg\xi_x$ or $r_\perp\gg\xi_\perp$) the positive and the negative density correlations are separated by a locus given by
\beq
{|x|\over\xi_x}=\left(r_\perp\over\xi_\perp\right)^\zeta\,.\label{Sepa_nonlinear}
\eeq
Note that (\ref{Sepa_linear1}) and (\ref{Sepa_nonlinear}) connect right at $|x|=\xi_x$.

The regions in $\br$-space with different signs of the density correlations are illustrated in \fig\ref{fig:rho_correlation}.

\section{Summary \& Outlook.}
Focusing on the ordered phase of a generic Malthusian flock in dimensions $d>2$, we have used dynamic renormalization group analysis to reveal a novel universality class that describes the system's hydrodynamic properties. In particular, the predicted scaling exponents were shown to converge to the known exact result in 2D. Our work highlights another instance in which an active system can be fundamentally different from known equilibrium and non-equilibrium systems. Looking ahead,
we believe that much analytical work is needed to verify whether novel universality class indeed underlie some of the phenomenology reported from simulation work \cite{ginelli_prl10,peshkov_prl12,mahault_prl18,nesbitt_a19}.

\onecolumngrid
\appendix
\section{One-loop RG calculation with $\mu_2$ set to zero\label{Sec:mu_2=0}}

In this appendix, we derive the dynamical renormalization group recursion relations for the special case of $\mu_2=0$. This restriction {\it immensely} simplifies the calculation, by making the propagators $G_{ij}$  and correlation functions $C_{ij}$ diagonal. It also proves to be sufficient to explore this region, since it contains the only stable fixed point in the problem, which we can find, and the exponents of which we can calculate, using this restricted approach. However, to demonstrate the stability of this fixed point against non-zero $\mu_2$, and to show that it is the {\it only} stable  fixed point (and, indeed, the only fixed point other than the unstable Gaussian fixed point), it is necessary to extend these calculations to non-zero $\mu_2$, which we do in the next Appendix.

For $\mu_2=0$, the EOM simplifies to:
\beqn
-\ii\omega\bu^\perp_i &=& -\left(\mu_1 k^2_\perp+\mu_xk_x^2 \right)\bu^\perp_i-
{\ii \lambda\over\left(\sqrt{2\pi}\right)^{d+1}} \int_{\tilde{\bq}} \left[ \bu_\perp(\tilde{\bq})\cdot ({\bk}_\perp-{\bq}_\perp)\right]  \bu^\perp_i(\tilde{\bk}-\tilde{\bq})+\bff^\perp_i
\ .
\eeqn
We  can formally solve this equation for $\bu$, which gives
\beqn
u^\perp_i(\tilde{\bk})=G_{ij}(\tilde{\bk})\left\{ f^\perp_j(\tilde{\bk})
-{\ii \lambda\over \left(\sqrt{2\pi}\right)^{d+1}}\int_{\tilde{\bq}} \left[ \bu_{\perp} ({\bk}_{\perp}-{\bq}_{\perp})\cdot {\bq}_{\perp}\right] u_j^{\perp}(\tilde{\bq})\right\}
\, ,
\label{gij0}
\eeqn
where
\beq
G_{ij}(\tilde{\bk}) =G_T(\tilde{\bk})\delta^{\perp}_{ij}= \frac{\delta^{\perp}_{ij}}{-\ii \omega +\mu_1 k_\perp^2 +\mu_x k_x^2}\,\label{gij0}
\eeq
is diagonal, as noted above.

We now apply the dynamical renormalization group procedure of \cite{FNS} to this equation. As described in section IV, the first step of this procedure consists of averaging the
above solution  over  the short wavelength components $\bff^>(\tilde{\bk})$ of the noise $\bff$, which gives a closed EOM  for $\bup^<(\tilde{\bk})$.
This step can be represented by graphs. The  basic rules for the graphical representation are illustrated in Fig. {\ref{FR}}. We will now evaluate these graphs, each of which can be interpreted as adding a term to the equation of motion for $\bup^<(\tilde{\bk})$. We begin with
the graph in  \fig \ref{fig:prop}(a).

\subsection{Renormalizations of the $\mu$'s}

\subsubsection{Graph in \fig \ref{fig:prop}(a)\label{sec:profiga0}}
The graph   in \fig \ref{fig:prop}(a) gives a contribution  $\Delta(\pp_tu^<_j)_{\mu,a}$ to the EOM  for $u_j^<(\tilde{\bk})$:
\beqn
 \Delta(\pp_tu^<_j)_{\mu,a}=-{2D\lambda^2k_u^{\perp}u^{\perp}_c(\tilde{\bk})\over \left(2\pi\right)^{d+1}}
\int_{\tilde{q}}^>(k_i^\perp-q_i^\perp) C_{iu}(\tilde{\bq})G_{jc}(\tilde{\bk}-\tilde{\bq})
\equiv -2D\lambda^2k_u^{\perp}u^{\perp}_c(\tilde{\bk})\left[ (I^{\mu,a}_1)_{cju}(\tilde{\bk}) - (I^{\mu,a}_2)_{cju}(\tilde{\bk})\right]\,,\nonumber\\\label{pa0}
\eeqn
where
\beq
\int_{\tilde{\bq}}^>\ \
\equiv\int_{\Lambda>{|\bq_\perp|>\Lambda e^{-\dd\ell}}}\dd^{d-1}q_\perp\int_{-\infty}^{\infty}\dd\Omega\int_{-\infty}^{\infty}\dd q_x
\label{int>def}
\,,
\eeq

\beqn
C_{iu}(\tilde{\bq})&\equiv& \mid G_T(\tilde{\bq})\mid^2\delta^{\perp}_{iu}= \frac{\delta^{\perp}_{iu}}{\omega^2 +\left(\mu_1 q_\perp^2 +\mu_x q_x^2\right)^2}\,,
\label{cau0}\\
( I^{\mu, a}_1)_{cju}(\tilde{\bk}) &\equiv&
{k_i^\perp \over \left(2\pi\right)^{d+1}}\int_{\tilde{q}}^>
C_{iu}(\tilde{\bq})G_{jc}(\tilde{\bk}-\tilde{\bq})\,,\label{ia10}
\\
 ( I^{\mu, a}_2)_{cju} (\tilde{\bk}) &\equiv&{1\over \left(2\pi\right)^{d+1}}\int_{\tilde{q}}^>
\label{ia20}
q_i^\perp C_{iu}(\tilde{\bq})G_{jc}(\tilde{\bk}-\tilde{\bq})\,.
\,.
\eeqn

 Since we are interested in terms only up to $O(k^2)$, since that is the order of the $\mu$ terms in the EOM, and (\ref{pa0}) already has an explicit factor $k^\perp_u$, we need only expand these integrals $(I^{5a}_{1,2})_{cju}$ up to linear order in $\bk$. Doing so, and
inserting (\ref{gij0},\ref{int>def}), and (\ref{cau0}) into (\ref{ia10}) and (\ref{ia20}), we obtain to this order
\beqn
( I^{\mu, a}_1)_{cju}  (\tilde{\bk})&=& {k_u^\perp{ \delta^\perp_{jc}}\over (2\pi)^{d+1}} \int_{\tilde{\bq} }^>  \mid G_T(\tilde{\bq})\mid^2G_T(-\tilde{\bq})=\frac{k_u^\perp}{16}
\frac{D\lambda^2}{\sqrt{\mu_x \mu_1^3}} \frac{S_{d-1}}{(2\pi)^{d-1}} \Lambda^{d-4}\dd \ell\,,\label{ia101}
\\
( I^{\mu, a}_2)_{cju}(\tilde{\bk}) &=&{ \delta^\perp_{jc}\over (2\pi)^{d+1}}\int_{\tilde{\bq} }^> q_u^\perp \mid G_T(\tilde{\bq})\mid^2G_T(\tilde{\bk}-\tilde{\bq})\nonumber
\\
&=&{ \delta^\perp_{jc}\over (2\pi)^{d+1}}\int_{\tilde{\bq}}^> q_u^\perp \mid G_T(\tilde{\bq})\mid^2\left[G_T(-\tilde{\bq})\right]^2 \left(2\mu_1\bq_\perp \cdot \bk_\perp+ 2\mu_x q_x k_x \right)\nonumber
\\
&=&
\frac{3k_u^\perp{ \delta^\perp_{jc}}}{64 (d-1)}\frac{D\lambda^2}{\sqrt{\mu_x \mu_1^3}} \frac{S_{d-1}}{(2\pi)^{d-1}} \Lambda^{d-4}\dd \ell\,,\label{ia201}
\eeqn
where we have only kept terms up to $O(k)$.  In the last equality in (\ref{ia201}), we have used the fact that the second ($2\mu_x$) term in the parenthesis is odd in $q_x$,  and so integrates to zero. We have also evaluated the first term by using the fact that
\beq
\int\dd\Xi_{\bq_\perp}  \,\,q^\perp_u\bq^\perp \cdot \bk_\perp=\int\dd\Xi_{\bq_\perp}  \,\,q^\perp_uq^\perp_\ell k^\perp_\ell=S_{d-1}\delta^\perp_{u\ell}k^\perp_\ell{q_\perp^2\over d-1}=S_{d-1}k^\perp_u{q_\perp^2\over d-1}\,,
\label{angleave2}
\eeq
where $\int\dd\Xi_{\bq_\perp}$ denotes an integral over the $d-1$-dimensional solid angle associated with $\bq_\perp$.

Inserting (\ref{ia101}) and (\ref{ia201}) into (\ref{pa0}),  we find:
\beq
\Delta(\pp_tu^<_j)_{\mu,a}=-\bigg({1\over 32} \frac{D\lambda^2}{\sqrt{\mu_x \mu_1^3}} \frac{S_{d-1}}{(2\pi)^{d-1}} \Lambda^{d-4}\dd \ell\left[4-{3\over (d-1)}\right]\bigg)=-\left[{1\over 8}-{3\over 32(d-1)}\right](g_1\mu_1\dd\ell )k^2u^<_j
\eeq
where in the second equality we have used our earlier definition (\ref{eq:def_g}) of $g_1$. Since this contribution to $\pp_tu^<_j$ has the same form as the $\mu_1$ term already present, we can absorb it into a renormalization of $\mu_1$:
\beq
(\delta\mu_1)_{ \mu,a}={1\over 32} \frac{D\lambda^2}{\sqrt{\mu_x \mu_1^3}} \frac{S_{d-1}}{(2\pi)^{d-1}} \Lambda^{d-4}\dd \ell\left[4-{3\over (d-1)}
\right]=\left[{1\over 8}-{3\over 32(d-1)}\right]g_1\mu_1\dd\ell\,.
\eeq

\subsubsection{Graph in \fig \ref{fig:prop}(b)}
The graph   in \fig \ref{fig:prop}(b) gives a contribution  $\Delta(\pp_tu^<_j)_{\mu,b}$ to the  EOM  for $u_j^<(\tilde{\bk})$:
\beqn
\Delta(\pp_tu^<_j)_{ \mu,b}=-{2\lambda^2Dk_u^{\perp}u_c^{\perp}(\tilde{\bk})\over (2\pi)^{d+1}}\int_{\tilde{\bq} }^> q_i^\perp C_{ju}(\tilde{\bq})G_{ic}(\tilde{\bk}-\tilde{\bq})
=-2\lambda^2 Dk_u^{\perp}u_b^{\perp}(\tilde{\bk}) ( I^{\mu,b})_{cju}(\tilde{\bk})\,,\label{pb0}
\eeqn
where
\beqn
( I^{\mu,b})_{cju}(\tilde{\bk})\equiv {1\over (2\pi)^{d+1}}\int_{\tilde{\bq}}^> q_i^\perp
C_{ju}(\tilde{\bq})G_{ic}(\tilde{\bk}-\tilde{\bq})\,.\label{ib01}
\eeqn

Inserting (\ref{gij0},\ref{cau0}) into (\ref{ib01}) we get
\beqn
( I^{\mu,b})_{cju}(\tilde{\bk})&=&{\delta_{uj}^\perp\over (2\pi)^{d+1}}\int_{\tilde{\bq}}^> q_c^\perp
\mid G_T(\tilde{\bq})\mid^2G_T(\tilde{\bk}-\tilde{\bq})\nonumber\\
&=&{\delta_{uj}^\perp\over (2\pi)^{d+1}} \int_{\tilde{\bq}}^> q_c^\perp
\mid G_T(\tilde{\bq})\mid^2\left[G(-\tilde{\bq})\right]^2 \left(2\mu_1\bq_\perp \cdot \bk_\perp+ 2\mu_x q_x k_x \right)
\\
&=&\frac{3\delta_{uj}^\perp k_c^\perp}{64 (d-1) } \frac{D\lambda^2}{\sqrt{\mu_x \mu_1^3}} \frac{S_{d-1}}{(2\pi)^{d-1}} \Lambda^{d-4}\dd \ell\, .\label{ib02}
\eeqn
where we have only kept terms up to $O(k)$, and we have again used (\ref{angleave2})  to evaluate the angular integrals, and thrown out odd terms that integrate to zero.

Inserting (\ref{ib02}) into (\ref{pb0}) we obtain a modification to the equation of motion for $u_j^<$:
\beq
\Delta(\pp_tu^<_j)_{\mu,b}=-\bigg(\frac{3 }{32 (d-1) } \frac{D\lambda^2}{\sqrt{\mu_x \mu_1^3}} \frac{S_{d-1}}{(2\pi)^{d-1}} \Lambda^{d-4}\dd \ell\bigg)k^\perp_jk^\perp_cu_c^\perp \,.
\label{5bcorr}
\eeq
From the form of this correction (namely, the fact that it is proportional to $k^\perp_jk^\perp_bu_b^\perp$), we can identify this as
 a correction to $\mu_2$ (which we remind the reader is the parameter whose bare value we took to be zero):
\beqn
(\delta\mu_2)_{ \mu, b}=\frac{3 }{32 (d-1) } \frac{D\lambda^2}{\sqrt{\mu_x \mu_1^3}} \frac{S_{d-1}}{(2\pi)^{d-1}} \Lambda^{d-4}\dd \ell\, .
\label{dmu2.1}
\eeqn
 Thus, it would appear at this point that, even starting as we have with a model in which $\mu_2=0$, we generate a non-zero $\mu_2$ upon renormalization. This proves to {\it not} be the case, at least to one loop order. Instead, to this order, (\ref{dmu2.1}) is exactly cancelled by other graphs, as we will now see.

\subsubsection{Graph in \fig \ref{fig:prop}(c)}
The graph   in \fig \ref{fig:prop}(c) gives a contribution $\Delta(\pp_tu^<_j)_{\mu,c}$ to the EOM  for $u_j^<(\tilde{\bk})$:
\beqn
\Delta(\pp_tu^<_j)_{\mu,c}={2\lambda^2Du^{\perp}_u(\tilde{\bk})\over (2\pi)^{d+1}}
\int_{\tilde{\bq} }^> (k_i^\perp-q_i^\perp)q^\perp_u
C_{i\ell}(\tilde{\bq})G_{j\ell}(\tilde{\bk}-\tilde{\bq})\equiv 2\lambda^2 Du^{\perp}_u(\tilde{\bk}) \left[(I^{\mu,c}_1)_{ju} (\tilde{\bk}) + (I^{\mu,c}_2)_{ju}(\tilde{\bk})\right]\,,
\label{pc0}
\eeqn
where
\beqn
(I^{\mu,c}_1)_{ju} (\tilde{\bk})&\equiv&{k_i^{\perp}\over (2\pi)^{d+1}}
\int_{\tilde{\bq} }^> q^\perp_u
C_{i\ell}(\tilde{\bq})G_{j\ell}(\tilde{\bk}-\tilde{\bq})\,\label{ic101}\\
(I^{\mu,c}_2)_{ju} (\tilde{\bk})&\equiv&-{1\over (2\pi)^{d+1}}
\int_{\tilde{\bq} }^> q^\perp_iq^\perp_u
C_{i\ell}(\tilde{\bq})G_{j\ell}(\tilde{\bk}-\tilde{\bq})\,.\label{ic201}
\eeqn

Inserting (\ref{cau0},\ref{gij0}) into (\ref{ic101}) leads to
\beqn
(I^{\mu,c}_1)_{ju}  (\tilde{\bk})&=& {k_j^\perp\over (2\pi)^{d+1}} \int_{\tilde{\bq}}^> q^\perp_u
\mid G_T(\tilde{\bq})\mid^2G_T(\tilde{\bk}-\tilde{\bq})\nonumber
\\
&=& {k_j^\perp\over (2\pi)^{d+1}} \int_{\tilde{\bq} }^> q_u^\perp
\mid G_T(\tilde{\bq})\mid^2\left[G_T(-\tilde{\bq})\right]^2
\left(2\mu_1\bq^\perp \cdot \bk^\perp+ 2\mu_x q_x k_x \right)
\nonumber\\
&=&\frac{3k_j^\perp k_u^\perp}{64 (d-1)}\frac{D\lambda^2}{\sqrt{\mu_x \mu_1^3}} \frac{S_{d-1}}{(2\pi)^{d-1}} \Lambda^{d-4}\dd \ell\ .
\label{ic102}
\eeqn
We deliberately leave $(I^{\mu,c}_2)_{ju}  (\tilde{\bk})$ untouched since will show later in next section that this piece is canceled out by that from \fig \ref{fig:prop}(d).

Inserting only this $(I^{\mu,c}_1)_{ju} $ contribution (\ref{ic102}) into (\ref{pc0}) leads to a  term in the equation of motion for $u_j^<$:
\beq
\Delta(\pp_tu^<_j)_{\mu,c}=\bigg(\frac{3 }{32 (d-1) } \frac{D\lambda^2}{\sqrt{\mu_x \mu_1^3}} \frac{S_{d-1}}{(2\pi)^{d-1}} \Lambda^{d-4}\dd \ell\bigg)k^\perp_jk^\perp_uu_u^\perp \,.
\label{5bcorr}
\eeq
which, as before,  can be interpreted as a correction to $\mu_2$:
\beq
\label{eq:zeromu2_3}
\delta\mu_2=-\frac{3}{32(d-1)}\frac{D\lambda^2}{\sqrt{\mu_x \mu_1^3}} \frac{S_{d-1}}{(2\pi)^{d-1}} \Lambda^{d-4}\dd \ell
\ .
\eeq
Note that this exactly cancels the contribution to $\mu_2$ from \fig \ref{fig:prop}b that we just calculated.

\subsubsection{Graph in \fig \ref{fig:prop}(d)}
The graph   in \fig \ref{fig:prop}(d) gives a contribution $\Delta(\pp_tu^<_j)_{\mu,d}$ to the EOM  for $u_j^<(\tilde{\bk})$:
\beqn
\Delta(\pp_tu^<_j)_{\mu,d}={2\lambda^2Du_u^{\perp}(\tilde{\bk})\over (2\pi)^{d+1}}\int_{\tilde{\bq} }^> q_i^\perp q^\perp_u
C_{j\ell}(\tilde{\bq})G_{i\ell}(\tilde{\bk}-\tilde{\bq})
\equiv 2\lambda^2 Du_u^{\perp}(\tilde{\bk}) (I^{\mu,d})_{uj}(\tilde{\bk})
\ ,
\eeqn
where
\beqn
(I^{\mu,d})_{uj}(\tilde{\bk})\equiv {1\over (2\pi)^{d+1}}\int_{\tilde{\bq} }^> q_i^\perp q^\perp_uC_{j\ell}(\tilde{\bq})G_{i\ell}(\tilde{\bk}-\tilde{\bq})\,.
\eeqn
This contribution cancels out the $I^{\mu,c}_2$contribution from \fig \ref{fig:prop}c above exactly, leaving no correction to $\mu_2$ at all, to one loop order. Thus, to this order, $\mu_2=0$ is a fixed point.

\subsection{Noise renormalization}
The graphs   in \fig \ref{fig:noise}(a) and  \fig \ref{fig:noise}(b) represent the following two corrections to the noise correlator $\left<f_\ell(\tilde\bk)f_u(-\tilde\bk)\right>$:
\beqn
\Delta\left<f_\ell(\tilde\bk)f_u(-\tilde\bk)\right>_{D,a}&=&{2\lambda^2D^2\over (2\pi)^{d+1}}\int_{\tilde{\bq} }^>q_i^\perp q_m^\perp C_{im}(\tilde{\bk}-\tilde{\bq})C_{\ell u}(\tilde{\bq})\ ,\label{pd01}\\
\Delta\left<f_\ell(\tilde\bk)f_u(-\tilde\bk)\right>_{D,b}&=&{2\lambda^2D^2\over (2\pi)^{d+1}}\int_{\tilde{\bq} }^>q_i^\perp (k_m^\perp -q_m^\perp ) C_{iu}(\tilde{\bk}-\tilde{\bq})C_{\ell m}(\tilde{\bq})\label{pd02}\ .
\eeqn
 Since the noise strength $D$ is the value of this correlation at $\bk={\bf 0}$,
we set $\tilde{\bk}=0$ in (\ref{pd01},\ref{pd02}) to get
\beqn
\Delta\left<f_\ell(\tilde\bk)f_u(-\tilde\bk)\right>_{D,a}&=&{2\lambda^2D^2\over (2\pi)^{d+1}}\int_{\tilde{\bq} }^>q_i^\perp q_m^\perp C_{im}(-\tilde{\bq})
C_{\ell u}(\tilde{\bq})\ ,\label{}\\
\Delta\left<f_\ell(\tilde\bk)f_u(-\tilde\bk)\right>_{D,b}&=&{2\lambda^2D^2\over (2\pi)^{d+1}}\int_{\tilde{\bq} }^>q_i^\perp (-q_m^\perp )
C_{iu}(-\tilde{\bq})C_{\ell m}(\tilde{\bq})\label{}\ .
\eeqn
Inserting  our expression (\ref{cau0})  for the correlation function into the above two formulae we get
\beqn
{\Delta\left<f_\ell(\tilde\bk)f_u(-\tilde\bk)\right>_{D,a}}&=&{2\lambda^2D^2\over (2\pi)^{d+1}}\delta_{\ell u}^\perp\int_{\tilde{\bq} }^>q_\perp^2 \mid G_T(\tilde{\bq})\mid^4={3\over 32}\frac{D^2\lambda^2}{\sqrt{\mu_x \mu_1^5}} \frac{S_{d-1}}{(2\pi)^{d-1}} \Lambda^{d-4}\dd\ell
\delta_{\ell u}^\perp\ ,\label{pd03}\\
{ \Delta\left<f_\ell(\tilde\bk)f_u(-\tilde\bk)\right>_{D,b}}
&=&-{2\lambda^2D^2\over (2\pi)^{d+1}}{\delta_{\ell u}^\perp\over d-1}\int_{\tilde{\bq} }^>q_\perp^2 \mid G_T(\tilde{\bq})\mid^4=-{3\over 32(d-1)}\frac{D^2\lambda^2}{\sqrt{\mu_x \mu_1^5}} \frac{S_{d-1}}{(2\pi)^{d-1}} \Lambda^{d-4}\dd\ell\delta_{\ell u}^\perp\label{pd04}\ ,
\eeqn
where in the first equality of (\ref{pd04}) we have again used the angular average  (\ref{Angleave3}) derived in appendix (\ref{sec:formulae}).

Adding these two pieces together,  and identifying the coefficient of $\delta_{\ell u}^\perp$  as a  correction to $D$ gives  the total correction $\delta D$ to $D$ to one loop order:
\beq
\delta D={3\over 32}\left(1-{1\over d-1}\right)\frac{D^2\lambda^2}{\sqrt{\mu_x \mu_1^5}} \frac{S_{d-1}}{(2\pi)^{d-1}} \Lambda^{d-4}\dd\ell={3\over 32}\left(1-{1\over d-1}\right)g_1D\dd\ell\,.
\eeq

\subsection{
 Summary of all corrections to one loop order in the $\mu_2=0$ limit}

 Adding up the results obtained in previous sections   gives the total one loop graphical corrections to the various parameters when $\mu_2=0$:
\beqn
&&\delta\mu_1=\left[{1\over 8}-{3\over 32(d-1)}\right]g_1\mu_1\dd\ell\,,\label{mu1graphmu2=0}\\
&&\delta D={3\over 32}\left(1-{1\over (d-1)}\right)g_1D\dd\ell\,,\label{Dgraphmu2=0}\\
&&\delta\mu_2=0\,,\label{mu2graphmu2=0}\\
&&\delta\mu_x=0\label{muxgraphmu2=0}\,.
\eeqn
Dividing both sides of each of these equations by $\dd\ell$, we obtain the graphical contributions to the recursion relations for the parameters of our model, in the special case $\mu_2=0$:
\beqn
&&\left({\dd\mu_1\over\dd\ell}\right)_{\rm graph}=\left[{1\over 8}-{3\over 32(d-1)}\right]g_1\mu_1\,,\label{mu1graphmu2=0rr}\\
&&\left({\dd D\over\dd\ell}\right)_{\rm graph}={3\over 32}\left({d-2\over d-1}\right)g_1D\,,\label{Dgraphmu2=0rr}\\
&&\left({\dd\mu_2\over\dd\ell}\right)_{\rm graph}=0\,,\label{mu2graphmu2=0rr}\\
&&\left({\dd\mu_x\over\dd\ell}\right)_{\rm graph}=0\label{muxgraphmu2=0rr}\,.
\eeqn

 The vanishing of the correction to $\mu_2$ at this one loop order, in the restricted model in which the bare $\mu_2=0$, tells us that at one loop order $\mu_2=0$ is a fixed point. It requires analysis at non-zero $\mu_2$, which we perform in the next appendix, to show that this $\mu_2=0$ fixed point is actually stable, and furthermore, is the only fixed point in the problem.

Since $g_2\equiv{\mu_2\over\mu_1}$ vanishes when $\mu_2=0$, our results (\ref{mu1graphmu2=0}-\ref{muxgraphmu2=0}) constrain the values of $G_{\mu_{1,2}, D}(g_2=0)$ in equations (\ref{eq:mu1}), (\ref{eq:mu2}), and (\ref{eq:D}) as
\beqn
G_{\mu_{1}}(g_2=0)&=&{1\over 8}-{3\over 32(d-1)}\,,\label{Gmu1mu2=0}\\
\bigg(\mu_2G_{\mu_{2}}(g_2=0)\bigg)_{\mu_2=0}&=&0\,,\label{Gmu2mu2=0}\\
G_{D}(g_2=0)&=&{3\over 32}\left({d-2\over d-1}\right)\,.\label{GDmu2=0}
\eeqn
These in turn fix the value of $G_{g_1}$ as
\beqn
&&G_{g_{1}}(\mu_2=0)=G_{D}(\mu_2=0)-{5\over2}G_{\mu_1}(\mu_2=0)={23-14d\over 64(d-1)}\,,\label{Gg1mu2=0}\\
\eeqn
which is the value that we used in our analysis of the fixed points and exponents in section (\ref{nonlin}).   Note that we can not, by this $\mu_2=0$ analysis, which only gives us equation (\ref{Gmu2mu2=0}), say anything about the behavior  of $G_{g_{2}}(g_2)=G_{\mu_{2}}(g_2)-G_{\mu_{1}}(g_2)$ in the limit $g_2\to0$, other than that $G_{g_{2}}(g_2) g_2\to0$ as $g_2\to0$, which can be satisfied if $G_{g_{2}}(g_2)$ approached {\it any} finite limit as $g_2\to0$.

 We obtain the same values for $G_{\mu_{1}, D}(g_2=0)$ and $G_{g_{1}}(\mu_2=0)$, albeit with much greater effort, by taking the tricky limit $\mu_2\to0$ of the recursion relations for the full problem with $\mu_2\ne0$, which we'll derive in the next appendix. This provides a reassuring check on the accuracy of those far more difficult calculations, to which we now, with some trepidation, turn.

\section{One-loop graphical corrections with non-zero $\mu_2$ }

We now turn to the calculation of the graphical corrections to the parameters for the full model with $\mu_2\ne0$.
The reasoning of this section is exactly the same as that of the previous section; the only difference is that the algebra
is complicated by the non-zero value of $\mu_2$.

The origin of this complication lies in the more complicated form of the propagators and correlation functions. Instead of the simple, diagonal expressions (\ref{gij0}) and (\ref{cau0}) that we have when $\mu_2=0$, we now, as shown in section (\ref{lin}), have, for the propagators:
\beq
G_{ij}(\tilde{\bk})\equiv L^\perp_{ij}(\bkp)G_L(\tilde{\bk}) +P^\perp_{ij}(\bkp)G_T(\tilde{\bk})\,,
\eeq
with
\beqn
G_L(\tilde{\bk}) &=& \frac{1}{-\ii \omega +\mu_L k_\perp^2 +\mu_x k_x^2} \ ,
\\
G_T(\tilde{\bk}) &=& \frac{1}{-\ii \omega +\mu_1 k_\perp^2 +\mu_x k_x^2} \ ,
\eeqn
and the longitudinal and transverse projection operators $L^\perp_{ij}(\bkp)$ and $P^\perp_{ij}(\bkp)$, respectively, defined as
and we have defined the ``longitudinal projection operator"
\beq
L^\perp_{ij}(\bkp)\equiv k^\perp_ik^\perp_j/k_\perp^2 \ ,
\label{Ldefapp}
\eeq
which projects any vector along $\bkp$, and
\beq
P^\perp_{ij}(\bkp)\equiv \delta^\perp_{ij}k^\perp_ik^\perp_j/k_\perp^2,
\label{Pdefapp}
\eeq
which projects any vector onto the space orthogonal to both the mean direction of flock motion $\hat{x}$ and $\bkp$.

We now also have similar decompositions for the correlation functions:
\beqn
C_{ij}(\tilde{\bk})\equiv
L_{ij}^\perp(\bk)|G_L(\tilde{\bk})|^2 +P_{ij}^\perp(\bk)|G_T(\tilde{\bk})|^2\,.
\eeqn

With these in hand, we'll now calculate the graphical corrections to the various parameters in the full model with $\mu_2\ne0$. Note that all of the graphs are exactly the same as those we evaluated in the previous section; all that will change is their values, because we are now taking $\mu_2\ne0$.

\subsection{Renormalizations of $\mu_{1,2,x}$}

\subsubsection{Graph in \fig \ref{fig:prop}(a)\label{sec:profiga}}
The graph   in \fig \ref{fig:prop}a gives a contribution $\Delta(\pp_tu^<_j)_{\mu,a}$ to the
EOM  for $u_j^<(\tilde{\bk})$:
\beqn
{\Delta(\pp_tu^<_j)_{\mu,a}}=-{2D\lambda^2k_u^{\perp}u^{\perp}_c(\tilde{\bk})\over \left(2\pi\right)^{d+1}}
\int_{\tilde{q}}^>(k_i^\perp-q_i^\perp) C_{iu}(\tilde{\bq})G_{jc}(\tilde{\bk}-\tilde{\bq})
\equiv -2D\lambda^2k_u^{\perp}u^{\perp}_c(\tilde{\bk})\left[ {(I^{\mu, a}_1)}_{cju}(\tilde{\bk}) - (
{(I^{\mu, a}_1)}_{cju}(\tilde{\bk})\right] \ ,\nonumber\\
\eeqn
where
\beqn
{(I^{\mu, a}_1)}_{cju} (\tilde{\bk}) &\equiv&
{k_i^\perp \over \left(2\pi\right)^{d+1}}\int_{\tilde{q}}^>
C_{iu}(\tilde{\bq})G_{jc}(\tilde{\bk}-\tilde{\bq})\,,
\\
{(I^{\mu, a}_2)}_{cju} (\tilde{\bk}) &\equiv&{1\over \left(2\pi\right)^{d+1}}\int_{\tilde{q}}^>
q_i^\perp C_{iu}(\tilde{\bq})G_{jc}(\tilde{\bk}-\tilde{\bq})\,,\label{I5adef}\,.
\eeqn

Note that the overall correction already has a common factor $k_u$
outside the loop integral. That means for both $ {(I^{\mu, a}_1)}_{cju}$ and $ {(I^{\mu, a}_2)}_{cju}$ we can set $\omega=0$ inside the loop integral since expanding the loop integral to $O(\omega)$ or higher orders in $\omega$ only gives terms irrelevant compared to $-\ii\omega\bup$, which is already present in the EOM (\ref{eq:main1}).  This also means  that we can obtain the renormalization of the $\mu$'s, which enter the EOM at $O(k^2)$, by setting
$\bk={\mathbf 0}$ inside the integral for ${(I^{\mu, a}_1)}_{cju}$,  and expanding the integral $ {(I^{\mu, a}_2)}_{cju}$ to $O(k)$.

Let's calculate ${(I^{\mu, a}_1)}_{cju}$ first. {Using the integrals and angular averages given in appendix  \ref{sec:formulae}, we obtain}
\beqn
{(I^{\mu, a}_1)}_{cju}(\tilde{\bk}) &=&{k_i^\perp K_d\dd \ell\over (2\pi)^{d+1}}
\int_{\tilde{\bq} }^> \left[
\mid G_L(\tilde{\bq})\mid^2L^\perp_{iu}({\bq}) +\mid G_T(\tilde{\bq})\mid^2P^\perp_{iu}({\bq})\right]
\left[G_L(-\tilde{\bq})L^\perp_{jc}(-{\bq}) +G_T(-\tilde{\bq})P^\perp_{jc}(-{\bq})\right]\nonumber
\\
&=&
{k_i^\perp \over (2\pi)^{d-1}}
\int_{\bq_\perp }^> \Bigg[
\frac{1}{16\sqrt{\mu_x \mu_L^3}} \frac{ L^\perp_{iu}({\bq})L^\perp_{jc}({\bq})}{q_\perp^3}  +
\frac{1}{16\sqrt{\mu_x \mu_1^3}} \frac{ P^\perp_{iu}({\bq})P^\perp_{jc}({\bq})}{q_\perp^3}
\nonumber\\
&&
+A(\mu_L,\mu_1)\frac{ L^\perp_{iu}({\bq})P^\perp_{jc}({\bq})}{q_\perp^3}
+A(\mu_1,\mu_L)\frac{ P^\perp_{iu}({\bq})L^\perp_{jc}({\bq})}{q_\perp^3}
\Bigg]\nonumber
\\
&=&
k_i^\perp K_d\Lambda^{d-4}\dd \ell\Bigg[
\frac{1}{16\sqrt{\mu_x \mu_L^3}}\frac{\Pi^\perp_{iujc}}{(d-1)(d+1)} +
\frac{1}{16\sqrt{\mu_x \mu_1^3}} \left(  \frac{ (d-3)\delta^\perp_{iu} \delta^\perp_{jc} }{d-1} +  \frac{\Pi^\perp_{iujc}}{(d-1)(d+1)}\nonumber
\right)\nonumber
\\
&&
+A(\mu_L,\mu_1)\left(  \frac{ \delta^\perp_{iu} \delta^\perp_{jc} }{d-1} - \frac{\Pi^\perp_{iujc}}{(d-1)(d+1)}
\right)
+A(\mu_1,\mu_L) \left(  \frac{ \delta^\perp_{iu} \delta^\perp_{jc} }{d-1} - \frac{\Pi^\perp_{iujc}}{(d-1)(d+1)}
\right)
\Bigg]\nonumber
\\
&=&
\frac{K_d \Lambda^{d-4}\dd \ell}{(d-1)(d+1)}
\Bigg[k_u^\perp \delta^\perp_{jc}
\left(\frac{1}{16\sqrt{\mu_x \mu_L^3}} +
\frac{d^2-2d-2}{16\sqrt{\mu_x \mu_1^3}}
+dA(\mu_L,\mu_1)
+dA(\mu_1,\mu_L) \right)\nonumber
\\
&&+k_j^\perp \delta^\perp_{cu}
\left(\frac{1}{8\sqrt{\mu_x \mu_L^3}} +
\frac{1}{8\sqrt{\mu_x \mu_1^3}}
-2A(\mu_L,\mu_1)
-2A(\mu_1,\mu_2) \right)
\Bigg]
\ ,
\eeqn
where the function $A(x,y)$ is defined in appendix \ref{sec:formulae}, and $\Pi_{iucd} \equiv \delta_{iu}\delta_{cd}+\delta_{ic}\delta_{ud}+\delta_{id}\delta_{uc}$.  (Note that the two ``$dA$"'s in the penultimate line above represent spatial dimension $d$ times $A$, {\it not} the differential of $A$.)
We have also equated $k_j^\perp \delta^\perp_{cu}$ with $k_c^\perp \delta^\perp_{ju}$ in the last step as they lead to the same correction
when  contracted with the prefactor $k_u^\perp$ outside the integral.

Now we turn to ${(I^{\mu, a}_2)}_{cju}$. We need to expand the integrand up to $O(k)$. Note that the integration of the zeroth-order part of the integrand gives 0 since it is odd in $q_{\perp}$ while the integration region is isotropic in $q_{\perp}$. Therefore we only keep the $O(k)$ part of
\beqn
 {(I^{\mu, a}_2)}_{cju}(\tilde{\bk}) &=&{1\over \left(2\pi\right)^{d+1}}
\int_{\tilde{\bq} }^> q_u^\perp
\mid G_L(\tilde{\bq})\mid^2
\left[G_L(\tilde{\bk}-\tilde{\bq})L^\perp_{jc}({\bk}-{\bq})+ G_T(\tilde{\bk}-\tilde{\bq})P^\perp_{jc}(\bk-{\bq})\right]\nonumber
\\
\nonumber
&=& {1\over \left(2\pi\right)^{d+1}}
\int_{\tilde{\bq} }^> q_u^\perp
\mid G_L(\tilde{\bq})\mid^2\Bigg\{\left(2\mu_L\bq_\perp \cdot \bk_\perp+ 2\mu_x q_x k_x \right) G_L(-\tilde{\bq})^2L^\perp_{jc}(\bq) +
\left(2\mu_1\bq_\perp \cdot \bk_\perp+ 2\mu_x q_x k_x \right) \times\nonumber\\ &&G_T(-\tilde{\bq})^2P^\perp_{jc}(\bq)
+  \frac{ 2L_{jc}^\perp(\bq) \bq_\perp \cdot \bk_\perp
 -k^\perp_jq^\perp_c-k^\perp_cq^\perp_j}{q_\perp^2}
\left[G_L(-\tilde{\bq})-G_T(-\tilde{\bq})\right]
\Bigg\}\nonumber
\\
&=& {1\over \left(2\pi\right)^{d-1}}\int_{\bq_\perp}^> \frac{q_u^\perp}{q_\perp^5}
\Bigg\{\left(2\mu_L\bq_\perp \cdot \bk_\perp+ 2\mu_x q_x k_x \right)\frac{3 L^\perp_{jc}(\bq) }{128\sqrt{\mu_x \mu_L^5}} +
\left(2\mu_1\bq_\perp \cdot \bk_\perp+ 2\mu_x q_x k_x \right) P^\perp_{jc}(\bq) B(\mu_L,\mu_1)\nonumber
\\
&&+  \left( 2L_{jc}^\perp(\bq) \bq_\perp \cdot \bk_\perp
 -k^\perp_jq^\perp_c-k^\perp_cq^\perp_j\right)
\left[\frac{1}{16\sqrt{\mu_x \mu_L^3}} -A(\mu_L,\mu_1)  \right]
\Bigg\}\nonumber
\\
&=& \frac{K_d\Lambda^{d-4}\dd \ell}{(d-1)(d+1)}
\Bigg\{\frac{3 \Pi_{mujc}^\perp k_m^\perp}{64\sqrt{\mu_x \mu_L^3}} +
2\mu_1 B(\mu_L,\mu_1) \left[(d+1)\delta^\perp_{jc} k_u^\perp - \Pi_{mujc}^\perp k_m^\perp\right]\nonumber
\\
&&+  \left[2 \Pi_{mujc}^\perp k_m^\perp-(d+1)(\delta^\perp_{cu} k_j^\perp +\delta^\perp_{ju} k_c^\perp) \right]
\left(\frac{1}{16\sqrt{\mu_x \mu_L^3}} -A(\mu_L,\mu_1) \right)
\Bigg\}\nonumber
\\
&=&
\frac{K_d \Lambda^{d-4}\dd \ell}{(d-1)(d+1)}
\Bigg[k_u^\perp \delta^\perp_{jc}
\left(\frac{11}{64\sqrt{\mu_x \mu_L^3}}
-2A(\mu_L,\mu_1)
+2d\mu_1B(\mu_L,\mu_1) \right)\nonumber
\\
&& +k_j^\perp \delta^\perp_{cu}
\left(\frac{7-4d}{32\sqrt{\mu_x \mu_L^3}}
-2(1-d) A(\mu_L,\mu_1)
-4\mu_1 B(\mu_L,\mu_1) \right)
\Bigg]
\ ,
\eeqn
where the function $B(x,y)$ is defined in appendix \ref{sec:formulae}.

The overall correction  to the equation of motion is thus:
\beqn
\Delta(\pp_tu^<_j)_{ \mu,a}&=&-2\lambda^2Dk_u^{\perp}u_c^{\perp}(\tilde{\bk})\left[ {(I^{\mu, a}_1)}_{cju}(\tilde{\bk}) - {(I^{\mu, a}_2)}_{cju}(\tilde{\bk})\right]\nonumber
\\
&=&
-\frac{2\lambda^2DK_d\Lambda^{d-4}\dd \ell k_u^{\perp}u_c^{\perp}(\tilde{\bk}) }{(d-1)(d+1)}
\times
\Bigg[k_u^\perp \delta^\perp_{jc}
\left(-\frac{7}{64\sqrt{\mu_x \mu_L^3}} +
\frac{d^2-2d-2}{16\sqrt{\mu_x \mu_1^3}}
+(d+2) A(\mu_L,\mu_1)
+d A(\mu_1,\mu_L)-\right.\nonumber\\
&&\left.2d\mu_1 B(\mu_L,\mu_1)\right)+k_j^\perp \delta^\perp_{cu}
\left(\frac{4d-3}{32\sqrt{\mu_x \mu_L^3}} +
\frac{1}{8\sqrt{\mu_x \mu_1^3}}
-2dA(\mu_L,\mu_1)
-2A(\mu_1,\mu_L)
+4\mu_1 B(\mu_L,\mu_1) \right)
\Bigg]\nonumber
\\
&=&
-\frac{2\lambda^2DK_d\Lambda^{d-4}\dd \ell }{(d-1)(d+1)}
\times
\Bigg[k_\perp^2u_j^{\perp}(\tilde{\bk})
\left(-\frac{7}{64\sqrt{\mu_x \mu_L^3}} +
\frac{d^2-2d-2}{16\sqrt{\mu_x \mu_1^3}}
+(d+2) A(\mu_L,\mu_1)
+d A(\mu_1,\mu_L)\right.\nonumber\\
&&\left.-2d\mu_1 B(\mu_L,\mu_1)\right)+k_j^\perp k_u^{\perp}u_u^{\perp}(\tilde{\bk})
\left(\frac{4d-3}{32\sqrt{\mu_x \mu_L^3}} +
\frac{1}{8\sqrt{\mu_x \mu_1^3}}-2dA(\mu_L,\mu_1)-2A(\mu_1,\mu_L)\right.\nonumber\\
&&\left.+4\mu_1 B(\mu_L,\mu_1) \right)
\Bigg]\nonumber\\
&=&
-\frac{2\mu_1 g_1\dd \ell}{(d-1)(d+1)}
\Bigg[k_\perp^2 u_j^{\perp}(\tilde{\bk})
\bigg(-\frac{7}{64}(1+g_2)^{-3/2} +
\frac{d^2-2d-2}{16}
+(d+2)\sqrt{\mu_x \mu_1^3}A(\mu_L,\mu_1)\nonumber\\
&&+d\sqrt{\mu_x \mu_1^3}A(\mu_1,\mu_L)-2d \sqrt{\mu_x \mu_1^5}B(\mu_L,\mu_1)\bigg)+k_j^\perp k_u^{\perp}u_u^{\perp}(\tilde{\bk})
\left(\frac{4d-3}{32}(1+g_2)^{-3/2} +\frac{1}{8}\right.\nonumber\\
&&\left.-2d\sqrt{\mu_x \mu_1^3}A(\mu_L,\mu_1)
-2\sqrt{\mu_x \mu_1^3}A(\mu_1,\mu_L)+4\sqrt{\mu_x \mu_1^5}B(\mu_L,\mu_1)\right)
\Bigg]\,.
\eeqn

\subsubsection{Graph in \fig \ref{fig:prop}(b)}
The graph   in \fig \ref{fig:prop}(b) gives a contribution $\Delta(\pp_tu^<_j)_{\mu,b}$ to the
EOM  for $u_j^<(\tilde{\bk})$:
\beqn
\Delta(\pp_tu^<_j)_{\mu,b}&=&-{2\lambda^2Dk_u^{\perp}u_c^{\perp}(\tilde{\bk})\over (2\pi)^{d+1}}\int_{\tilde{\bq} }^> q_i^\perp C_{ju}(\tilde{\bq})G_{ic}(\tilde{\bk}-\tilde{\bq})
=-2\lambda^2 Dk_u^{\perp}u_c^{\perp}(\tilde{\bk}) {(I^{\mu, b})}_{cju}(\tilde{\bk})\,,
\eeqn
where
\beqn
 {(I^{\mu, b})}_{cju}(\tilde{\bk})\equiv {1\over (2\pi)^{d+1}}\int_{\tilde{\bq}}^> q_i^\perp
C_{ju}(\tilde{\bq})G_{ic}(\tilde{\bk}-\tilde{\bq})\,.
\eeqn

Again, since there's already a factor $k_c^{\perp}$ outside the loop integral, we can set $\omega=0$ in the integrand. Also we only need to expand the integrand up to $O(k)$ and keep the $O(k)$ part, since the integration of the zeroth-order part  is odd in $q^\perp_i$, and hence vanishes.  Therefore, we have
\beqn
 {(I^{\mu, b})}_{cju}(\tilde{\bk})&=&{1\over (2\pi)^{d+1}}\int_{\tilde{\bq} }^> q_i^\perp
\left[\mid G_L(\tilde{\bq})\mid^2L_{ju}(\bq)+\mid G_T(\tilde{\bq})\mid^2P_{ju}(\bq)\right]\times\nonumber
\\
&&
\left[\left(2\mu_L\bq_\perp \cdot \bk_\perp+ 2\mu_x q_x k_x \right) G_L(-\tilde{\bq})^2L^\perp_{ic}(\bq) +  \frac{ 2L_{ic}^\perp(\bq) \bq_\perp \cdot \bk_\perp
 -k^\perp_iq^\perp_c-k^\perp_cq^\perp_i}{q_\perp^2}
\left[G_L(-\tilde{\bq})-G_T(-\tilde{\bq})\right]
\right]\nonumber
\\
&=&{1\over (2\pi)^{d+1}}\int_{\tilde{\bq} }^> \Bigg\{
G_L(\tilde{\bq})G_L(-\tilde{\bq})L_{ju}(\bq)\times\nonumber
\\
&&
\left[\left(2\mu_L\bq_\perp \cdot \bk_\perp+ 2\mu_x q_x k_x \right) G_L(-\tilde{\bq})^2q_c^\perp  +  \frac{ 2q_c^\perp  \bq_\perp \cdot \bk_\perp
 -k^\perp_iq_i^\perp q^\perp_c-k^\perp_cq_\perp^2}{q_\perp^2}
\left[G_L(-\tilde{\bq})-G_T(-\tilde{\bq})\right]
\right]\nonumber
\\
&&
+G_T(\tilde{\bq})G_T(-\tilde{\bq})P_{ju}(\bq)\times\nonumber
\\
&&
\left[\left(2\mu_L\bq_\perp \cdot \bk_\perp+ 2\mu_x q_x k_x \right) G_L(-\tilde{\bq})^2q_c^\perp  +  \frac{ 2q_c^\perp  \bq_\perp \cdot \bk_\perp
 -k^\perp_iq_i^\perp q^\perp_c-k^\perp_cq_\perp^2}{q_\perp^2}
\left[G_L(-\tilde{\bq})-G_T(-\tilde{\bq})\right]
\right]
\Bigg\}\nonumber
\\
&=& {1\over (2\pi)^{d-1}}\int_{\bq_\perp}^>\frac{1}{q_\perp^5} \Bigg\{ \left(2\mu_L\bq_\perp \cdot \bk_\perp+ 2\mu_x q_x k_x \right) \frac{3q_c^\perp L^\perp_{ju}(\bq)}{128\sqrt{\mu_x \mu_L^5}}   +  \left( 2q_c^\perp
\bq_\perp \cdot \bk_\perp
 -k^\perp_iq_i^\perp q^\perp_c-k^\perp_cq_\perp^2\right)\times\nonumber\\
&&\left(\frac{L^\perp_{ju}(\bq)}{16\sqrt{\mu_x \mu_L^3}} -L^\perp_{ju}(\bq)A(\mu_L,\mu_1)\right)
+\left(2\mu_L\bq_\perp \cdot \bk_\perp+ 2\mu_x q_x k_x \right)  q_c^\perp P^\perp_{ju}(\bq)B(\mu_1,\mu_L) + \nonumber\\
&&(2q_c^\perp  \bq_\perp \cdot \bk_\perp
 -k^\perp_iq_i^\perp q^\perp_c-k^\perp_cq_\perp^2)\left(P^\perp_{ju}(\bq)A(\mu_1,\mu_L)-\frac{P^\perp_{ju}(\bq)}{16\sqrt{\mu_x \mu_1^3}} \right)
\Bigg\}\nonumber
\\
&=& \frac{K_d\Lambda^{d-4}\dd \ell}{d-1} \Bigg\{  \frac{3 \Pi^\perp_{mcju}k^\perp_m }{64\sqrt{\mu_x \mu_L^3}(d+1)}   +  \left( \frac{ \Pi^\perp_{mcju}k^\perp_m }{d+1}-\delta^\perp_{ju}k^\perp_c
\right)
 \left(\frac{1}{16\sqrt{\mu_x \mu_L^3}} -A(\mu_L,\mu_1)\right)\nonumber
\\
&&
+  \frac{2\mu_L [(d+1)\delta_{ju}k^\perp_c -\Pi^\perp_{mcju}k^\perp_m]B(\mu_1,\mu_L)} { (d+1)}
+   \left( (2-d)\delta_{ju}^\perp k_c^\perp - \frac{ \Pi^\perp_{mcju}k^\perp_m }{d+1}+\delta^\perp_{ju}k^\perp_c
\right)\times\nonumber\\
&&\left(A(\mu_1,\mu_L)-\frac{1}{16\sqrt{\mu_x \mu_1^3}} \right)
\Bigg\}\nonumber\\
&=&\frac{ K_d\Lambda^{d-4}\dd \ell}{(d-1)(d+1)} \Bigg\{
k^\perp_u \delta^\perp_{jc} \Bigg[
\frac{7}{64\sqrt{\mu_x \mu_L^3}}   +   \frac{1}{16\sqrt{\mu_x \mu_1^3}}
-A(\mu_L,\mu_1)
-A(\mu_1,\mu_L)-2\mu_LB(\mu_1,\mu_L)
\Bigg]\nonumber
\\
&&k^\perp_c \delta^\perp_{ju} \Bigg[
\frac{5-2d}{32\sqrt{\mu_x \mu_L^3}}   +   \frac{d^2-2d-1}{16\sqrt{\mu_x \mu_1^3}} - (d^2-2d-1) A(\mu_1,\mu_L)
-(1-d) A(\mu_L,\mu_1) +\nonumber\\
&&2(d-1) \mu_LB(\mu_1,\mu_L)
\Bigg]
\Bigg\}
\ .
\eeqn

The overall correction $\Delta(\pp_tu^<_j)_{\mu,b}$ to the  EOM  for $u_j^<(\tilde{\bk})$ is therefore :
\beqn
 \Delta(\pp_tu^<_j)_{\mu,b}&=&-2\lambda^2 Dk_u^{\perp}u_c^{\perp}(\tilde{\bk}) {(I^{\mu, b})}_{cju}(\tilde{\bk})\nonumber\\
&=&-\frac{ 2\lambda^2 D K_d\Lambda^{d-4}\dd \ell k_u^{\perp}u_c^{\perp}(\tilde{\bk})  }{(d-1)(d+1)} \Bigg\{
k^\perp_u \delta^\perp_{jc} \Bigg[
\frac{7}{64\sqrt{\mu_x \mu_L^3}}   +   \frac{1}{16\sqrt{\mu_x \mu_1^3}}
-A(\mu_L,\mu_1)
-A(\mu_1,\mu_L)-2\mu_LB(\mu_1,\mu_L)
\Bigg]\nonumber
\\
&&k^\perp_c \delta^\perp_{ju} \Bigg[
\frac{5-2d}{32\sqrt{\mu_x \mu_L^3}}   +   \frac{d^2-2d-1}{16\sqrt{\mu_x \mu_1^3}} - (d^2-2d-1) A(\mu_1,\mu_L)
-(1-d) A(\mu_L,\mu_1) +\nonumber\\
&&2(d-1) \mu_LB(\mu_1,\mu_L)
\Bigg]
\Bigg\}\nonumber\\
&=&-\frac{ 2\mu_1g_1\dd \ell}{(d-1)(d+1)} \Bigg\{
k_\perp^2u_j^{\perp}(\tilde{\bk})  \Bigg[
\frac{7}{64} (1+g_2)^{-3/2}   +   \frac{1}{16}
-\sqrt{\mu_x \mu_1^3}A(\mu_L,\mu_1)-\sqrt{\mu_x \mu_1^3}A(\mu_1,\mu_L)-\nonumber\\
&&2\sqrt{\mu_x \mu_1^3}\mu_L B(\mu_1,\mu_L)
\Bigg]+k^\perp_jk^\perp_cu_c^{\perp}(\tilde{\bk})  \Bigg[
\frac{5-2d}{32} (1+g_2)^{-3/2}   +   \frac{d^2-2d-1}{16} -\nonumber\\
&&(d^2-2d-1)\sqrt{\mu_x \mu_1^3}A(\mu_1,\mu_L)
- (1-d)\sqrt{\mu_x \mu_1^3}A(\mu_L,\mu_1)+ 2(d-1)\sqrt{\mu_x \mu_1^3}\mu_L B(\mu_1,\mu_L)
\Bigg]
\Bigg\}
\ .
\eeqn

\subsubsection{Graph in \fig \ref{fig:prop}(c)}
The graph   in \fig \ref{fig:prop}(c) gives a contribution $\Delta(\pp_tu^<_j)_{\mu,c}$ to the  EOM  for $u_j^<(\tilde{\bk})$:
\beqn
 \Delta(\pp_tu^<_j)_{\mu,c}&=&{2\lambda^2Du^{\perp}_u(\tilde{\bk})\over (2\pi)^{d+1}}
\int_{\tilde{\bq} }^> (k_i^\perp-q_i^\perp)q^\perp_u
C_{i\ell}(\tilde{\bq})G_{j\ell}(\tilde{\bk}-\tilde{\bq})\equiv 2\lambda^2 Du^{\perp}_u(\tilde{\bk}) \left[ {(I^{\mu, c}_1)}_{ju}(\tilde{\bk}) +{(I^{\mu, c}_2)}_{ju}(\tilde{\bk})\right]\,,
\eeqn
where
\beqn
{(I^{\mu, c}_1)}_{ju} (\tilde{\bk})&\equiv&{k_i^{\perp}\over (2\pi)^{d+1}}
\int_{\tilde{\bq} }^> q^\perp_u
C_{i\ell}(\tilde{\bq})G_{j\ell}(\tilde{\bk}-\tilde{\bq})\\
{(I^{\mu, c}_2)}_{ju}(\tilde{\bk})&\equiv&-{1\over (2\pi)^{d+1}}
\int_{\tilde{\bq} }^> q^\perp_iq^\perp_u
C_{i\ell}(\tilde{\bq})G_{j\ell}(\tilde{\bk}-\tilde{\bq})
\eeqn

Let's calculate ${(I^{\mu, c}_1)}_{ju}$ first. Again,
since there's already a factor $k_i^{\perp}$ outside the integral, we can set $\omega=0$ in the integrand. Also we only need to expand the integrand up to $O(k)$ and keep the $O(k)$ part, since the integration of the zeroth-order part gives 0. Therefore, we have 
\beqn
{(I^{\mu, c}_1)}_{ju}(\tilde{\bk})&=&{k_i^\perp\over (2\pi)^{d+1} }\int_{\tilde{\bq}}^>q^\perp_u  \left[
\mid G_L(\tilde{\bq})\mid^2L^\perp_{i\ell}({\bq}) +\mid G_T(\tilde{\bq})\mid^2P^\perp_{i\ell}({\bq})\right]
 \Bigg\{
G_L(-\tilde{\bq}) \frac{ 2L^\perp_{j\ell}({\bq}) q^\perp_mk^\perp_m-k^\perp_jq^\perp_\ell-k^\perp_\ell q^\perp_j}{q_\perp^2}\nonumber
\\
&&+G_L(-\tilde{\bq})^2 L^\perp_{j\ell}({\bq})
\left(2\mu_Lq^\perp_mk^\perp_m+ 2\mu_x q_x k_x \right)
 -G_T(-\tilde{\bq}) \frac{ 2L^\perp_{j\ell}({\bq}) q^\perp_mk^\perp_m-k^\perp_jq^\perp_\ell-k^\perp_\ell q^\perp_j}{q_\perp^2}\nonumber
\\
&&+G_T(-\tilde{\bq})^2 P^\perp_{j\ell}({\bq})
\left(2\mu_1q^\perp_mk^\perp_m+ 2\mu_x q_x k_x \right)
\Bigg\}\nonumber
\\
&=&{k_i^\perp\over (2\pi)^{d+1}} \int_{\tilde{\bq} }^>q^\perp_u
\Bigg\{
\mid G_L(\tilde{\bq})\mid^2
 \Bigg[
G_L(-\tilde{\bq}) \frac{ 2L^\perp_{ij}({\bq}) q^\perp_mk^\perp_m-k^\perp_jq^\perp_i-L^\perp_{i\ell}({\bq})k^\perp_\ell q^\perp_j}{q_\perp^2}
\nonumber\\
&&+G_L(-\tilde{\bq})^2 L^\perp_{ij}({\bq})
\left(2\mu_Lq^\perp_mk^\perp_m \right)
 -G_T(-\tilde{\bq}) \frac{ 2L^\perp_{ij}({\bq}) q^\perp_mk^\perp_m-k^\perp_jq^\perp_i-k^\perp_\ell q^\perp_jL^\perp_{i\ell}({\bq})}{q_\perp^2}
\Bigg]\nonumber
\\
&&
    +\mid G_T(\tilde{\bq})\mid^2) \Bigg[-G_L(-\tilde{\bq}) \frac{ k^\perp_\ell q^\perp_jP^\perp_{i\ell}({\bq})}{q_\perp^2}+G_T(-\tilde{\bq}) \frac{ k^\perp_\ell q^\perp_jP^\perp_{i\ell}({\bq})}{q_\perp^2}
    +G_T(-\tilde{\bq})^2 P^\perp_{ij}({\bq})
    \left(2\mu_1q^\perp_mk^\perp_m \right)
    \Bigg]
\Bigg\}\nonumber
\\
&=&{k_i^\perp\over(2\pi)^{d-1}} \int_{\bq_\perp}^>\frac{q^\perp_u}{q_\perp^5}
\Bigg\{
\frac{2L^\perp_{ij}({\bq}) q^\perp_mk^\perp_m-k^\perp_jq^\perp_i-L^\perp_{i\ell}({\bq})k^\perp_\ell q^\perp_j}{16\sqrt{\mu_x \mu_L^3}}
+ \frac{3
\left(\mu_Lq^\perp_mk^\perp_m \right)L^\perp_{ij}({\bq})}{64\sqrt{\mu_x \mu_L^5}}\nonumber
\\
&&
 -[2L^\perp_{ij}({\bq}) q^\perp_mk^\perp_m-k^\perp_jq^\perp_i-k^\perp_\ell q^\perp_jL^\perp_{i\ell}({\bq})]
A(\mu_L,\mu_1)\nonumber
\\
&&
  - k^\perp_\ell q^\perp_jP^\perp_{i\ell}({\bq})A(\mu_1,\mu_L) +\frac{k^\perp_\ell q^\perp_jP^\perp_{i\ell}({\bq})}{16\sqrt{\mu_x \mu_1^3}}
  +\frac{3P^\perp_{ij}({\bq})
  	\left(\mu_1q^\perp_mk^\perp_m \right) }{64\sqrt{\mu_x \mu_1^5}}
\Bigg\}\nonumber
\\
&=& \frac{K_d \Lambda^{d-4}\dd \ell k_i^\perp}{(d-1)(d+1)} \Bigg\{
\frac{\Pi^\perp_{miju}k^\perp_m -(d+1) \delta_{iu} k^\perp_j}{16\sqrt{\mu_x \mu_L^3}}
+ \frac{3\mu_L \Pi^\perp_{miju}k^\perp_m}{64\sqrt{\mu_x \mu_L^5}}
 -[\Pi^\perp_{miju}k^\perp_m -(d+1) \delta_{iu} k^\perp_j]A(\mu_L,\mu_1)\nonumber
 \\
&&
  +\left(-\Pi^\perp_{miju}k^\perp_m +(d+1) \delta^\perp_{ju} k^\perp_i\right)\left(-A(\mu_1,\mu_L)  +\frac{1}{16\sqrt{\mu_x \mu_1^3}} \right)+\left(-\Pi^\perp_{miju}k^\perp_m +(d+1) \delta^\perp_{ij} k^\perp_u\right)\frac{3}{64 \sqrt{\mu_x \mu_1^3}}
\Bigg\}
\nonumber\\
&=& \frac{K_d\Lambda^{d-4}\dd \ell }{(d-1)(d+1)} \Bigg\{
k^2_\perp \delta_{ju}^\perp \Bigg[
\frac{7}{64\sqrt{\mu_x \mu_L^3}}
 -A(\mu_L,\mu_1)
-dA(\mu_1,\mu_L) +\frac{4d-3}{64\sqrt{\mu_x \mu_1^3}} \Bigg]\nonumber
\\
&&+k_j^\perp k_u^\perp \Bigg[
\frac{5-2d}{32\sqrt{\mu_x \mu_L^3}}
 +(d-1) A(\mu_L,\mu_1)
+2A(\mu_1,\mu_L) +\frac{3d-11}{64\sqrt{\mu_x \mu_1^3}} \Bigg]
\Bigg\}
\ .
\eeqn

Now we turn to ${(I^{\mu, c}_2)}_{ju}(\tilde{\bk})$. In appendix
\ref{sec:figd} we show that the sum of ${(I^{\mu, c}_2)}_{ju}(\tilde{\bk})$
and ${(I^{\mu, d})}_{ju}(\tilde{\bk})$, which is introduced in evaluating
Fig. \ref{fig:prop}(d), is at most of $O(k^2)$. This means we can set
$\omega=0$ and focus on the $O(k^2)$ part when evaluating
${(I^{\mu, c}_2)}_{ju}(\tilde{\bk})$ and ${(I^{\mu, d})}_{ju}(\tilde{\bk})$,
since their lower order parts (i.e., $O(1)$ and $O(k)$) all cancel out.
We will use this knowledge in the following calculations.

\beqn
\nonumber
{(I^{\mu, c}_2)}_{ju}(\tilde{\bk})
&=&-{1\over (2\pi)^{d+1}}\int_{\tilde{\bq} }^> q^\perp_u q_\ell^\perp
G_L(\tilde{\bq})G_L(-\tilde{\bq})
\Bigg\{-\left(\mu_L k_\perp^2 +\mu_xk_x^2\right)G_L(-\tilde{\bq})^2L^\perp_{j\ell}({\bq})+\left(4\mu_L^2 (\bq_\perp \cdot  \bk_\perp)^2 +4\mu_x^2 q_x^2k_x^2\right)\times \nonumber\\ &&G_L(-\tilde{\bq})^3L^\perp_{j\ell}({\bq})
+\left[L^\perp_{j\ell}({\bq}) \left(-\frac{k_\perp^2}{q_\perp^2} +\frac{4(\bq_\perp \cdot \bk_\perp)^2}{q_\perp^4}\right) +\frac{k^\perp_jk^\perp_\ell}{q_\perp^2} - \frac{2\bq_\perp \cdot \bk_\perp (k^\perp_jq^\perp_\ell+k^\perp_\ell q^\perp_j)}{q_\perp^4}\right]G_L(-\tilde{\bq})\nonumber
\\
&&+2\mu_L\bq_\perp \cdot \bk_\perp G_L(-\tilde{\bq})^2 \left(\frac{ 2L^\perp_{j\ell}({\bq})\bq_\perp \cdot \bk_\perp -k^\perp_jq^\perp_\ell-k^\perp_\ell q^\perp_j}{q_\perp^2}\right)\nonumber
\\
&&-\left[L^\perp_{j\ell}({\bq})\left(-\frac{k_\perp^2}{q_\perp^2} +\frac{4(\bq_\perp \cdot \bk_\perp)^2}{q_\perp^4}\right) +\frac{k^\perp_jk^\perp_\ell}{q_\perp^2} - \frac{2\bq_\perp \cdot \bk_\perp (k^\perp_jq^\perp_\ell+k^\perp_\ell q^\perp_j)}{q_\perp^4}\right]G_T(-\tilde{\bq})\nonumber
\\
&&-2\mu_1\bq_\perp \cdot \bk_\perp
G_T(-\tilde{\bq})^2 \left(\frac{ 2L^\perp_{j\ell}({\bq}) \bq_\perp \cdot \bk_\perp -k^\perp_jq^\perp_\ell-k^\perp_\ell q^\perp_j}{q_\perp^2}\right)\Bigg\}
\label{Ic2hard}
\ ,
\eeqn
where we have discarded terms linear in $q_x$ since the integration of these terms gives 0.
Performing the $\Omega$ and $q_x$ integrals in (\ref{Ic2hard}), as described in detail in appendix (\ref{sec:formulae}), we have
\beqn
{(I^{\mu, c}_2)}_{ju}(\tilde{\bk})
&=&-{1\over (2\pi)^{d-1}}\int_{\bq_\perp}^>\frac{q^\perp_u q_\ell^\perp}{q_\perp^5}
\Bigg\{-\frac{3\delta_{j\ell}\left(\mu_L k_\perp^2 +\mu_x k_x^2\right)  }{128\sqrt{\mu_x \mu_L^5}} + \frac{\left(5\mu_L^2 q^\perp_m q^\perp_n k^\perp_m k^\perp_n +\mu_x \mu_Lk_x^2 q_\perp^2\right)\delta_{j\ell}}{128\sqrt{\mu_x \mu_L^7} q_\perp^2}\nonumber
\\
&&+\left[\delta_{j\ell} \left(-k_\perp^2 +\frac{4(\bq_\perp \cdot \bk_\perp)^2}{q_\perp^2}\right) +k^\perp_jk^\perp_\ell - \frac{2\bq_\perp \cdot \bk_\perp (k^\perp_jq^\perp_\ell+k^\perp_\ell q^\perp_j)}{q_\perp^2}\right]\frac{1}{16\sqrt{\mu_x \mu_L^3}}\nonumber
\\
&&+\frac{3\mu_L\bq_\perp \cdot \bk_\perp}{64\sqrt{\mu_x \mu_L^5}}  \left(\frac{ 2\delta_{j\ell} \bq_\perp \cdot \bk_\perp -k^\perp_jq^\perp_\ell-k^\perp_\ell q^\perp_j}{q_\perp^2}\right)\nonumber
\\
&&-\left[\delta_{j\ell} \left(-k_\perp^2 +\frac{4(\bq_\perp \cdot \bk_\perp)^2}{q_\perp^2}\right) +k^\perp_jk^\perp_\ell - \frac{2\bq_\perp \cdot \bk_\perp (k^\perp_jq^\perp_\ell+k^\perp_\ell q^\perp_j)}{q_\perp^2}\right]A(\mu_L,\mu_1)\nonumber
\\
&&-2\mu_1\bq_\perp \cdot \bk_\perp B(\mu_L,\mu_1) \left(\frac{ 2\delta_{j\ell} \bq_\perp \cdot \bk_\perp -k^\perp_jq^\perp_\ell-k^\perp_\ell q^\perp_j}{q_\perp^2}\right)\Bigg\}\nonumber
\\
&=&-\frac{K_d\Lambda^{d-4}\dd \ell}{(d-1)(d+1)}
\Bigg\{-\frac{3(d+1)\delta^\perp_{ju}\left(\mu_L k_\perp^2 +\mu_xk_x^2\right)  }{128\sqrt{\mu_x \mu_L^5}} + \frac{5\mu_L^2 \Pi^\perp_{mnju} k^\perp_mk^\perp_n +(d+1)\delta^\perp_{ju}\mu_x \mu_L
k_x^2}{128\sqrt{\mu_x \mu_L^7}}\nonumber
\\
&&+\left(
-(d+1)\delta_{ju}^\perp k_\perp^2 +2 \Pi^\perp_{mnju} k^\perp_mk^\perp_n- (d+1) k^\perp_jk^\perp_u
\right)\left(\frac{1}{16\sqrt{\mu_x \mu_L^3}}-A(\mu_L,\mu_1) \right)\nonumber
\\
&&+
\left(
\Pi^\perp_{mnju} k^\perp_mk^\perp_n- (d+1) k^\perp_jk^\perp_u
\right)\left(
\frac{3\mu_L}{64\sqrt{\mu_x \mu_L^5}} -2\mu_1 B(\mu_L,\mu_1)\right)\Bigg\}\nonumber
\\
\nonumber
&=&-\frac{K_d\Lambda^{d-4}\dd \ell}{(d-1)(d+1)}
\Bigg\{-k_x^2 \delta^\perp_{ju} \frac{(d+1)\sqrt{\mu_x} }{64\sqrt{\mu_L^5}}+k_\perp^2 \delta^\perp_{ju} \Bigg[
\frac{16-11d}{128\sqrt{\mu_x \mu_L^3}}+
(d-1)A(\mu_L,\mu_1)-2\mu_1B(\mu_L,\mu_1)
\Bigg]
\\
&&+
k^\perp_jk^\perp_u \left[ \frac{20-7d}{64 \sqrt{\mu_x \mu_L^3}}
+(d-3)A(\mu_L,\mu_1)
+2(d-1)\mu_1 B(\mu_L,\mu_1) \right]
\Bigg\}
\ .
\eeqn

The total contribution $\Delta(\pp_tu^<_j)_{\mu,c}$ of the graph Fig. \ref{fig:prop}(c) to the equation of motion is therefore:
\beqn
 \Delta(\pp_tu^<_j)_{\mu,c}&=&2\lambda^2 D u_u^{\perp}(\tilde{\bk})\left[{(I^{\mu, c}_1)}_{ju} (\tilde{\bk}) + {(I^{\mu, c}_2)}_{ju}(\tilde{\bk})\right]
\nonumber\\
\nonumber
&=&\frac{2\lambda^2 D K_d\Lambda^{d-4}u_u^{\perp}(\tilde{\bk})\dd \ell}{(d-1)(d+1)}\Bigg\{k_x^2 \delta^\perp_{ju} \frac{(d+1)\sqrt{\mu_x} }{64\sqrt{\mu_L^5}}+
k^2_\perp \delta_{ju}^\perp \Bigg[
\frac{11d-2}{128\sqrt{\mu_x \mu_L^3}}-dA(\mu_L,\mu_1)+2\mu_1 B(\mu_L,\mu_1)
\nonumber\\
&& -dA(\mu_1,\mu_L)+\frac{4d-3}{64\sqrt{\mu_x \mu_1^3}} \Bigg]+k_j^\perp k_u^\perp \Bigg[
\frac{3d-10}{64\sqrt{\mu_x \mu_L^3}}
+2A(\mu_L,\mu_1)
+2(1-d)\mu_1 B(\mu_L,\mu_1)\nonumber\\
&&+2A(\mu_1,\mu_L)+\frac{3d-11}{64\sqrt{\mu_x \mu_1^3}} \Bigg]
\Bigg\}\nonumber\\
%
\nonumber
&=&\frac{2\mu_1g_1\dd \ell}{(d-1)(d+1)}\Bigg\{k_x^2 u_j^{\perp}(\tilde{\bk}) \frac{(d+1)\sqrt{\mu_x} }{64\sqrt{\mu_L^5}}+
k^2_\perp u_j^{\perp}(\tilde{\bk})  \Bigg[
\frac{11d-2}{128}   (1+g_2)^{-3/2}
-d\sqrt{\mu_x \mu_1^3}A(\mu_L,\mu_1)\nonumber\\
&&+2\sqrt{\mu_x \mu_1^5}B(\mu_L,\mu_1)
-d\sqrt{\mu_x \mu_1^3}A(\mu_1,\mu_L)+\frac{4d-3}{64} \Bigg]
+k_j^\perp k_u^\perp u_u^{\perp}(\tilde{\bk})\Bigg[
\frac{3d-10}{64}    (1+g_2)^{-3/2}\nonumber\\
&&+2\sqrt{\mu_x \mu_1^3}A(\mu_L,\mu_1)+2(1-d)\sqrt{\mu_x \mu_1^5}B(\mu_L,\mu_1)
+2\sqrt{\mu_x \mu_1^3}A(\mu_1,\mu_L)+\frac{3d-11}{64} \Bigg]
\Bigg\}
\ .
\eeqn

\subsubsection{Graph in \fig \ref{fig:prop}(d) \label{sec:figd}}
 The graph   in \fig \ref{fig:prop}(d) gives a contribution  $\Delta(\pp_tu^<_j)_{\mu,d}$ to the  EOM  for $u_j^<(\tilde{\bk})$:
\beqn
 \Delta(\pp_tu^<_j)_{\mu,d}&=&
{2\lambda^2Du_u^{\perp}(\tilde{\bk})\over (2\pi)^{d+1}}\int_{\tilde{\bq} }^> q_i^\perp q^\perp_u
C_{j\ell}(\tilde{\bq})G_{i\ell}(\tilde{\bk}-\tilde{\bq})
\equiv 2\lambda^2 Du_u^{\perp}(\tilde{\bk}){(I^{\mu, d})}_{ju}(\tilde{\bk})
\ ,
\eeqn
where
\beqn
{(I^{\mu, d})}_{ju}(\tilde{\bk})\equiv {1\over (2\pi)^{d+1}}\int_{\tilde{\bq} }^> q_i^\perp q^\perp_u
C_{j\ell}(\tilde{\bq})G_{i\ell}(\tilde{\bk}-\tilde{\bq})\,.
\eeqn

First let's show that the sum of ${(I^{\mu, c}_2)}_{ju}(\tilde{\bk})$ and ${(I^{\mu, d})}_{ju}(\tilde{\bk})$ is at most of $O(k^2)$.
The integrand of ${(I^{\mu, c})_2}_{ju}(\tilde{\bk})$ can be rewritten as
\beqn
&&- q^\perp_iq^\perp_uC_{i\ell}(\tilde{\bq})G_{j\ell}(\tilde{\bk}-\tilde{\bq})\nonumber\\
&=&- q^\perp_u \mid G_L(\tilde{\bq})\mid^2 q^\perp_\ell G_{j\ell}(\tilde{\bk}-\tilde{\bq})
\nonumber\\
&=&- q^\perp_u \mid G_L(\tilde{\bq})\mid^2 \left(q^\perp_\ell-k^\perp_\ell\right) G_{j\ell}(\tilde{\bk}-\tilde{\bq})- q^\perp_uk^\perp_\ell \mid G_L(\tilde{\bq})\mid^2 G_{j\ell}(\tilde{\bk}-\tilde{\bq})
\nonumber\\
&=&- q^\perp_u \mid G_L(\tilde{\bq})\mid^2 \left(q^\perp_j-k^\perp_j\right) G_L(\tilde{\bk}-\tilde{\bq})- q^\perp_uk^\perp_\ell \mid G_L(\tilde{\bq})\mid^2 G_{j\ell}(\tilde{\bk}-\tilde{\bq})
\nonumber\\
&=&- q^\perp_u q^\perp_j\mid G_L(\tilde{\bq})\mid^2 G_L(\tilde{\bk}-\tilde{\bq})+k^\perp_j q^\perp_u \mid G_L(\tilde{\bq})\mid^2 G_L(\tilde{\bk}-\tilde{\bq})-k^\perp_\ell q^\perp_u\mid G_L(\tilde{\bq})\mid^2 G_{j\ell}(\tilde{\bk}-\tilde{\bq})\,;
\eeqn
The integrand of ${(I^{\mu, d})}_{ju}(\tilde{\bk})$ can be rewritten as
\beqn
&&q^\perp_iq^\perp_uC_{j\ell}(\tilde{\bq})G_{i\ell}(\tilde{\bk}-\tilde{\bq})\nonumber\\
&=&q^\perp_u C_{j\ell}(\tilde{\bq}) (q^\perp_i- k^\perp_i)G_{i\ell}(\tilde{\bk}-\tilde{\bq})
+k^\perp_i q^\perp_u C_{j\ell}(\tilde{\bq})G_{i\ell}(\tilde{\bk}-\tilde{\bq})\nonumber\\
&=&q^\perp_u C_{j\ell}(\tilde{\bq}) (q^\perp_\ell- k^\perp_\ell)G_L(\tilde{\bk}-\tilde{\bq})
+k^\perp_i q^\perp_u C_{j\ell}(\tilde{\bq})G_{i\ell}(\tilde{\bk}-\tilde{\bq})
\nonumber\\
&=&q^\perp_u q^\perp_j\mid G_L(\tilde{\bq})\mid^2 G_L(\tilde{\bk}-\tilde{\bq})
- k^\perp_\ell q^\perp_u C_{j\ell}(\tilde{\bq}) G_L(\tilde{\bk}-\tilde{\bq})
+k^\perp_i q^\perp_u C_{j\ell}(\tilde{\bq})G_{i\ell}(\tilde{\bk}-\tilde{\bq})\,.
\eeqn
Clearly the zeroth-order parts of the two integrands cancel out. Thus the sum of the two integrands is of $O(k)$. Furthermore, it is easy to see that the $O(k)$ parts vanish after the integration. This proves our earlier claim.

In the following calculations we will focus on the $O(k^2)$ parts. Again we will omit terms linear in $q_x$ since the integration of these terms gives 0.
\beqn
{(I^{\mu, d})}_{ju}(\tilde{\bk}) &=&{1\over (2\pi)^{d+1}}\int_{\tilde{\bq} }^> q_i^\perp q^\perp_u\Bigg\{\mid G_L(\tilde{\bq})\mid^2 L^\perp_{\ell j}({\bq})
\Bigg[
-\left(\mu_L k_\perp^2 +\mu_xk_x^2\right)G_L(-\tilde{\bq})^2L^\perp_{i\ell}(\bq)+\left(4\mu_L^2 (\bq_\perp \cdot \bk_\perp)^2 +4\mu_x^2 q_xk_x^2\right)\nonumber\\
&&\times G_L(-\tilde{\bq})^3L^\perp_{i\ell}(\bq)+\left[L^\perp_{i\ell}(\bq) \left(-\frac{k_\perp^2}{q_\perp^2} +\frac{4(\bq_\perp \cdot \bk_\perp)^2}{q_\perp^4}\right) +\frac{k^\perp_ik^\perp_\ell}{q_\perp^2} - \frac{2\bq_\perp \cdot \bk_\perp (k^\perp_iq^\perp_\ell+k^\perp_\ell q^\perp_i)}{q_\perp^4}\right]\times\nonumber\\
&&\left(G_L(-\tilde{\bq})-G_T(-\tilde{\bq})\right)
+2\mu_L\bq_\perp \cdot \bk_\perp G_L(-\tilde{\bq})^2 \left(\frac{ 2L^\perp_{i\ell}(\bq) \bq_\perp \cdot \bk_\perp -k^\perp_iq^\perp_\ell-k^\perp_\ell q^\perp_i}{q_\perp^2}\right)
\nonumber\\
&&-2\mu_1\bq_\perp \cdot \bk_\perp G_T(-\tilde{\bq})^2 \left(\frac{ 2L^\perp_{i\ell}(\bq) \bq_\perp \cdot \bk_\perp -k^\perp_iq^\perp_\ell-k^\perp_\ell q^\perp_i}{q_\perp^2}\right)
\Bigg]+\mid G_T(\tilde{\bq})\mid^2P^\perp_{\ell j}({\bq})\times\nonumber
 \\
 \nonumber
 && \Bigg[
 -\left(\mu_1 k_\perp^2 +\mu_xk_x^2\right)G_T(-\tilde{\bq})^2P^\perp_{i\ell}(\bq)
 -\left[(4\mu_1^2 (\bq_\perp \cdot \bk_\perp)^2 +4\mu_x^2 q_x^2k_x^2+4 \mu_x\mu_1(\bq_\perp \cdot \bk_\perp)q_xk_x\right]\times\nonumber\\ &&G_T(-\tilde{\bq})^3P^\perp_{i\ell}(\bq)+\left(\frac{k^\perp_ik^\perp_\ell}{q_\perp^2} - \frac{2\bq_\perp \cdot \bk_\perp k^\perp_\ell q^\perp_i}{q_\perp^4}\right)\left(G_L(-\tilde{\bq})-G_T(-\tilde{\bq})\right)
\nonumber\\
&&-2\mu_L\bq_\perp \cdot \bk_\perp G_L(-\tilde{\bq})^2 \left(\frac{ k^\perp_\ell q^\perp_i}{q_\perp^2}\right)
+2\mu_1\bq_\perp \cdot \bk_\perp G_T(-\tilde{\bq})^2 \left(\frac{k^\perp_\ell q^\perp_i}{q_\perp^2}\right)
\Bigg]\Bigg\}\nonumber\\
&=&{1\over (2\pi)^{d-1}}\int_{\bq_\perp}^>\frac{q_i^\perp q^\perp_u}{q_\perp^5} \Bigg\{
-\frac{3L^\perp_{ij}(\bq)\left(\mu_L k_\perp^2 +\mu_xk_x^2\right)}{128\sqrt{\mu_x \mu_L^5}} + \frac{\left(5\mu_L^2   q^\perp_m q^\perp_n k^\perp_m k^\perp_n  +\mu_x \mu_L k_x^2q_\perp^2\right)L^\perp_{ij}(\bq)}{128\sqrt{\mu_x \mu_L^7}q_\perp^2}
\nonumber\\ \nonumber
&&+\left[L^\perp_{ij}(\bq) \left(-k_\perp^2 +\frac{4(\bq_\perp \cdot \bk_\perp)^2}{q_\perp^2}\right) +k^\perp_ik^\perp_\ell L^\perp_{\ell j}(\bq) - \frac{2\bq_\perp \cdot \bk_\perp (k^\perp_iq^\perp_j+k^\perp_\ell q^\perp_iL^\perp_{\ell j}(\bq))}{q_\perp^2}\right]\times
\nonumber\\
&&\left(\frac{1}{16\sqrt{\mu_x \mu_L^3}}-A(\mu_L,\mu_1)  \right)
+2\bq_\perp \cdot \bk_\perp \left( \frac{3\mu_L}{128\sqrt{\mu_x \mu_L^5}}
-\mu_1B(\mu_L,\mu_1) \right) \times\nonumber\\
&&\left(\frac{ 2L^\perp_{ij}(\bq) \bq_\perp \cdot \bk_\perp -k^\perp_iq^\perp_j-k^\perp_\ell q^\perp_iL^\perp_{\ell j}(\bq))}{q_\perp^2}\right)
\Bigg]+ P^\perp_{\ell j}({\bq})\left(k^\perp_ik^\perp_\ell - \frac{2\bq_\perp \cdot \bk_\perp k^\perp_\ell q^\perp_i}{q_\perp^2}\right)\times\nonumber\\
&&\left(A(\mu_1,\mu_L)-\frac{1}{16\sqrt{\mu_x \mu_1^3}}\right)
-2\bq_\perp \cdot \bk_\perp \left[
\mu_LB(\mu_1,\mu_L)-\frac{3\mu_1}{128\sqrt{\mu_x \mu_1^5}}\right] \left(\frac{ k^\perp_\ell q^\perp_i}{q_\perp^2}\right)P^\perp_{\ell j}({\bq})
 \Bigg\}\nonumber
 \\
 &=&\frac{K_d\Lambda^{d-4}\dd \ell}{(d-1)(d+1)} \Bigg\{
-\frac{3\delta^\perp_{ju}(d+1)\left(\mu_L k_\perp^2 +\mu_xk_x^2\right)}{128\sqrt{\mu_x \mu_L^5}} + \frac{5\mu_L^2
	\Pi^\perp_{mnju} k^\perp_mk^\perp_n
	 + \mu_x \delta^\perp_{ju}(d+1) \mu_L k_x^2}{128\sqrt{\mu_x \mu_L^7}}\nonumber
\\
&&+\left(-\delta^\perp_{ju}(d+1)k_\perp^2 +\Pi^\perp_{mnju} k^\perp_mk^\perp_n  \right)\left(\frac{1}{16\sqrt{\mu_x \mu_L^3}}-A(\mu_L,\mu_1)\right)
 + \left( -(d+1)k^\perp_uk^\perp_j +\Pi^\perp_{mnju} k^\perp_mk^\perp_n \right)\times
 \nonumber\\
&&\left(A(\mu_1,\mu_L)-\frac{1}{16\sqrt{\mu_x \mu_1^3}}\right)
-2\left( (d+1)k^\perp_uk^\perp_j -\Pi^\perp_{mnju} k^\perp_mk^\perp_n \right) \left( \mu_LB(\mu_1,\mu_L)-\frac{3}{128\sqrt{\mu_x \mu_1^3}}\right)\Bigg\}\ .
\eeqn

 The  total contribution $\Delta(\pp_tu^<_j)_{\mu,d}$ of the graph Fig. \ref{fig:prop}(d) to the equation of motion is therefore:
\beqn
\Delta(\pp_tu^<_j)_{\mu,d}&=& 2\lambda^2 D u_u^{\perp}(\tilde{\bk}){(I^{\mu, d})}_{ju}(\tilde{\bk})
\\
&=&\frac{2\lambda^2 Du_u^{\perp}(\tilde{\bk}) K_d\Lambda^{d-4}\dd \ell}{(d-1)(d+1)} \Bigg\{-
\delta^\perp_{ju}k_x^2\frac{(d+1)\sqrt{\mu_x} }{64\sqrt{\mu_L^5}}
+\delta^\perp_{ju}k_\perp^2\left[
\frac{2-11d }{128\sqrt{\mu_x \mu_L^3}}
+dA(\mu_L,\mu_1)+A(\mu_1,\mu_L)\right.\nonumber\\
&&\left.-\frac{7}{64\sqrt{\mu_x \mu_1^3}}+2\mu_LB(\mu_1,\mu_L)
\right]+k^\perp_uk^\perp_j\left[
\frac{13}{64\sqrt{\mu_x \mu_L^3}}-2A(\mu_L,\mu_1)
+(1-d) A(\mu_1,\mu_L)\right.\nonumber\\
&&\left.+\frac{7d-7}{64\sqrt{\mu_x \mu_1^3}}
+2(1-d) \mu_LB(\mu_1,\mu_L)
\right]
 \Bigg\}\nonumber\\
&=&\frac{2\mu_1g_1\dd \ell}{(d-1)(d+1)} \Bigg\{-
k_x^2u_j^{\perp}(\tilde{\bk})\frac{(d+1)\sqrt{\mu_x} }{64\sqrt{\mu_L^5}}
+k_\perp^2u_j^{\perp}(\tilde{\bk})\left[
\frac{2-11d }{128} (1+g_2)^{-3/2}
+d\sqrt{\mu_x \mu_1^3}A(\mu_L,\mu_1) +\right.\nonumber\\
&&\left.\sqrt{\mu_x \mu_1^3}A(\mu_1,\mu_L)
-\frac{7}{64}
+2\sqrt{\mu_x \mu_1^3}\mu_L B(\mu_1,\mu_L)
\right]+k^\perp_j k^\perp_uu_u^{\perp}(\tilde{\bk})\left[\frac{13}{64} (1+g_2)^{-3/2}
-2\sqrt{\mu_x \mu_1^3}A(\mu_L,\mu_1)\right.\nonumber\\
&&\left.+(1-d)\sqrt{\mu_x \mu_1^3}A(\mu_1,\mu_L)
+\frac{7(d-1)}{64}+2(1-d)\sqrt{\mu_x \mu_1^3} \mu_LB(\mu_1,\mu_L)
\right]
\Bigg\}
\ .
\eeqn

\subsection{Overall propagator renormalization}

Summing over all the contributions from the one-loop diagrams in Fig. \ref{fig:prop}, we find two types of terms: $k_{\perp}^2u_j^{\perp}(\tilde{\bk})$ and $k_j^\perp k_u^{\perp}u_u^{\perp}(\tilde{\bk})$. The sum of the coefficients of the former gives the correction to $-\mu_1$;  that of the latter gives the correction to $-\mu_2$. There is no correction to $\mu_x$ to one-loop order,  since no terms proportional to $k^2_x u_j$ survive to this order.

Thus the graphical correction  $\delta\mu_1$ to $\mu_1$ is
\beqn
\delta\mu_1&=&\frac{2\mu_1g_1\dd \ell}{(d-1)(d+1)}\Bigg\{
\left(-\frac{7}{64}(1+g_2)^{-3/2} +
\frac{d^2-2d-2}{16}
+(d+2)\sqrt{\mu_x \mu_1^3}A(\mu_L,\mu_1)
+d\sqrt{\mu_x \mu_1^3}A(\mu_1,\mu_L)-\right.\nonumber\\
&&\left.2d\sqrt{\mu_x \mu_1^5} B(\mu_L,\mu_1)\right)+\Bigg[
\frac{7}{64} (1+g_2)^{-3/2}   +   \frac{1}{16}
-\sqrt{\mu_x \mu_1^3}A(\mu_L,\mu_1)-\sqrt{\mu_x \mu_1^3}A(\mu_1,\mu_L)-2\sqrt{\mu_x \mu_1^3}\mu_LB(\mu_1,\mu_L)
\Bigg]\nn
\\
&&-\Bigg[
\frac{11d-2}{128}   (1+g_2)^{-3/2}
-d\sqrt{\mu_x \mu_1^3}A(\mu_L,\mu_1)
+2\sqrt{\mu_x \mu_1^5}B(\mu_L,\mu_1)
-d\sqrt{\mu_x \mu_1^3}A(\mu_1,\mu_L)+\frac{4d-3}{64} \Bigg]\nn
\\
&&-\left[
\frac{2-11d }{128} (1+g_2)^{-3/2}
+d\sqrt{\mu_x \mu_1^3}A(\mu_L,\mu_1) +\sqrt{\mu_x \mu_1^3}A(\mu_1,\mu_L)
-\frac{7}{64}
+2\sqrt{\mu_x \mu_1^3}\mu_L B(\mu_1,\mu_L)
\right]
\Bigg\}\nn
\\
&=&\frac{2\mu_1g_1\dd \ell}{(d-1)(d+1)}\Bigg\{\frac{2d^2-6d+3}{32}
+(1+d)\sqrt{\mu_x \mu_1^3}A(\mu_L,\mu_1)
+2(d-1)\sqrt{\mu_x \mu_1^3}A(\mu_1,\mu_L)\nn
\\
&&
-2(d+1)\sqrt{\mu_x \mu_1^5}B(\mu_L,\mu_1)
-4\sqrt{\mu_x \mu_1^3}\mu_L B(\mu_1,\mu_L)
\Bigg\}
\ ,
\label{mu1rr}
\eeqn
where
\beqn
\sqrt{\mu_x \mu_1^3}A(\mu_L,\mu_1)&=&\frac{1}{4g_2} \left( \frac{\sqrt{2}}{\sqrt{2+g_2}}-\frac{1}{\sqrt{1+g_2}}\right)
\\
\sqrt{\mu_x \mu_1^3}A(\mu_1,\mu_L)&=&\frac{1}{4g_2} \left( 1-\frac{\sqrt{2}}{\sqrt{2+g_2}}\right)
\\
\sqrt{\mu_x \mu_1^5}B(\mu_L,\mu_1)&=& \frac{2(2+g_2)^{3/2}-\sqrt{2(1+g_2)} (4+g_2) }{8\sqrt{1+g_2} (2+g_2)^{3/2}g_2^2}
\\
\sqrt{\mu_x \mu_1^3}\mu_LB(\mu_1,\mu_L)&=& \left(1+g_2\right)\frac{2(2+g_2)^{3/2}-\sqrt{2} (4+3g_2) }{8 (2+g_2)^{3/2}g_2^2}
\ .
\label{ABdefs}
\eeqn
 Using these, we can, after considerable algebra,
rewrite (\ref{mu1rr}) as
\beq
\delta\mu_1=g_1\mu_1G_{\mu_1}(g_2)\dd \ell \,,
\label{mu1corr}
\eeq
with $G_{\mu_1}(g_2)$ given by (\ref{Gen1}).
 The graphical correction  $\delta\mu_2$ to $\mu_2$ is
\beqn
 \delta\mu_2&=&
\frac{2\mu_2g_1\dd \ell}{(d-1)(d+1)g_2}  \Bigg\{
\left(\frac{4d-3}{32}(1+g_2)^{-3/2} +
\frac{1}{8}
-2d\sqrt{\mu_x \mu_1^3}A(\mu_L,\mu_1)
-2\sqrt{\mu_x \mu_1^3}A(\mu_1,\mu_L)
+4\sqrt{\mu_x \mu_1^5}B(\mu_L,\mu_1)
 \right)\nn
\\
&&
+\Bigg[
\frac{5-2d}{32} (1+g_2)^{-3/2}   +   \frac{d^2-2d-1}{16} -(d^2-2d-1)\sqrt{\mu_x \mu_1^3}A(\mu_1,\mu_L)
-  (1-d)\sqrt{\mu_x \mu_1^3}A(\mu_L,\mu_1) \nn
\\
&&+ 2(d-1)\sqrt{\mu_x \mu_1^3}\mu_L B(\mu_1,\mu_L)
\Bigg]-\Bigg[
\frac{3d-10}{64}    (1+g_2)^{-3/2}
+2\sqrt{\mu_x \mu_1^3}A(\mu_L,\mu_1)
+2(1-d)\sqrt{\mu_x \mu_1^5}B(\mu_L,\mu_1)+\nn\\
&&2\sqrt{\mu_x \mu_1^3}A(\mu_1,\mu_L) +\frac{3d-11}{64} \Bigg]\nn
-\left[
\frac{13}{64} (1+g_2)^{-3/2}
-2\sqrt{\mu_x \mu_1^3}A(\mu_L,\mu_1)
+(1-d)\sqrt{\mu_x \mu_1^3}A(\mu_1,\mu_L)+\right.\nn\\
&&\left.\frac{7(d-1)}{64}
+2(1-d) \sqrt{\mu_x \mu_1^3}\mu_LB(\mu_1,\mu_L)
\right]
\Bigg\}\nn
\\
\nonumber
&=&\frac{2\mu_2g_1\dd \ell}{(d-1)(d+1)g_2}  \Bigg[\frac{1+d}{64}(1+g_2)^{-3/2} +
\frac{2d^2-9d+11}{32}
-(d+1)\sqrt{\mu_x \mu_1^3}A(\mu_L,\mu_1)
-\nn\\
&&(d^2-3d+4)\sqrt{\mu_x \mu_1^3}A(\mu_1,\mu_L)+2(d+1)\sqrt{\mu_x \mu_1^5}B(\mu_L,\mu_1)+4(d-1)\sqrt{\mu_x \mu_1^3} \mu_LB(\mu_1,\mu_L)
	\Bigg]\nn
	\\
	&=&g_1\mu_2G_{\mu_2}(g_2)
	\ ,
\eeqn
 where $G_{\mu_2}(g_2)$ is given in (\ref{Gen2}).

\subsection{Noise renormalization}
\subsubsection{Graph in \fig \ref{fig:noise}(a)}
The graph   in \fig \ref{fig:noise}(a)  represents the following correction to the noise  correlator $\left<f_\ell(\tilde\bk)f_u(-\tilde\bk)\right>$:
\beqn
 \Delta\left<f_\ell(\tilde\bk)f_u(-\tilde\bk)\right>_{D,a}&=&
{2\lambda^2D^2\over (2\pi)^{d+1}}\int_{\tilde{\bq} }^>q_i^\perp q_m^\perp C_{im}(\tilde{\bk}-\tilde{\bq})C_{\ell u}(\tilde{\bq})
\equiv 2 \lambda^2 D^2 (I^{D,a})_{\ell u}(\tilde{\bk})
 \ ,\label{CorrD1}
\eeqn
where
\beq
{(I^{D,a})}_{\ell u}(\tilde{\bk})\equiv{1\over (2\pi)^{d+1}}\int_{\tilde{\bq} }^>q_i^\perp q_m^\perp C_{im}(\tilde{\bk}-\tilde{\bq})C_{\ell u}(\tilde{\bq})\,.
\eeq

   Since the noise strength $D$ is the value of this correlation at $\bk={\bf 0}$,  we can evaluate ${(I^{D,a})}_{\ell u}(\tilde{\bk})$ at $\tilde{\bk}=0$. This gives
\beqn
{(I^{D,a})}_{\ell u}({\tilde{\bf 0}})&=&{1\over (2\pi)^{d+1}}\int_{\tilde{\bq} }^>q_i^\perp q_m^\perp C_{im}(-\tilde{\bq})C_{\ell u}(\tilde{\bq})\nonumber
\\
&=&{1\over (2\pi)^{d+1}}\int_{\tilde{\bq} }^>q_i^\perp q_m^\perp \left[
\mid G_L(\tilde{\bq})\mid^2L^\perp_{im}({\bq}) +\mid G_T(\tilde{\bq})\mid^2P^\perp_{im}({\bq})\right]
 \left[\mid G_L(\tilde{\bq})\mid^2L^\perp_{\ell u}({\bq}) +\mid G_T(\tilde{\bq})\mid^2P^\perp_{\ell u}({\bq})
\right]\nonumber
\\
&=&{1\over (2\pi)^{d+1}}\int_{\tilde{\bq} }^>q_\perp^2\mid G_L(\tilde{\bq})\mid^2
\left[ \mid G_L(\tilde{\bq})\mid^2L^\perp_{\ell u}({\bq}) +\mid^2 G_T(\tilde{\bq})\mid^2P^\perp_{\ell u}({\bq})\right]\nonumber
\\
&=&{1\over (2\pi)^{d-1}}\int_{\bq_\perp }^>\frac{1}{q_\perp^3}\left[\frac{3}{64\sqrt{\mu_x \mu_L^5}} L^\perp_{\ell u}({\bq}) + \frac{1}{ 4\sqrt{\mu_x} (\mu_L-\mu_1)^2} \left(\frac{1}{\sqrt{\mu_L}} +\frac{1}{\sqrt{\mu_1}}-\frac{2\sqrt{2}}{\sqrt{\mu_L+\mu_1}}  \right) P^\perp_{\ell u}({\bq})\right]\nonumber\\
&=& \left[\frac{3}{64\sqrt{\mu_x \mu_L^5}} \frac{1}{d-1}+
\frac{1}{ 4\sqrt{\mu_x} (\mu_L-\mu_1)^2} \left(\frac{1}{\sqrt{\mu_L}} +\frac{1}{\sqrt{\mu_1}}-\frac{2\sqrt{2}}{\sqrt{\mu_L+\mu_1}}  \right) \left(1-\frac{1}{d-1}\right)\right]\frac{\delta^\perp_{\ell u}S_{d-1}}{(2\pi)^{d-1}} \Lambda^{d-4}\dd \ell
 \ .\nonumber\\
\eeqn

 Identifying the coefficient of $\delta_{cu}^\perp$ as a  correction $(\delta D)_{D,a}$ to $D$ gives:
\beqn
(\delta D)_{D,a}&=&2 \left[\frac{3}{64\sqrt{\mu_x \mu_L^5}} \frac{1}{d-1}+
\frac{1}{ 4\sqrt{\mu_x} (\mu_L-\mu_1)^2} \left(\frac{1}{\sqrt{\mu_L}} +\frac{1}{\sqrt{\mu_1}}-\frac{2\sqrt{2}}{\sqrt{\mu_L+\mu_1}}  \right) \left(1-\frac{1}{d-1}\right)\right]\frac{D^2\lambda^2 S_{d-1}}{(2\pi)^{d-1}} \Lambda^{d-4}\dd \ell
\nonumber\\
&=&
2 g_1Dd\ell\left[\frac{3}{64(1+g_2)^{5/2}} \left(\frac{1}{d-1}\right)+
\frac{1}{ 4g_2^2} \left(\frac{1}{\sqrt{1+g_2}} +1-2\sqrt{2\over2+g_2}  \right) \left(\frac{d-2}{d-1}\right)\right]
\,.
\eeqn

\subsubsection{Graph in \fig \ref{fig:noise}(b)}
The second correction to $D$ comes from the diagram in \fig \ref{fig:noise}(b), which represents
 the following correction to the noise  correlator $\left<f_\ell(\tilde\bk)f_u(-\tilde\bk)\right>$:
\beqn
\Delta\left<f_\ell(\tilde\bk)f_u(-\tilde\bk)\right>_{D,b}&=&{2\lambda^2D^2\over (2\pi)^{d+1}}\int_{\tilde{\bq} }^>q_i^\perp (k_m^\perp -q_m^\perp ) C_{iu}(\tilde{\bk}-\tilde{\bq})C_{\ell m}(\tilde{\bq}) \equiv 2\lambda^2D^2 (I^{D,b})_{\ell u}(\tilde{\bk})\label{CorrD2}
\ ,
\eeqn
where
\beqn
(I^{D,b})_{\ell u}(\tilde{\bk})\equiv {1\over (2\pi)^{d+1}}\int_{\tilde{\bq} }^>q_i^\perp (k_m^\perp -q_m^\perp ) C_{iu}(\tilde{\bk}-\tilde{\bq})C_{\ell m}(\tilde{\bq})\,.
\eeqn

Again we only need to calculate $(I^{D,b})_{\ell u}({\tilde{\bf 0}})$ to get the relevant correction to the noise strength $D$.
\beqn
(I^{D,b})_{\ell u}({\tilde{\bf 0}}) &=& -{1\over (2\pi)^{d+1}}\int_{\tilde{\bq} }^>q_i^\perp q_m^\perp  C_{iu}(\tilde{\bq})C_{\ell m}(\tilde{\bq})\nonumber
\\
&=& -{1\over (2\pi)^{d+1}}\int_{\tilde{\bq} }^>q_i^\perp q_m^\perp \left[
\mid G_L(\tilde{\bq})\mid^2L^\perp_{iu}({\bq}) +\mid G_T(\tilde{\bq})\mid^2 P^\perp_{iu}({\bq})\right]
 \left[\mid G_L(\tilde{\bq})\mid^2L^\perp_{\ell m}({\bq}) +\mid G_T(\tilde{\bq})\mid^2P^\perp_{\ell m}({\bq})
\right]
\nonumber\\
&=&-{1\over (2\pi)^{d+1}}\int_{\tilde{\bq} }^>q_\ell^\perp q_u^\perp
\mid G_L(\tilde{\bq})\mid^4\nonumber\\
&=&-\frac{3}{64\sqrt{\mu_x \mu_L^5}}{1\over (2\pi)^{d-1}} \int_{\bq_\perp}^>\frac{q_\ell^\perp q_u^\perp}{q_\perp^5}\nonumber
\\
&=&-\frac{3\delta^\perp
_{\ell u}}{64(d-1)}{1\over\sqrt{\mu_x \mu_L^5}}\frac{S_{d-1}}{(2\pi)^{d-1}} \Lambda^{d-4}\dd \ell
\eeqn

Identifying the coefficient of $\delta_{cu}^\perp$ as a  correction $(\delta D)_{D,b}$ to $D$ gives:
\beq
 (\delta D)_{D,b}=- \frac{3}{32(d-1)}\frac{D^2\lambda^2}{\sqrt{\mu_x \mu_L^5}} \frac{S_{d-1}}{(2\pi)^{d-1}} \Lambda^{d-4}\dd \ell =- \frac{3}{32(d-1)}\frac{g_1D\dd \ell}{(1+g_2)^{5/2}}
\,.
\eeq

Combining the two corrections to $D$ from the two diagrams in Fig. \ref{fig:noise}, we obtain the total  one loop correction  $\delta D$ to $D$:
\beq
\delta D=(\delta D)_{6a}+(\delta D)_{6b}=\frac{g_1D\dd \ell}{2g_2^2} \frac{(d-2)}{(d-1)}
\left[ 1+\frac{1}{\sqrt{1+g_2}}  -\frac{2\sqrt{2}}{\sqrt{2+g_2}}\right]=g_1D\dd \ell G_D(g_2)\,,
\eeq
where $ G_D(g_2)$ is given by (\ref{eq:GD0}).

\subsection{ Summary of all corrections to one loop order}

Adding up the results obtained in previous sections gives the total one loop graphical corrections to the various parameters:

\beqn
&&\delta\mu_1=g_1\mu_1G_{\mu_1}(g_2)\dd\ell\,,\label{mu1graphmu2=0}\\
&&\delta\mu_2=g_1\mu_2G_{\mu_2}(g_2)\dd\ell\,,\label{mu2graphmu2=0}\\
&&\delta D=g_1DG_D(g_2)\dd\ell\,,\label{Dgraphmu2=0}\\
&&\delta\mu_x=0\label{muxgraphmu2=0}\,.
\eeqn
Dividing both sides of each of these equations by $\dd\ell$, we obtain the graphical contributions to the recursion relations for the parameters of our model:
\beqn
&&\left({\dd\mu_1\over\dd\ell}\right)_{\rm graph}=g_1\mu_1G_{\mu_1}(g_2)\,,\label{mu1graphmu2=0rr}\\
&&\left({\dd\mu_2\over\dd\ell}\right)_{\rm graph}=g_1\mu_2G_{\mu_2}(g_2)\,,\label{mu2graphmu2=0rr}\\
&&\left({\dd D\over\dd\ell}\right)_{\rm graph}=g_1DG_D(g_2)\,,\label{Dgraphmu2=0rr}\\
&&\left({\dd\mu_x\over\dd\ell}\right)_{\rm graph}=0\label{muxgraphmu2=0rr}\,,
\eeqn
where the functions $G_{\mu_1}(g_2)$, $G_{\mu_2}(g_2)$, and $G_{D}(g_2)$ are given respectively by equations (\ref{Gen1}),  (\ref{Gen2}), and (\ref{eq:GD0}) of section (\ref{nonlin}). Combining these with the RG rescalings discussed in that section lead to the full recursion relations (\ref{eq:D})-(\ref{eq:mu2}) of that section.

It is also a straightforward, but tedious, exercise in the application of l'Hopital's rule to show that, in the limit $g_2\to0$, all of the apparent singularities at small $g_2$ in $G_{\mu_1}(g_2)$, $G_{\mu_2}(g_2)$, and $G_{D}(g_2)$ exactly cancel, leaving precisely the finite results (\ref{Gmu1mu2=0})-(\ref{GDmu2=0}) obtained in the previous appendix for the $\mu_2=0$ case.

\section{Useful Formulae\label{sec:formulae}}

In this appendix, we summarize the integrals needed for the graphical calculations done in the preceding appendices. Throughout this section, we will for convenience define the wavevector dependent dampings:
\beq
\Gamma_L\equiv {1\over\mu_Lq_\perp^2+\mu_xq_x^2} \,,\,\,\,\,\,\,\,\Gamma_T\equiv {1\over\mu_Tq_\perp^2+\mu_xq_x^2} \ ,
\label{gdef}
\eeq
where we remind the reader of our definition $\mu_L\equiv\mu_1+\mu_2$.

 We begin with:

\subsection{Integrations over $\Omega$ and $q_x$}
\beqn
{1\over (2\pi)^{d+1}} \int_{-\infty}^{\infty}\dd q_x\int_{-\infty}^{\infty}\dd\Omega\,\,\,
G_L(\tilde{\bq})G_L(-\tilde{\bq})&=&{1\over (2\pi)^{d+1}}\int_{-\infty}^{\infty}\dd q_x\int_{-\infty}^{\infty}{\dd\Omega\over\Omega^2+\Gamma_L^2(\bq)}\nonumber\\
&=&{1\over (2\pi)^{d}}\int_{-\infty}^{\infty}\frac{\dd q_x}{2\Gamma_L(\bq)}
\nonumber\\
&=&{1\over (2\pi)^{d-1}}\frac{1}{4\sqrt{\mu_x \mu_L}} {1\over q_\perp}\ ,
\eeqn
where both the integrals over $\Omega$ and $q_x$ can be done either by simple complex contour
techniques, or by even simpler trigonometric substitutions. The same statement applies to all of the integrals that follow here; we will therefore simply quote the results for the remainder:

\beq
{1\over (2\pi)^{d+1}} \int_{-\infty}^{\infty}\dd q_x\int_{-\infty}^{\infty}\dd\Omega\,\,\,G_L(\tilde{\bq})G_L(-\tilde{\bq})^2
={1\over (2\pi)^d} \int_{-\infty}^{\infty}\dd q_x\,\,\,\frac{1}{4\Gamma_L(\bq)^2}=\frac{1}{16\sqrt{\mu_x \mu_L^3}}{1\over (2\pi)^{d-1}} \frac{1}{q_\perp^3}
\ .
\eeq

\beq
{1\over (2\pi)^{d+1}}\int_{-\infty}^{\infty}\dd q_x\int_{-\infty}^{\infty}\dd\Omega\,\,\,G_L(\tilde{\bq})G_L(-\tilde{\bq})^3
={1\over (2\pi)^d} \int_{-\infty}^{\infty}\dd q_x\,\,\,\frac{1}{8\Gamma_L(\bq)^3} = \frac{3}{128\sqrt{\mu_x \mu_L^5}}{1\over (2\pi)^{d-1}} \frac{1}{q_\perp^5}
\ .
\eeq

\beq
{1\over (2\pi)^{d+1}} \int_{-\infty}^{\infty}\dd q_x\int_{-\infty}^{\infty}\dd\Omega\,\,\,G_L(\tilde{\bq})G_L(-\tilde{\bq})^4
={1\over (2\pi)^d} \int_{-\infty}^{\infty}\dd q_x\,\,\,\frac{1}{16\Gamma_L(\bq)^4} = \frac{5}{512\sqrt{\mu_x \mu_L^7}} {1\over (2\pi)^{d-1}}\int_{\bq_\perp}\frac{1}{q_\perp^7}
\ .
\eeq

\beq
{1\over (2\pi)^{d+1}} \int_{-\infty}^{\infty}\dd q_x\int_{-\infty}^{\infty}\dd\Omega\,\,\,q_x^2 G_L(\tilde{\bq})G_L(-\tilde{\bq})^4
={1\over (2\pi)^d} \int_{-\infty}^{\infty}\dd q_x\,\,\,\frac{q_x^2}{16\Gamma_L(\bq)^4} = \frac{1}{512\sqrt{\mu_x^3 \mu_L^5}}{1\over (2\pi)^{d-1}} \frac{1}{q_\perp^5}
\ .
\eeq

\beq
{1\over (2\pi)^{d+1}}\int_{\tilde{\bq} }
G_L(\tilde{\bq})^2 G_L(-\tilde{\bq})^2
={1\over (2\pi)^d}\int_{-\infty}^{\infty}\dd q_x\,\,\,\frac{1}{4\Gamma_L(\bq)^3}=\frac{3}{64\sqrt{\mu_x \mu_L^5}} {1\over (2\pi)^{d-1}}\int_{\bq_\perp}\frac{1}{q_\perp^5}
\ .
\eeq

\beq
{1\over (2\pi)^{d+1}} \int_{-\infty}^{\infty}\dd q_x\int_{-\infty}^{\infty}\dd\Omega\,\,\,G_T(\tilde{\bq})G_T(-\tilde{\bq})^2
={1\over (2\pi)^d} \int_{-\infty}^{\infty}\dd q_x\,\,\,\frac{1}{4\Gamma_T(\bq)^2}=\frac{1}{16\sqrt{\mu_x \mu_1^3}}{1\over (2\pi)^{d-1}} \frac{1}{q_\perp^3}
\ .
\eeq

\beq
{1\over (2\pi)^{d+1}} \int_{-\infty}^{\infty}\dd q_x\int_{-\infty}^{\infty}\dd\Omega\,\,\,G_T(\tilde{\bq})G_T(-\tilde{\bq})^3
={1\over (2\pi)^d}\int_{-\infty}^{\infty}\dd q_x\,\,\,\frac{1}{8\Gamma_T(\bq)^3} = \frac{3}{128\sqrt{\mu_x \mu_1^5}}{1\over (2\pi)^{d-1}} \frac{1}{q_\perp^5}
\ .
\eeq

\beq
{1\over (2\pi)^{d+1}}\int_{-\infty}^{\infty}\dd q_x\int_{-\infty}^{\infty}\dd\Omega\,\,\,G_T(\tilde{\bq})G_T(-\tilde{\bq})^4
={1\over (2\pi)^d} \int_{-\infty}^{\infty}\dd q_x\,\,\,\frac{1}{16\Gamma_T(\bq)^4} = \frac{5}{512\sqrt{\mu_x \mu_1^7}} {1\over (2\pi)^{d-1}}\int_{\bq_\perp}\frac{1}{q_\perp^7}
\ .
\eeq

\beq
{1\over (2\pi)^{d+1}} \int_{-\infty}^{\infty}\dd q_x\int_{-\infty}^{\infty}\dd\Omega\,\,\,q_x^2 G_T(\tilde{\bq})G_T(-\tilde{\bq})^4
={1\over (2\pi)^d} \int_{-\infty}^{\infty}\dd q_x\,\,\,\frac{q_x^2}{16\Gamma_T(\bq)^4} = \frac{1}{512\sqrt{\mu_x^3 \mu_1^5}}{1\over (2\pi)^{d-1}} \frac{1}{q_\perp^5}
\ .
\eeq

\beq
{1\over (2\pi)^{d+1}}\int_{\tilde{\bq} }
G_T(\tilde{\bq})^2 G_T(-\tilde{\bq})^2
={1\over (2\pi)^d} \int_{-\infty}^{\infty}\dd q_x\,\,\,\frac{1}{4\Gamma_T(\bq)^3}=\frac{3}{64\sqrt{\mu_x \mu_1^5}} {1\over (2\pi)^{d-1}}\int_{\bq_\perp}\frac{1}{q_\perp^5}
\ .
\eeq

\beqn
{1\over (2\pi)^{d+1}}\int_{\tilde{\bq} }
G_L(\tilde{\bq})G_L(-\tilde{\bq})G_T(\tilde{\bq})G_T(-\tilde{\bq}) &=& {1\over (2\pi)^d} \int_{-\infty}^{\infty}\dd q_x\,\,\,\frac{1}{2\Gamma_L(\bq)\Gamma_T(\bq) (\Gamma_L(\bq)+\Gamma_T(\bq))}\nn
\\
&=& \frac{1}{2\sqrt{\mu_x} (\mu_L-\mu_1)^2} \left(\frac{1}{\sqrt{\mu_L}} +\frac{1}{\sqrt{\mu_1}}-\frac{2\sqrt{2}}{\sqrt{\mu_L+\mu_1}}  \right) {1\over (2\pi)^{d-1}}\int_{\bq_\perp}\frac{1}{q_\perp^5}\label{lltt}
\ .\nonumber\\
\\
{1\over (2\pi)^{d+1}} \int_{-\infty}^{\infty}\dd q_x\int_{-\infty}^{\infty}\dd\Omega\,\,\,G_L(\tilde{\bq})G_L(-\tilde{\bq})G_T(-\tilde{\bq})
&=&{1\over (2\pi)^d} \int_{-\infty}^{\infty}\dd q_x\,\,\,\frac{1}{4\Gamma_L(\bq) (\Gamma_L(\bq)+\Gamma_T(\bq))}\nn
\\
&=&\frac{1}{4 \sqrt{\mu_x} (\mu_L-\mu_1)}
\left( \frac{\sqrt{2}}{\sqrt{\mu_L+\mu_1}} -\frac{1}{\sqrt{\mu_L}}\right)
{1\over (2\pi)^{d-1}} \frac{1}{q_\perp^3} \nn
 \\
 &\equiv&A(\mu_L,\mu_1){1\over (2\pi)^{d-1}}\int_{\bq_\perp}\frac{1}{q_\perp^3}
 \ .
 \\
{1\over (2\pi)^{d+1}} \int_{-\infty}^{\infty}\dd q_x\int_{-\infty}^{\infty}\dd\Omega\,\,\,G_L(\tilde{\bq})G_L(-\tilde{\bq})G_T(-\tilde{\bq})^2
&=&{1\over (2\pi)^d} \int_{-\infty}^{\infty}\dd q_x\,\,\,\frac{1}{2\Gamma_L(\bq) (\Gamma_L(\bq)+\Gamma_T(\bq))^2}\nn
\\
&=&\frac{2(\mu_L+\mu_1)^{3/2}-\sqrt{2\mu_L} (\mu_L+3\mu_1)}{8\sqrt{\mu_x\mu_L}(\mu_L+\mu_1)^{3/2}(\mu_L-\mu_1)^2} {1\over (2\pi)^{d-1}}\int_{\bq_\perp}\frac{1}{q_\perp^5}\nn
\\
&\equiv& B(\mu_L,\mu_1) {1\over (2\pi)^{d-1}}\int_{\bq_\perp}\frac{1}{q_\perp^5}
\ .
\eeqn

\subsection{ Integrals over $\bq_\perp$}

 After performing the integrals over $\Omega$ and $q_x$, our last step is performing the remaining integral over $\bq_\perp$ in
\beq
\int_{\tilde{\bq}}^>\ \
\equiv\int_{\Lambda>{|\bq_\perp|>\Lambda e^{-d\ell}}}\dd^{d-1}q_\perp\int_{-\infty}^{\infty}\dd\Omega\int_{-\infty}^{\infty}\dd q_x
\,.
\eeq
The simplest such integral that arises is:
\beq
{1\over (2\pi)^{d-1}}\int_{\Lambda>{|\bq_\perp|>\Lambda e^{-d\ell}}}\frac{\dd^{d-1}q_\perp}{q_\perp^3}
={1\over (2\pi)^{d-1}}\int_{\Lambda e^{-d\ell}}^\Lambda dq_\perp \int\dd\Xi_{\bq_\perp}q_\perp^{d-5}
\ ,
\eeq
where $\int\dd\Xi_{\bq_\perp}$ denotes an integral over the $d-1$-dimensional solid angle associated with $\bq_\perp$.
Since the integrand is independent of the direction of $\bq_\perp$, we can do this integral trivial; it simply gives a multiplicative factor of $S_{d-1}$, the surface area of a unit $d-1$-dimensional sphere. Thus we have
\beq
{1\over (2\pi)^{d-1}}\int_{\Lambda>{|\bq_\perp|>\Lambda e^{-\dd\ell}}}\frac{\dd^{d-1}q_\perp}{q_\perp^3}={S_{d-1}\over (2\pi)^{d-1}}\int_{\Lambda e^{-\dd\ell}}^\Lambda dq_\perp q_\perp^{d-5}=\frac{S_{d-1}}{(2\pi)^{d-1}}\Lambda^{d-4}\dd \ell
\label{0q}
\ ,
\eeq
where in the last step we have used the fact that $\dd\ell$ is infinitesimal to write $1-e^{-\dd\ell}=\dd\ell$.

A slightly harder integral that arises in our calculations  is
\beq
I_{iu}={1\over (2\pi)^{d-1}}\int_{\bq_\perp }
\frac{q^\perp_iq^\perp_u}{q_\perp^5}
\label{2q}
\ .
\eeq
This can, however, be done using symmetry arguments. Since the integrand is odd if $i\ne u$, this integral can only be non-zero if $i=u$. Furthermore, by spherical symmetry, if $i=u$, we should get the same value for this integral regardless of the value of $i$. Hence, this integral must be proportional to $\delta_{iu}^\perp$:
\beq
I_{iu}=A\delta_{iu}^\perp\ ,
\label{Itens}
\eeq
where the constant $A$ remains to be determined. This can be done by taking the trace of (\ref{Itens}) over $iu$,, which gives
\beq
I_{ii}=A(d-1)\ .
\label{Itrace}
\eeq
On the other hand, taking the trace over $iu$ of (\ref{2q}) gives
\beq
I_{ii}={1\over (2\pi)^{d-1}}\int_{\bq_\perp }
\frac{1}{q_\perp^3}=\frac{S_{d-1}}{(2\pi)^{d-1}}\Lambda^{d-4}\dd \ell
\label{2qtrace}
\ ,
\eeq
where in the second equality we have used (\ref{0q}).

Equating (\ref{2qtrace}) and (\ref{Itrace}) and solving for $A$ gives
\beq
A=\frac{S_{d-1}}{(d-1)(2\pi)^{d-1}}\Lambda^{d-4}\dd \ell \ ,
\label{Asol}
\eeq
whence it follows from (\ref{Itens}) that
\beq
{1\over (2\pi)^{d-1}}\int_{\bq_\perp }
\frac{q^\perp_iq^\perp_u}{q_\perp^5}
=\frac{\delta_{iu}}{(d-1)} \frac{S_{d-1}}{(2\pi)^{d-1}}\Lambda^{d-4}\dd \ell
\ .\label{Angleave3}
\eeq

Very similar reasoning can be used to show that
\beqn
{1\over (2\pi)^{d-1}}\int_{\bq_\perp }
 \frac{q^\perp_iq^\perp_uq^\perp_jq^\perp_\ell}{q_\perp^7}
 &=&{1\over (2\pi)^{d-1}}\int_{\bq_\perp }
 \frac{q^\perp_iq^\perp_uq^\perp_jq^\perp_\ell}{q_\perp^7}=
 \frac{\delta_{iu}\delta_{jc}+\delta_{ij}\delta_{uc}+\delta_{ic}\delta_{ju}}{(d-1)(d+1)}
 {1\over (2\pi)^{d-1}}\int \dd q q^{d-5}\nn
 \\
 &=&\frac{\delta_{iu}\delta_{jc}+\delta_{ij}\delta_{uc}+\delta_{ic}\delta_{ju}}{(d-1)(d+1)} \frac{S_{d-1}}{(2\pi)^{d-1}}\Lambda^{d-4}\dd \ell\nn
 \\
 &\equiv &\frac{\Pi_{iujc}}{(d-1)(d+1)} \frac{S_{d-1}}{(2\pi)^{d-1}}\Lambda^{d-4}\dd \ell
 \ .
\eeqn

\subsection{Expansions with respect to $\bk$}
At many points in our clculations, we have to expand the popagators and projection operators in powers of the external wavevector. These expansions are:
\beqn
L_{i\ell}(\bk-\bq)&=& \frac{(k_i-q_i)(k_b-q_b)}{|\bk-\bq|^2} = \frac{(k_i-q_i)(k_b-q_b)}{q^2} \left(1+\frac{2\bq \cdot \bk -k^2}{q^2}
+\frac{4(\bq \cdot \bk)^2}{q^4} +\cO(k^3)
\right)\nn
\\
&=& L_{i\ell}(\bq) +  \frac{ 2L_{i\ell}(\bq) \bq \cdot \bk -k_iq_b-k_bq_i}{q^2}
+L_{i\ell}(\bq) \left(-\frac{k^2}{q^2} +\frac{4(\bq \cdot \bk)^2}{q^4}\right) +\frac{k_ik_\ell}{q^2}\nn
\\
&&- \frac{2\bq \cdot \bk (k_iq_b+k_bq_i)}{q^4} +\cO(k^3)
\ .
\\
P_{i\ell}(\bk-\bq)&=& P_{i\ell}(\bq) -  \frac{ 2L_{i\ell}(\bq) \bq \cdot \bk -k_iq_b-k_bq_i}{q^2}
-L_{i\ell}(\bq) \left(-\frac{k^2}{q^2} +\frac{4(\bq \cdot \bk)^2}{q^4}\right) -\frac{k_ik_\ell}{q^2}\nn
\\
&&+ \frac{2\bq \cdot \bk (k_iq_b+k_bq_i)}{q^4} +\cO(k^3)
\ .
\eeqn

\beqn
\nonumber
 G_L(\tilde{\bk}-\tilde{\bq})
&=& \frac{1}{\ii \omega +\mu_L (\bk_\perp-\bq_\perp)^2 +\mu_x (k_x-q_x)^2}
\\
\nonumber
&=&\frac{1}{\ii (\omega-\Omega) +\Gamma(\bq)} \Bigg(1+\frac{2\mu_L\bq_\perp \cdot \bk_\perp+ 2\mu_x q_x k_x -\mu_L k_\perp^2 -\mu_xk_x^2}{\ii  (\omega-\Omega) +\Gamma(\bq)}+\nn\\
&&\frac{4\mu_L^2 (\bq_\perp \cdot \bk_\perp )^2 +4\mu_x^2 q_x^2k_x^2+4 \mu_x\mu_L(\bq_\perp \cdot \bk_\perp)q_x k_x}{(\ii (\omega-\Omega) +\Gamma(\bq))^2} +\cO(k^3)\Bigg)\nn\\
&=&G_L(-\tilde{\bq}) +\left(2\mu_L\bq_\perp \cdot \bk_\perp+ 2\mu_x q_x k_x \right) G_L(-\tilde{\bq})^2-\left(\mu_L k_\perp^2 +\mu_xk_x^2\right)G_L(-\tilde{\bq})^2
+\nn\\
&&\left[(4\mu_L^2 (\bq_\perp \cdot \bk_\perp)^2 +4\mu_x^2 q_x^2k_x^2+4 \mu_x\mu_L(\bq_\perp \cdot \bk_\perp)q_x k_x\right] G_L(-\tilde{\bq})^3 +\cO(k^3)
\ .
\eeqn

\beqn
G_L(\tilde{\bk}-\tilde{\bq})L^\perp_{i\ell}(\bk-\bq)&=&
G_L(-\tilde{\bq})L^\perp_{i\ell}(\bq)
+\left(2\mu_L\bq_\perp \cdot \bk_\perp+ 2\mu_x q_x k_x \right) G_L(-\tilde{\bq})^2L^\perp_{i\ell}(\bq) +  \nn\\
&&\frac{ 2L_{i\ell}^\perp(\bq) \bq_\perp \cdot \bk_\perp -k^\perp_iq^\perp_b-k^\perp_bq^\perp_i}{q_\perp^2}G_L(-\tilde{\bq})
-\left(\mu_L k_\perp^2 +\mu_xk_x^2\right)G_L(-\tilde{\bq})^2L^\perp_{i\ell}(\bq)\nn
    \\
    &&+\left[(4\mu_L^2 (\bq_\perp \cdot \bk_\perp)^2 +4\mu_x^2 q_x^2k_x^2+4 \mu_x\mu_L(\bq_\perp \cdot \bk_\perp)q_xk_x\right] G_L(-\tilde{\bq})^3L^\perp_{i\ell}(\bq)\nn
\\
&&+\left[L^\perp_{i\ell}(\bq) \left(-\frac{k_\perp^2}{q_\perp^2} +\frac{4(\bq_\perp \cdot \bk_\perp)^2}{q_\perp^4}\right) +\frac{k^\perp_ik^\perp_\ell}{q_\perp^2} - \frac{2\bq_\perp \cdot \bk_\perp (k^\perp_iq^\perp_b+k^\perp_bq^\perp_i)}{q_\perp^4}\right]G_L(-\tilde{\bq})\nn
\\
&&+\left(2\mu_L\bq_\perp \cdot \bk_\perp+ 2\mu_x q_x k_x \right) G_L(-\tilde{\bq})^2 \left(\frac{ 2L^\perp_{i\ell}(\bq) \bq_\perp \cdot \bk_\perp -k^\perp_iq^\perp_b-k^\perp_bq^\perp_i}{q_\perp^2}\right)
\ .
\eeqn

\beqn
G_T(\tilde{\bk}-\tilde{\bq})P^\perp_{i\ell}(\bk-\bq)&=&
G_T(-\tilde{\bq})P^\perp_{i\ell}(\bq)
+\left(2\mu_1\bq_\perp \cdot \bk_\perp+ 2\mu_x q_x k_x \right) G_T(-\tilde{\bq})^2P^\perp_{i\ell}(\bq) -\nonumber\\
&&\frac{ 2L_{i\ell}^\perp(\bq) \bq_\perp \cdot \bk_\perp -k^\perp_iq^\perp_b-k^\perp_bq^\perp_i}{q_\perp^2}G_T(-\tilde{\bq})
-\left(\mu_1 k_\perp^2 +\mu_xk_x^2\right)G_T(-\tilde{\bq})^2P^\perp_{i\ell}(\bq)\nn
    \\
    &&-\left[(4\mu_1^2 (\bq_\perp \cdot \bk_\perp)^2 +4\mu_x^2 q_x^2k_x^2+4 \mu_x\mu_1(\bq_\perp \cdot \bk_\perp)q_xk_x\right] G_T(-\tilde{\bq})^3P^\perp_{i\ell}(\bq)\nn
\\
&&-\left[L^\perp_{i\ell}(\bq) \left(-\frac{k_\perp^2}{q_\perp^2} +\frac{4(\bq_\perp \cdot \bk_\perp)^2}{q_\perp^4}\right) +\frac{k^\perp_ik^\perp_\ell}{q_\perp^2} - \frac{2\bq_\perp \cdot \bk_\perp (k^\perp_iq^\perp_b+k^\perp_bq^\perp_i)}{q_\perp^4}\right]G_T(-\tilde{\bq})\nn
\\
&&-\left(2\mu_1\bq_\perp \cdot \bk_\perp+ 2\mu_x q_x k_x \right) G_T(-\tilde{\bq})^2 \left(\frac{ 2L^\perp_{i\ell}(\bq) \bq_\perp \cdot \bk_\perp -k^\perp_iq^\perp_b-k^\perp_bq^\perp_i}{q_\perp^2}\right)
\ .
\eeqn

\twocolumngrid

\begin{acknowledgments}
LC acknowledges support by the National Science Foundation of China (under Grant No. 11874420). JT thanks  The Higgs Centre for Theoretical Physics at the University of Edinburgh for their hospitality and support while this work was in progress.
\end{acknowledgments}


\begin{thebibliography}{7}
	\expandafter\ifx\csname natexlab\endcsname\relax\def\natexlab#1{#1}\fi
	\expandafter\ifx\csname bibnamefont\endcsname\relax
	\def\bibnamefont#1{#1}\fi
	\expandafter\ifx\csname bibfnamefont\endcsname\relax
	\def\bibfnamefont#1{#1}\fi
	\expandafter\ifx\csname citenamefont\endcsname\relax
	\def\citenamefont#1{#1}\fi
	\expandafter\ifx\csname url\endcsname\relax
	\def\url#1{\texttt{#1}}\fi
	\expandafter\ifx\csname urlprefix\endcsname\relax\def\urlprefix{URL }\fi
	\providecommand{\bibinfo}[2]{#2}
	\providecommand{\eprint}[2][]{\url{#2}}
	


		\bibitem{Active1}
	S.~Ramaswamy, The mechanics and statics of active matter. Ann. Rev. Condens. Matt. Phys. {\bf 1}, 323-345 (2010).
	
			\bibitem{Active2}
 M.C.~Marchetti, J.F.~Joanny, S.~Ramaswamy, T.B.~Liverpool, J.~Prost, M.~Rao,  and R.A.~Simha, Hydrodynamics of soft active matter, {\it Rev. Mod. Phys.} {\bf 85}, 1143-1188 (2013).
 	
		\bibitem{Active3}
 C.~Bechinger, R.~Di Leonardo, H.~L\"{o}wen, C.~Reichhardt, G.~Volpe, and G.~Volpe, Active particles in complex and crowded environments, Rev. Mod. Phys. {\bf 88}, 045006 (2016).
 	
		\bibitem{Active4}
	F.~Schweitzer, {\it Brownian Agents and Active Particles: Collective Dynamics in the Natural and Social Sciences}. Springer Series in Synergetics (Springer, New York, 2003).

	
        \bibitem{Vicsek}
T. Vicsek, A. Czir\'{o}k, E. Ben-jacob, I. Cohen, and O. Shochet, Novel type of phase transition in a system of self-Driven particles. Phys.\ Rev.\ Lett. {\bf 75}, 1226 (1995).

\bibitem{TT1}
	J.~Toner, and Y.~Tu, Long-range order in a two-dimensional dynamical XY model: how birds fly together. Phys.\ Rev.\ Lett. {\bf 75}, 4326 (1995).

\bibitem{Chate1}
G. Gr\'{e}goire and H. Chat\'{e}, Onset of Collective and Cohesive Motion. Phys.\ Rev.\ Lett. {\bf 92}, 025702 (2004).

\bibitem{Chate2}
H. Chat\'{e}, F. Ginelli, G. Gr\'{e}goire, and F. Raynaud, Collective motion of self-propelled particles interacting without cohesion. Phys.\ Rev.\ E {\bf 77}, 046113 (2008).



	
		
	\bibitem{TT3}
	J.~Toner, and Y.~Tu, Flocks, herds, and schools: a quantitative theory of flocking. Phys.\ Rev.\ E {\bf 58,} 4828(1998).
	
	\bibitem{birdrev} J.\ Toner,   Y. Tu, and S.
Ramaswamy, Hydrodynamics and phases of flocks,  Ann.\ Phys.\  {\bf 318}, 170 (2005).





\bibitem{NL} J. Toner, A reanalysis of the hydrodynamic theory of fluid, polar-ordered flocks,
Phys. Rev. E {\bf 86}, 031918 (2012).

\bibitem{chen_njp_2018}
	L.~Chen, C.~F.~Lee, and J.~Toner, Incompressible polar active fluids in the moving phase in dimensions $d>2$. New J.~Phys. {\bf 20}, 113035 (2018)
	
 \bibitem{chen_nc_2016}	
    L.~Chen, C.~F.~Lee, and J.~Toner, Mapping two-dimensional polar active fluids to
    two-dimensional soap and one-dimensional sandblasting. Nat.~Commun. {\bf 7}, 12215 (2016).
	
	\bibitem[{\citenamefont{Toner}(2012)}]{Malthus}
	\bibinfo{author}{\bibfnamefont{J.}~\bibnamefont{Toner}},
	{Birth, Death, and Flight: A Theory of Malthusian Flocks.}
	\bibinfo{journal}{Physical Review Letters} \textbf{\bibinfo{volume}{108}},
	\bibinfo{pages}{088102} (\bibinfo{year}{2012}).

\bibitem{lattice}   One could eliminate crystal field effects altogether by allowing the spins to perform diffusive motion (that is uncoupled to the spins' directions) confined  to the unit cells around their lattice sites so that the spin spatial distribution approximates a continuum at long time.

\bibitem{TT2}
		Y. Tu, M. Ulm, $\&$ J. Toner,  Sound waves and the absence of galilean invariance in flocks.
		{\it Phys.\ Rev.\ Lett.} {\bf 80}, 4819 (1998).
	


\bibitem{Malthus_1789}
T.\ R.\ Malthus, An Essay on the Principle of Population, edited by J. Johnson (St. Paul's Churchyard, London, 1798).

\bibitem{short} 	L.~Chen, C.~F.~Lee, and J.~Toner, {
 associated letter submitted to PRL; preprint: arXiv:2001.01300.}

\bibitem{GNF}
F. Ginelli,  The Physics of the Vicsek model, EPJ ST {\bf 225}, 2099 (2016).

{
\bibitem{toner_jcp19}
J. Toner, Giant number fluctuations in dry active polar fluids: A shocking analogy with lightning rods. J. Chem. Phys. {\bf 150}, 154120 (2019).
}


\bibitem{FNS}
D.~Forster, D.R.~Nelson and M.J.~Stephen, Large-distance and long-time properties
of a randomly stirred fluid.
Phys. Rev. A {\bf 16,} 732 (1977).

\bibitem{foot1}
The diffusion coefficient $\mu_2$ is kept fixed simply by flowing to zero, as it can readily be seen to do for the choice of $z$ that we are making here.


\bibitem{Ma}
Ma S-K 2000 {\it Modern Theory of Critical Phenomena} (Boulder, CO:Westview)

\bibitem{TIMF}
{
D. R. Nelson, Crossover scaling functions and
renormalization-group trajectory integrals, Phys.
Rev. B 11, 3504 (1975).
}

\bibitem{irrel} This argument can easily be extended to include any one of the infinite number of {\it irrelevant} parameters that one could add to our starting model (e.g., terms involving, say, $\nabla^4\bu$). The precise value of $\ell^*$ could also depend on these parameters; all other dependence on these parameters would drop out of the problem, since they would renormalize to zero on the right hand side of (\ref{TIMF}).
	

	
	
	
	
	\bibitem[{\citenamefont{Ginelli and Chat{\'{e}}}(2010)}]{ginelli_prl10}
	\bibinfo{author}{\bibfnamefont{F.}~\bibnamefont{Ginelli}} \bibnamefont{and}
	\bibinfo{author}{\bibfnamefont{H.}~\bibnamefont{Chat{\'{e}}}}, {
	Relevance of Metric-Free Interactions in Flocking Phenomena.}
	\bibinfo{journal}{Physical Review Letters}
	 \textbf{\bibinfo{volume}{105}},
	\bibinfo{pages}{168103} (\bibinfo{year}{2010}).
	
	\bibitem[{\citenamefont{Peshkov et~al.}(2012)\citenamefont{Peshkov, Ngo,
			Bertin, Chat{\'{e}}, and Ginelli}}]{peshkov_prl12}
	\bibinfo{author}{\bibfnamefont{A.}~\bibnamefont{Peshkov}},
	\bibinfo{author}{\bibfnamefont{S.}~\bibnamefont{Ngo}},
	\bibinfo{author}{\bibfnamefont{E.}~\bibnamefont{Bertin}},
	\bibinfo{author}{\bibfnamefont{H.}~\bibnamefont{Chat{\'{e}}}},
	\bibnamefont{and} \bibinfo{author}{\bibfnamefont{F.}~\bibnamefont{Ginelli}},
	{Continuous Theory of Active Matter Systems with Metric-Free Interactions.}
	\bibinfo{journal}{Physical Review Letters} \textbf{\bibinfo{volume}{109}},
	\bibinfo{pages}{098101} (\bibinfo{year}{2012}).
	
	\bibitem[{\citenamefont{Mahault et~al.}(2018)\citenamefont{Mahault, Jiang,
			Bertin, Ma, Patelli, Shi, and Chat{\'{e}}}}]{mahault_prl18}
	\bibinfo{author}{\bibfnamefont{B.}~\bibnamefont{Mahault}},
	\bibinfo{author}{\bibfnamefont{X.-c.} \bibnamefont{Jiang}},
	\bibinfo{author}{\bibfnamefont{E.}~\bibnamefont{Bertin}},
	\bibinfo{author}{\bibfnamefont{Y.-q.} \bibnamefont{Ma}},
	\bibinfo{author}{\bibfnamefont{A.}~\bibnamefont{Patelli}},
	\bibinfo{author}{\bibfnamefont{X.-q.} \bibnamefont{Shi}}, \bibnamefont{and}
	\bibinfo{author}{\bibfnamefont{H.}~\bibnamefont{Chat{\'{e}}}},
	{
	Self-Propelled Particles with Velocity Reversals and Ferromagnetic Alignment: Active Matter Class with Second-Order Transition to Quasi-Long-Range Polar Order.}
	\bibinfo{journal}{Physical Review Letters} \textbf{\bibinfo{volume}{120}},
	\bibinfo{pages}{258002} (\bibinfo{year}{2018}).
	
	\bibitem[{\citenamefont{Nesbitt et~al.}(2019)\citenamefont{Nesbitt, Pruessner,
			and Lee}}]{nesbitt_a19}
	\bibinfo{author}{\bibfnamefont{D.}~\bibnamefont{Nesbitt}},
	\bibinfo{author}{\bibfnamefont{G.}~\bibnamefont{Pruessner}},
	\bibnamefont{and} \bibinfo{author}{\bibfnamefont{C.~F.} \bibnamefont{Lee}},
{Uncovering novel phase transitions in dense dry polar active fluids using a lattice Boltzmann method. Preprint: arxiv:1902.00530.}
	

	
	





\end{thebibliography}
\end{document}